\documentclass[submission, Phys]{SciPost}

\makeatletter
\@ifundefined{iflatexml}{%
    \newif\iflatexml
    \@ifundefined{lx@ApplyFunction}{}{\latexmltrue}%
}{}
\makeatother
\usepackage{iftex}

\usepackage[utf8]{inputenc}
\usepackage[T1]{fontenc}
\usepackage{lmodern}
\iflatexml\else
    \usepackage{newunicodechar}
    \newunicodechar{‐}{-}
    \newunicodechar{–}{\textendash}
\fi
\usepackage{amsmath}
\usepackage{amssymb}
\usepackage{bm}
\usepackage{mathtools}
\iflatexml
    \usepackage{keyval}
    \makeatletter
    \define@key{dmath}{compact}[]{}
    \define@key{dmath}{label}{\label{#1}}
    \makeatother
    \newenvironment{dmath}[1][]{\begin{equation}\setkeys{dmath}{#1}}{\end{equation}}
    \newenvironment{dmath*}[1][]{\begin{equation*}\setkeys{dmath}{#1}}{\end{equation*}}
\else
    \usepackage{breqn}
\fi
\usepackage[expansion=false]{microtype}
\iflatexml
    \newcommand{\orcidlink}[1]{\href{https://orcid.org/#1}{ORCID}}
    \def\doi#1{\href{https://doi.org/#1}{doi:#1}}
\else
    \usepackage{orcidlink}
\fi
\hypersetup{colorlinks}

\iflatexml
    \let\citep\cite
\else
    \newcommand{\citep}[1]{\cite{#1}}
\fi

\iflatexml
\ifdefined\formalizationhtmlmarkers
\newcommand{\formalizationhtmlmarker}[1]{\texttt{FORMALIZATION-MARK:#1}}
\else
\newcommand{\formalizationhtmlmarker}[1]{}
\fi
\newenvironment{formalized}[2]{}{}
\else
\definecolor{formalizednumber}{HTML}{365F66}
\newsavebox{\formalizedtagbox}
\newcommand{\formalizationhtmlmarker}[1]{}
\makeatletter
\renewcommand{\set@label}[2]{%
    \protected@edef\@currentlabel{#2}%
    \hyper@makecurrent{#1}%
    \Hy@raisedlink{\hyper@anchorstart{\@currentHref}\hyper@anchorend}%
}
\ifpdftex
    \newcommand{\formalizedtagannotation}[1]{%
        \pdfannot width \wd\formalizedtagbox
            height \ht\formalizedtagbox
            depth \dp\formalizedtagbox {#1}%
    }
\else
    \newcommand{\formalizedtagannotation}[1]{%
    }
\fi
\newcommand{\formalizedtagtooltip}[3]{%
    \begingroup
    \sbox{\formalizedtagbox}{#1}%
    \formalizedtagannotation{/Subtype /Widget /TU (#2) /T (formalization-#3)
        /C [] /FT /Btn /F 768 /Ff 65536 /H /N /BS << /W 0 >>}%
    \usebox{\formalizedtagbox}%
    \endgroup
}
\newenvironment{formalized}[2]{%
    \def\formalizedlabel{#1}%
    \def\formalizedtooltip{#2}%
    \renewcommand{\tagform@}[1]{%
        \maketag@@@{\formalizedtagtooltip{%
            \textcolor{formalizednumber}{(\ignorespaces##1\unskip\@@italiccorr)}%
        }{\formalizedtooltip}{\formalizedlabel}}%
    }%
    \renewcommand{\eqnumform}[1]{%
        \formalizedtagtooltip{%
            \textcolor{formalizednumber}{(\ignorespaces##1\unskip\@@italiccorr)}%
        }{\formalizedtooltip}{\formalizedlabel}%
    }%
}{}
\makeatother
\fi

\begin{document}

\begin{center}{\Large \textbf{
    Transmission representation of the Josephson free energy beyond the
    short-junction limit
}}\end{center}

\begin{center}
    Anton R. Akhmerov\,\orcidlink{0000-0001-8031-1340}$^{\star}$
\end{center}

\begin{center}
    Kavli Institute of Nanoscience, Delft University of Technology,
    P.O. Box 4056, 2600 GA Delft, The Netherlands\\[4pt]
    $^{\star}$\href{mailto:anharmonicity@antonakhmerov.org}{anharmonicity@antonakhmerov.org}
\end{center}

\begin{center}
    October 13, 2026
\end{center}
\section*{Abstract}

Beenakker's formula,
\(U(\varphi)-U(0)=\Delta\sum_n\bigl(1-\sqrt{1-\tau_n\sin^2(\varphi/2)}\bigr)\),
expresses the zero-temperature energy-phase relation of a short Josephson
junction through its normal-state transmission eigenvalues. We generalize this
transmission representation beyond the short-junction limit and express the
phase-dependent free energy of a time-reversal-symmetric \(s\)-wave junction
as a Matsubara sum over channels with frequency-dependent transmissions. The
representation constrains the small-phase expansion of the free energy. In
particular, we prove that the normalized ratio of its quartic and quadratic
coefficients satisfies
\(-1\le\alpha_r\le7/80\)
at finite temperature. We also construct a four-site junction with
\(\alpha_r>0\), which disproves the conjectured upper bound \(\alpha_r\le0\).
So that these results can
be trusted without repeating the derivations, every numbered equation is
formally verified in the Lean proof assistant.

\begin{quote}\textbf{AI disclosure.} I formulated the problem and directed the research. I produced both the research reported here and the manuscript text using GPT-5.2 through GPT-6.1, in the Codex harness and in their Pro versions through ChatGPT, and Claude Fable~5.1 and Claude Opus~5.5 in Claude Code. App.~\ref{appendix-development-workflow} describes how these models contributed. I have fully validated both the scientific content and the writing, for which I take full responsibility.\end{quote}

\section{Introduction}\label{introduction}

Josephson's original work \citep{Josephson62} on superconductive
tunnelling gave a universal sinusoidal current-phase relation, and
therefore a cosine energy-phase relation, for a tunnel junction.
Superconductor--normal--superconductor (SNS) junctions are much more
varied. Their energy-phase relations depend on the full phase-dependent
spectrum of the weak link and generally do not admit such a universal
form. A major simplification reappears in the short-junction limit.
There, Beenakker showed \citep{Beenakker91Universal} that the
energy-phase relation is
{}
\begin{formalized}{eq:beenakker-short}{Rendered from Lean 4 theorem RmtNlin.AllChannels.manuscriptFiniteShortJunctionEnergy_eq_beenakker (RmtNlin/AllChannelViews.lean:24). 2 direct premises; rendered forms and their uses are in the generated assumption ledger. 7 project definitions are exposed in the assumptions and definitions ledger. Zero direct premises does not mean unrestricted model objects.}
\begin{dmath}[compact,label={eq:beenakker-short}]
U_{\mathrm{short}}(\varphi{}) - U_{\mathrm{short}}(0) = \Delta{}\,\sum_{n}\left(1 - \sqrt{1 - \tau{}_{n}\,\sin^{2} \frac{\varphi{}}{2}}\right),
\end{dmath}
\end{formalized}%
\noindent
where \(n\) labels the channels and \(\tau_n\) are the normal-state
transmission eigenvalues of the weak link.
Equation~\eqref{eq:beenakker-short} gives the full Josephson potential as a
sum over independent transport channels. This potential underlies weak-link
qubit models \citep{Larsen15, Lange15, Kringhj18}: after quantizing the
phase, its non-quadratic terms give the qubit anharmonicity.

General Green-function and scattering-matrix formulations apply at arbitrary
junction length \citep{Furusaki91,Titov01}, but to our knowledge they do not
reduce the phase dependence to a sum over independent scalar channels. The
phase-dependent spectrum away from the short-junction limit is not parametrized
by a fixed set of normal-state transmission eigenvalues, so the local form of
the free energy,
\(F(\varphi)-F(0)=c_2\varphi^2+c_4\varphi^4+O(\varphi^6)\), does not by itself expose
the signs or allowed range of \(c_2\) and \(c_4\). Titov, Jacquod, and
Beenakker showed \citep{Titov01} in a scattering approach that a
noninteracting time-reversal-symmetric disordered junction has a
Josephson-energy minimum at \(\varphi=0\), so \(c_2\ge0\). The quartic
coefficient \(c_4\), or equivalently the normalized ratio
\(\alpha_r=12c_4/c_2\), is harder to constrain. Based on numerical
perturbation theory of random-matrix Hamiltonians, we and our
collaborators conjectured \citep{Fatemi25} that this ratio lies
between \(-1\) and \(0\).

The short-junction formula raises the question of whether a channel description
survives at finite junction length. We show that it does,
but the channel weights become functions of Matsubara frequency rather
than fixed normal-state transmissions. We derive this representation for
finite-dimensional time-reversal-symmetric \(s\)-wave Bogoliubov--de Gennes
Hamiltonians, both with and without \(SU(2)\) spin rotation symmetry. It reduces to
Beenakker's formula in the short-junction limit, and near that limit it gives
the first finite-length correction for any normal region whose scattering
depends smoothly on energy near the Fermi level; its logarithmic term, which suppresses the
response in previously studied models, can have either sign. Away from that limit, the same
representation gives an infinite family of constraints on higher-order
nonlinearities. The frequency dependence of the transmission channels bounds
the normalized quartic coefficient by
\(7/80\)
at finite temperature and by
\((16 - 5\,\pi)/[4\,(4 - \pi)]
\approx
0.085\)
at zero temperature.
Both bounds leave room for positive \(c_4\), and we find a four-site
junction with \(\alpha_r>0\). This rules out the older conjectured upper bound
\(\alpha_r\le0\). Every numbered equation in this manuscript is printed from a
statement checked in the Lean proof assistant \citep{Moura21, Community20}, and
App.~\ref{appendix-verification} states what this verification covers.
App.~\ref{appendix-development-workflow} describes how AI models contributed to
these results.

\section{Derivation of the transmission
representation}\label{derivation-of-the-transmission-representation}

The derivation has three main steps. We first convert the
phase-dependent part of the free energy into a Matsubara sum and reduce
each term to finite-dimensional channel spaces obtained by factoring the pairing
blocks. Particle-hole symmetry fixes the allowed form of those
factorizations. We then impose time-reversal symmetry and \(s\)-wave
pairing, which add a transpose relation to the matrix obtained after
this reduction. In the two-terminal case, this relation pairs the
relevant generalized eigenvalues with opposite signs and yields the
final compact formula.

\subsection{Problem definition}\label{problem-definition}

We work in the conjugate-hole basis and split the pairing by
superconductor. The Bogoliubov--de Gennes Hamiltonian is
{}
\begin{formalized}{eq:bdg-hamiltonian}{Rendered from Lean 4 theorem RmtNlin.manuscriptBdgHamiltonian (RmtNlin/ParticleHoleSymmetry.lean:122). 3 project definitions are exposed in the assumptions and definitions ledger. Zero direct premises does not mean unrestricted model objects.}
\begin{dmath}[compact,label={eq:bdg-hamiltonian}]
H = \begin{pmatrix}h&\Delta\\-\eta{}\,\Delta^{\ast}&-h^{\ast}\end{pmatrix}.
\end{dmath}
\end{formalized}%
\noindent
Here \(h\) is the normal-state Hamiltonian and
\(\Delta = \sum_{i}e^{i{}\varphi{}_{i}}\,\Delta{}_{i}\) is the
superconducting pairing. The index \(i\) labels different
superconductors, so each phase \(\varphi_i\) enters only through its own
pairing block. The superconducting pairing is constrained by
particle-hole symmetry of the Hamiltonian, which in this basis is
{}
\begin{formalized}{eq:particle-hole-symmetry}{Rendered from Lean 4 theorem RmtNlin.manuscriptParticleHoleSymmetry (RmtNlin/ParticleHoleSymmetry.lean:134). 1 direct premise; rendered forms and their uses are in the generated assumption ledger. 8 project definitions are exposed in the assumptions and definitions ledger. Zero direct premises does not mean unrestricted model objects.}
\begin{dmath}[compact,label={eq:particle-hole-symmetry}]
\mathcal P\,H\,\mathcal P^{-1} = -H,\qquad \mathcal P = \begin{pmatrix}0&1\\\eta{}&0\end{pmatrix}\,K,
\end{dmath}
\end{formalized}%
\noindent
with \(K\) complex conjugation. The sign \(\eta=-1\) corresponds to the
spin-singlet, \(SU(2)\)-invariant case, while \(\eta=+1\) denotes the case without
\(SU(2)\) spin-rotation symmetry. Hermiticity of this
particle-hole-symmetric block, applied separately to the independent phase
factors, gives
{}
\begin{formalized}{eq:pairing-transpose-symmetry}{Rendered from Lean 4 theorem RmtNlin.manuscriptPairingTransposeSymmetry (RmtNlin/PairingFactorization.lean:511). 2 direct premises; rendered forms and their uses are in the generated assumption ledger. 4 project definitions are exposed in the assumptions and definitions ledger. Zero direct premises does not mean unrestricted model objects.}
\begin{dmath}[compact,label={eq:pairing-transpose-symmetry}]
\Delta{}^{T} = -\eta{}\,\Delta{}.
\end{dmath}
\end{formalized}%
\noindent
Each pairing block \(\Delta_i\) typically acts only on the degrees of
freedom that belong to superconductor \(i\), so it is naturally
rank-deficient. We therefore use a decomposition that reveals that rank
while keeping the symmetry \(\Delta_i^T=-\eta\Delta_i\) explicit: the
Takagi factorization for \(\eta=-1\) and the skew-Takagi, or Youla,
factorization for \(\eta=+1\), the analogues of the singular-value
decomposition for complex symmetric and antisymmetric matrices. In both cases,
{}
\begin{formalized}{eq:pairing-channel-factorization}{Rendered from Lean 4 theorem RmtNlin.manuscriptPairingChannelFactorization (RmtNlin/PairingFactorization.lean:456). 2 direct premises; rendered forms and their uses are in the generated assumption ledger. 8 project definitions are exposed in the assumptions and definitions ledger. Zero direct premises does not mean unrestricted model objects.}
\begin{dmath}[compact,label={eq:pairing-channel-factorization}]
\Delta{}_{i} = L\,Q\,L^{T},\qquad Q = \begin{cases}\textstyle\bigoplus i\sigma_y & \text{if } \eta{} = 1\\ 1 & \text{otherwise}\end{cases}.
\end{dmath}
\end{formalized}%
\noindent
The auxiliary space on which \(Q\) acts is the channel space for
superconductor \(i\). For \(\eta=+1\), this space splits into \(2\times2\)
blocks on which \(Q\) acts as \(i\sigma_y\). For several
superconductors, the total channel space is the direct sum of these spaces,
and the individual maps combine into a single map \(L\) from it to electron
space. The factorization on the total channel space is
{}
\begin{formalized}{eq:global-pairing-factorization}{Rendered from Lean 4 theorem RmtNlin.manuscriptGlobalPairingFactorization (RmtNlin/PairingFactorization.lean:489). 1 project definitions are exposed in the assumptions and definitions ledger. Zero direct premises does not mean unrestricted model objects.}
\begin{dmath}[compact,label={eq:global-pairing-factorization}]
\Delta_{i} = L\,P_{i}\,Q\,L^{T}.
\end{dmath}
\end{formalized}%
\noindent
Here, the \(P_i\) are the orthogonal projectors onto the
channel spaces of the individual superconductors, with \(\sum_i P_i=1\) and
\([P_i,Q]=0\).

Because \(\Delta=\sum_i e^{i\varphi_i}\Delta_i\), this
factorization puts the superconducting phases on the channel-space side
of the problem:
{}
\begin{formalized}{eq:delta-factorization}{Rendered from Lean 4 theorem RmtNlin.manuscriptTerminalPairingSum_eq_channelPhase (RmtNlin/PairingFactorization.lean:558). 3 project definitions are exposed in the assumptions and definitions ledger. Zero direct premises does not mean unrestricted model objects.}
\begin{dmath}[compact,label={eq:delta-factorization}]
\Phi(e^{i\varphi{}}) = \sum_{i}e^{i{}\varphi{}_{i}}\,P_{i},\qquad \Delta = L\,\Phi(e^{i\varphi{}})\,Q\,L^{T}.
\end{dmath}
\end{formalized}%
\noindent
Our goal is to compute the finite-temperature free energy
{}
\begin{formalized}{eq:free-energy}{Rendered from Lean 4 theorem RmtNlin.manuscriptPositiveSpectrumFreeEnergy (RmtNlin/FoundationalCorrespondence.lean:317). 1 direct premise; rendered forms and their uses are in the generated assumption ledger. 3 project definitions are exposed in the assumptions and definitions ledger. Zero direct premises does not mean unrestricted model objects.}
\begin{dmath}[compact,label={eq:free-energy}]
F_{\beta{}}(H) = -\frac{1}{\beta{}}\,\operatorname{Tr} \log\left(2\,\cosh \frac{\beta{}\,H}{2}\right) = -\frac{1}{\beta{}}\,\sum_{k}\log\left(2\,\cosh \frac{\beta{}\,\varepsilon_{k}\left(H\right)}{2}\right),
\end{dmath}
\end{formalized}%
\noindent
and its zero-temperature limit,
{}
\begin{formalized}{eq:ground-state-energy}{Rendered from Lean 4 theorem RmtNlin.manuscriptPositiveSpectrumGroundEnergy (RmtNlin/FoundationalCorrespondence.lean:332). 1 direct premise; rendered forms and their uses are in the generated assumption ledger. 5 project definitions are exposed in the assumptions and definitions ledger. Zero direct premises does not mean unrestricted model objects.}
\begin{dmath}[compact,label={eq:ground-state-energy}]
\lim_{\beta{}\to\infty}F_{\beta{}}(H)=U(H),\qquad U(H) = -\frac{1}{2}\,\operatorname{Tr} \lvert H\rvert.
\end{dmath}
\end{formalized}%
\noindent
We apply the standard Matsubara summation approach and use the product
identity
{}
\begin{formalized}{eq:scalar-cosh-product}{Rendered from Lean 4 theorem RmtNlin.manuscriptFermionicMatsubaraProduct (RmtNlin/MatsubaraProduct.lean:200). 1 direct premise; rendered forms and their uses are in the generated assumption ledger. 3 project definitions are exposed in the assumptions and definitions ledger. Zero direct premises does not mean unrestricted model objects.}
\begin{dmath}[compact,label={eq:scalar-cosh-product}]
\cosh \frac{\beta{}\,a}{2} = \prod_{m=0}^{\infty}\left(1 + \frac{a^{2}}{\omega_{m}^{2}}\right),
\end{dmath}
\end{formalized}%
\noindent
with positive fermionic Matsubara frequencies
\(\omega_m=(2m+1)\pi/\beta\). For these positive frequencies,
\(\det(\omega_m^2+H^2)=|\det(i\omega_m-H)|^2\).
With \(H_0=H(\varphi=0)\), the product identity gives
the Matsubara sum for the free-energy difference:
{}
\begin{formalized}{eq:free-energy-matsubara}{Rendered from Lean 4 theorem RmtNlin.manuscriptPositiveSpectrumFreeEnergyMatsubara (RmtNlin/FoundationalCorrespondence.lean:346). 3 direct premises; rendered forms and their uses are in the generated assumption ledger. 9 project definitions are exposed in the assumptions and definitions ledger. Zero direct premises does not mean unrestricted model objects.}
\begin{dmath}[compact,label={eq:free-energy-matsubara}]
F_{\beta{}}(H) - F_{\beta{}}(H_{0}) = -\frac{2}{\beta{}}\,\sum_{m=0}^{\infty}\log \left\lvert \frac{\det(i{}\,\omega_{m} - H)}{\det(i{}\,\omega_{m} - H_{0})}\right\rvert.
\end{dmath}
\end{formalized}%
\noindent
In the limit \(\beta\to\infty\), the Matsubara sum converges to the
corresponding integral, which gives the zero-temperature formula
{}
\begin{formalized}{eq:logdet}{Rendered from Lean 4 theorem RmtNlin.manuscriptPositiveSpectrumLogDetIntegral (RmtNlin/FoundationalCorrespondence.lean:358). 2 direct premises; rendered forms and their uses are in the generated assumption ledger. 6 project definitions are exposed in the assumptions and definitions ledger. Zero direct premises does not mean unrestricted model objects.}
\begin{dmath}[compact,label={eq:logdet}]
U(H) - U(H_{0}) = -\frac{1}{\pi}\,\int_{0}^{\infty}\log \left\lvert \frac{\det(i{}\,\omega{} - H)}{\det(i{}\,\omega{} - H_{0})}\right\rvert\,\mathrm d\omega{}.
\end{dmath}
\end{formalized}%
\noindent
\subsection{Scattering form of the Matsubara
formula}\label{scattering-form-of-the-matsubara-formula}

From now on we consider an SNS junction with left and right superconductors
and choose a gauge with phases \(0\) and \(\varphi\), so the phase dependence
is
{}
\begin{formalized}{eq:sns-phase-dependence}{Rendered from Lean 4 theorem RmtNlin.manuscriptSnsPhaseDependence (RmtNlin/TwoTerminalDeterminant.lean:171). 7 project definitions are exposed in the assumptions and definitions ledger. Zero direct premises does not mean unrestricted model objects.}
\begin{dmath}[compact,label={eq:sns-phase-dependence}]
\Delta{} = \Delta{}_{L} + e^{i{}\varphi{}}\,\Delta{}_{R},\qquad \Phi(e^{i{}\varphi{}}) = P_{L} + e^{i{}\varphi{}}\,P_{R}.
\end{dmath}
\end{formalized}%
\noindent
To expose the channel structure of the free energy, we reduce the full
BdG determinant \(\det[i\omega-H(\varphi)]\) to a scattering problem
on channel space.
Using the factorization in Eq.~\eqref{eq:delta-factorization}, we apply the
virtual-leads construction of Ref. \citep{Waintal24} and introduce
auxiliary currents on the superconducting channels. For a vector
satisfying \([i\omega-H(\varphi)](\psi_e,\psi_h)^T=0\), we define
{}
\begin{formalized}{eq:auxiliary-current-definitions}{Rendered from Lean 4 theorem RmtNlin.manuscriptAuxiliaryCurrentDefinitions (RmtNlin/FoundationalCorrespondence.lean:388). 9 project definitions are exposed in the assumptions and definitions ledger. Zero direct premises does not mean unrestricted model objects.}
\begin{dmath}[compact,label={eq:auxiliary-current-definitions}]
j_{\mathrm{eh}} = \frac{1}{\sqrt{2}}\,(i{}\,L^{\dagger}\,\psi{}_{e} - \Phi(e^{i{}\varphi{}})\,Q\,L^{T}\,\psi{}_{h}),\qquad j_{\mathrm{he}} = \frac{1}{\sqrt{2}}\,(L^{\dagger}\,\psi{}_{e} - i{}\,\Phi(e^{i{}\varphi{}})\,Q\,L^{T}\,\psi{}_{h}).
\end{dmath}
\end{formalized}%
\noindent
With these definitions, the BdG equations become
{}
\begin{formalized}{eq:matsubara-linear-system}{Rendered from Lean 4 definition RmtNlin.manuscriptMatsubaraLinearSystem (RmtNlin/FoundationalCorrespondence.lean:437). 8 project definitions are exposed in the assumptions and definitions ledger. Zero direct premises does not mean unrestricted model objects.}
\begin{dmath}[compact,label={eq:matsubara-linear-system}]
\mathcal M_{\omega{}}\,\begin{pmatrix}\psi{}_{e}\\\psi{}_{h}\\j_{\mathrm{eh}}\\j_{\mathrm{he}}\end{pmatrix} = 0,
\end{dmath}
\end{formalized}%
\noindent
with
{}
\begin{formalized}{eq:expanded-matsubara-matrix}{Rendered from Lean 4 theorem RmtNlin.manuscriptConcreteExpandedMatsubaraMatrix (RmtNlin/FoundationalCorrespondence.lean:447). 14 project definitions are exposed in the assumptions and definitions ledger. Zero direct premises does not mean unrestricted model objects.}
\begin{dmath}[compact,label={eq:expanded-matsubara-matrix}]
\mathcal M_{\omega{}} = \begin{pmatrix}i{}\,\omega{} - h + i{}\,L\,L^{\dagger}&0&0&-i{}\,\sqrt{2}\,L\\0&i{}\,\omega{} + h^{\ast} + i{}\,L^{\ast}\,L^{T}&i{}\,\sqrt{2}\,L^{\ast}\,Q^{\dagger}\,\Phi(e^{-i{}\varphi{}})&0\\-i{}\,L^{\dagger}&0&\frac{1}{\sqrt{2}}&i{}\,\frac{1}{\sqrt{2}}\\0&i{}\,\Phi(e^{i{}\varphi{}})\,Q\,L^{T}&i{}\,\frac{1}{\sqrt{2}}&\frac{1}{\sqrt{2}}\end{pmatrix}.
\end{dmath}
\end{formalized}%
\noindent
This enlarged system lets us eliminate either set of
variables. Eliminating the auxiliary currents returns the original BdG
block, because the lower-right block has determinant \(1\):
{}
\begin{formalized}{eq:current-elimination-determinant}{Rendered from Lean 4 theorem RmtNlin.manuscriptConcreteExpandedMatsubaraDeterminant (RmtNlin/FoundationalCorrespondence.lean:1013). 4 direct premises; rendered forms and their uses are in the generated assumption ledger. 18 project definitions are exposed in the assumptions and definitions ledger. Zero direct premises does not mean unrestricted model objects.}
\begin{dmath}[compact,label={eq:current-elimination-determinant}]
\det \mathcal M_{\omega{}} = \det(i{}\,\omega{} - H).
\end{dmath}
\end{formalized}%
\noindent
Eliminating the BdG amplitudes instead leaves a determinant entirely in
channel space:
{}
\begin{formalized}{eq:psi-elimination-determinant}{Rendered from Lean 4 theorem RmtNlin.manuscriptConcreteWavefunctionElimination (RmtNlin/FoundationalCorrespondence.lean:1045). 5 direct premises; rendered forms and their uses are in the generated assumption ledger. 22 project definitions are exposed in the assumptions and definitions ledger. Zero direct premises does not mean unrestricted model objects.}
\begin{dmath}[compact,label={eq:psi-elimination-determinant}]
\det \mathcal M_{\omega{}} = (\det A_{\omega{}})\,\det(-A_{\omega{}}^{\ast})\,\det \begin{pmatrix}\frac{1}{\sqrt{2}}&i{}\,\frac{1}{\sqrt{2}}\,S\,Q^{\dagger}\\-\eta{}\,i{}\,\frac{1}{\sqrt{2}}\,\Phi(e^{i{}\varphi{}})\,Q\,S^{\ast}\,\Phi(e^{-i{}\varphi{}})&\frac{1}{\sqrt{2}}\end{pmatrix},
\end{dmath}
\end{formalized}%
\noindent
where
{}
\begin{formalized}{eq:reduced-scattering}{Rendered from Lean 4 theorem RmtNlin.manuscriptScatteringPencilAndReduced (RmtNlin/ResolventVelocity.lean:588). 4 project definitions are exposed in the assumptions and definitions ledger. Zero direct premises does not mean unrestricted model objects.}
\begin{dmath}[compact,label={eq:reduced-scattering}]
A_{\omega{}} = i{}\,\omega{} - h + i{}\,L\,L^{\dagger},\qquad S_{\omega{}} = (1 - 2\,i{}\,L^{\dagger}\,A_{\omega{}}^{-1}\,L)\,Q.
\end{dmath}
\end{formalized}%
\noindent
Since \(A_\omega\) is independent of \(\varphi\), its determinant
cancels in the ratio between phase \(\varphi\) and phase \(0\).
Comparing the two eliminations and using \([Q,\Phi(e^{i\varphi})]=0\) and
\((Q^\dagger)^2=-\eta\) then gives
{}
\begin{formalized}{eq:eta-det-ratio}{Rendered from Lean 4 theorem RmtNlin.manuscriptPhysicalEtaDetRatioView (RmtNlin/PhysicalDeterminantViews.lean:13). 6 direct premises; rendered forms and their uses are in the generated assumption ledger. 28 project definitions are exposed in the assumptions and definitions ledger. Zero direct premises does not mean unrestricted model objects.}
\begin{dmath}[compact,label={eq:eta-det-ratio}]
\frac{\det[i{}\,\omega{} - H(\varphi{})]}{\det[i{}\,\omega{} - H(0)]} = \frac{\det(1 - \eta{}\,\Phi(e^{-i{}\varphi{}})\,S_{\omega{}}\,\Phi(e^{i{}\varphi{}})\,S_{\omega{}}^{\ast})}{\det(1 - \eta{}\,S_{\omega{}}\,S_{\omega{}}^{\ast})}.
\end{dmath}
\end{formalized}%
\noindent
Equation~\eqref{eq:eta-det-ratio} is the Matsubara scattering
determinant of Brouwer and Beenakker \citep{Brouwer97}; Repin, Chen, and
Nazarov used its multiterminal form to study the low-frequency response
tensor \citep{Repin19}.

\subsection{Time-reversal-symmetric \(s\)-wave junctions}\label{time-reversal-symmetric-s-wave-junctions}

In a time-reversal-symmetric \(s\)-wave junction, the scattering matrix
obeys an additional transpose symmetry. To derive this relation, we
consider the time-reversal symmetry \(\mathcal T=U_TK\), where
{}
\begin{formalized}{eq:time-reversal-operator-square}{Rendered from Lean 4 theorem RmtNlin.manuscriptTimeReversalOperator (RmtNlin/PairingFactorization.lean:412). 1 direct premise; rendered forms and their uses are in the generated assumption ledger. 4 project definitions are exposed in the assumptions and definitions ledger. Zero direct premises does not mean unrestricted model objects.}
\begin{dmath}[compact,label={eq:time-reversal-operator-square}]
U_T = \begin{cases}1 & \text{if } \eta{} = -1\\ \textstyle\bigoplus i\sigma_y & \text{otherwise}\end{cases}.
\end{dmath}
\end{formalized}%
\noindent
At zero phase, time-reversal symmetry imposes
{}
\begin{formalized}{eq:trs-blocks}{Rendered from Lean 4 definition RmtNlin.manuscriptTimeReversalInvariantBlocks (RmtNlin/TRSScattering.lean:36). 2 project definitions are exposed in the assumptions and definitions ledger. Zero direct premises does not mean unrestricted model objects.}
\begin{dmath}[compact,label={eq:trs-blocks}]
U_{T}\,h^{\ast}\,U_{T}^{\dagger} = h,\qquad U_{T}\,\Delta{}_{i}^{\ast}\,U_{T}^{T} = \Delta{}_{i}.
\end{dmath}
\end{formalized}%
\noindent
The case \(\eta=-1\) has \(U_T=1\), so \(\Delta_i\) is real
symmetric. The case \(\eta=+1\) has \(U_T^T=-U_T\)
and \(\Delta_i^T=-\Delta_i\). The \(s\)-wave condition makes
\(\Delta_i U_T^\dagger\) positive semidefinite. A square-root
factorization in a doubled channel space, real for \(\eta=-1\) and
self-dual for \(\eta=+1\), gives a channel basis satisfying
{}
\begin{formalized}{eq:trs-channel-basis-choice}{Rendered from Lean 4 theorem RmtNlin.manuscriptTrsAdaptedChannelBasis (RmtNlin/FoundationalCorrespondence.lean:1095). 1 direct premise; rendered forms and their uses are in the generated assumption ledger. 6 project definitions are exposed in the assumptions and definitions ledger. Zero direct premises does not mean unrestricted model objects.}
\begin{dmath}[compact,label={eq:trs-channel-basis-choice}]
\Delta{}\,U_{T}^{\dagger} = L\,L^{\dagger},\qquad U_{T}\,L^{\ast} = L\,Q.
\end{dmath}
\end{formalized}%
\noindent
In this channel basis, the product \(LL^\dagger\) obeys
{}
\begin{formalized}{eq:trs-lldagger-transpose-chain}{Rendered from Lean 4 theorem RmtNlin.manuscriptTrsChannelGramTransposeChain (RmtNlin/TRSScattering.lean:107). 1 direct premise; rendered forms and their uses are in the generated assumption ledger. 8 project definitions are exposed in the assumptions and definitions ledger. Zero direct premises does not mean unrestricted model objects.}
\begin{dmath}[compact,label={eq:trs-lldagger-transpose-chain}]
U_{T}\,(L\,L^{\dagger})^{T}\,U_{T}^{\dagger} = U_{T}\,L^{\ast}\,L^{T}\,U_{T}^{\dagger} = L\,Q\,Q^{\dagger}\,L^{\dagger} = L\,L^{\dagger},
\end{dmath}
\end{formalized}%
\noindent
so \(U_T A_\omega^T U_T^\dagger=A_\omega\). The
scattering matrix \(S=S_\omega\) then satisfies
{}
\begin{formalized}{eq:trs-symmetry}{Rendered from Lean 4 theorem RmtNlin.manuscriptReducedScatteringFamilyTranspose (RmtNlin/ResolventVelocity.lean:180). 5 direct premises; rendered forms and their uses are in the generated assumption ledger. 16 project definitions are exposed in the assumptions and definitions ledger. Zero direct premises does not mean unrestricted model objects.}
\begin{dmath}[compact,label={eq:trs-symmetry}]
S^{T} = Q^{T} - 2\,i{}\,Q^{T}\,L^{T}\,(A_{\omega{}}^{-1})^{T}\,L^{\ast} = Q^{T} - 2\,i{}\,Q^{T}\,Q^{\dagger}\,L^{\dagger}\,A_{\omega{}}^{-1}\,L\,Q = -\eta{}\,Q + 2\,i{}\,\eta{}\,L^{\dagger}\,A_{\omega{}}^{-1}\,L\,Q = -\eta{}\,S.
\end{dmath}
\end{formalized}%
\noindent
Using Eq.~\eqref{eq:trs-symmetry} removes the remaining sign that depends on
the symmetry class and transforms the ratio in
Eq.~\eqref{eq:eta-det-ratio} into
{}
\begin{formalized}{eq:trs-det-ratio}{Rendered from Lean 4 theorem RmtNlin.manuscriptPhysicalTrsDetRatio (RmtNlin/PhysicalDeterminantViews.lean:52). 9 direct premises; rendered forms and their uses are in the generated assumption ledger. 36 project definitions are exposed in the assumptions and definitions ledger. Zero direct premises does not mean unrestricted model objects.}
\begin{dmath}[compact,label={eq:trs-det-ratio}]
\frac{\det[i{}\,\omega{} - H(\varphi{})]}{\det[i{}\,\omega{} - H(0)]} = \frac{\det(1 + \Phi(e^{-i{}\varphi{}})\,S_{\omega{}}^{\dagger}\,\Phi(e^{i{}\varphi{}})\,S_{\omega{}})}{\det(1 + S_{\omega{}}^{\dagger}\,S_{\omega{}})}.
\end{dmath}
\end{formalized}%
\noindent
\subsection{Transmission
representation}\label{transmission-representation}

To extract one scalar factor per channel from the determinant ratio
in Eq.~\eqref{eq:trs-det-ratio}, we separate the left and right channel
spaces by introducing the signature \(J\):
{}
\begin{formalized}{eq:two-terminal-projectors}{Rendered from Lean 4 theorem RmtNlin.manuscriptCanonicalTwoTerminalProjectors (RmtNlin/TwoTerminalDeterminant.lean:131). 9 project definitions are exposed in the assumptions and definitions ledger. Zero direct premises does not mean unrestricted model objects.}
\begin{dmath}[compact,label={eq:two-terminal-projectors}]
J = P_{L} - P_{R},\qquad \Phi(e^{i{}\varphi{}}) = P_{L} + e^{i{}\varphi{}}\,P_{R} = \frac{1 + e^{i{}\varphi{}}}{2} + \frac{1 - e^{i{}\varphi{}}}{2}\,J.
\end{dmath}
\end{formalized}%
\noindent
The orthogonal projectors satisfy \(P_L+P_R=1\) on the total channel
space. We write \(r_L=\operatorname{rank}P_L\) and
\(r_R=\operatorname{rank}P_R\) for the dimensions of the left and right
channel spaces. After this substitution, the determinant depends on two
phase-independent matrices:
{}
\begin{formalized}{eq:d-and-f-definitions}{Rendered from Lean 4 theorem RmtNlin.manuscriptTwoTerminalDAndF (RmtNlin/PencilPairing.lean:37). 2 project definitions are exposed in the assumptions and definitions ledger. Zero direct premises does not mean unrestricted model objects.}
\begin{dmath}[compact,label={eq:d-and-f-definitions}]
D = 1 + S_{\omega{}}^{\dagger}\,S_{\omega{}},\qquad F = J + S_{\omega{}}^{\dagger}\,J\,S_{\omega{}}.
\end{dmath}
\end{formalized}%
\noindent
Because \(\det(P_L+e^{i\varphi}P_R)=e^{i r_R\varphi}\), the ratio of
determinants becomes
{}
\begin{formalized}{eq:det-ratio-df}{Rendered from Lean 4 theorem RmtNlin.manuscriptPhysicalDetRatioDF (RmtNlin/PhysicalDeterminantViews.lean:70). 9 direct premises; rendered forms and their uses are in the generated assumption ledger. 41 project definitions are exposed in the assumptions and definitions ledger. Zero direct premises does not mean unrestricted model objects.}
\begin{dmath}[compact,label={eq:det-ratio-df}]
\frac{\det[i{}\,\omega{} - H(\varphi{})]}{\det[i{}\,\omega{} - H(0)]} = e^{-i{}\,r_R\,\varphi{}}\,\frac{\det\left(\frac{1 + e^{i{}\varphi{}}}{2}\,D + \frac{1 - e^{i{}\varphi{}}}{2}\,F\right)}{\det D}.
\end{dmath}
\end{formalized}%
\noindent
Since \(D=1+S_\omega^\dagger S_\omega\succ0\), we
evaluate the determinant ratio in Eq.~\eqref{eq:det-ratio-df} in the
generalized eigenbasis
\(Fu=\mu Du\):
{}
\begin{formalized}{eq:d-f-generalized-eigenvalue-product}{Rendered from Lean 4 theorem RmtNlin.manuscriptPhysicalDetRatioEigenvalues (RmtNlin/PhysicalDeterminantViews.lean:94). 9 direct premises; rendered forms and their uses are in the generated assumption ledger. 46 project definitions are exposed in the assumptions and definitions ledger. Zero direct premises does not mean unrestricted model objects.}
\begin{dmath}[compact,label={eq:d-f-generalized-eigenvalue-product}]
\frac{\det[i{}\,\omega{} - H(\varphi{})]}{\det[i{}\,\omega{} - H(0)]} = e^{-i{}\,r_R\,\varphi{}}\,\prod_{j}\left(\frac{1 + e^{i{}\varphi{}}}{2} + \frac{1 - e^{i{}\varphi{}}}{2}\,\mu_{j}\right),
\end{dmath}
\end{formalized}%
\noindent
where the \(\mu_j\) are the generalized eigenvalues of the pair
\((F,D)\). The matrices \(D\pm F\) are positive semidefinite:
{}
\begin{formalized}{eq:d-plus-minus-f-positivity}{Rendered from Lean 4 theorem RmtNlin.manuscriptTwoTerminalDPlusMinusFPositivity (RmtNlin/PencilPositivity.lean:76). 1 direct premise; rendered forms and their uses are in the generated assumption ledger. 6 project definitions are exposed in the assumptions and definitions ledger. Zero direct premises does not mean unrestricted model objects.}
\begin{dmath}[compact,label={eq:d-plus-minus-f-positivity}]
D + F = 2\,(P_{L} + S_{\omega{}}^{\dagger}\,P_{L}\,S_{\omega{}}) \succeq 0,\qquad D - F = 2\,(P_{R} + S_{\omega{}}^{\dagger}\,P_{R}\,S_{\omega{}}) \succeq 0.
\end{dmath}
\end{formalized}%
\noindent
Thus \(-D\preceq F\preceq D\), and since \(D\succ0\), the generalized
eigenvalues obey \(-1\le\mu_j\le1\).

We now show that each nonendpoint eigenvalue is paired with its
negative. For a generalized eigenvector \(u\) at \(\mu\), with
\(|\mu|<1\), the eigenvalue equation reads
{}
\begin{formalized}{eq:paired-spectrum-eqn}{Rendered from Lean 4 theorem RmtNlin.manuscriptTwoTerminalGeneralizedEigenvector_iff_pairingResidual (RmtNlin/PencilPairing.lean:123). 5 project definitions are exposed in the assumptions and definitions ledger. Zero direct premises does not mean unrestricted model objects.}
\begin{dmath}[compact,label={eq:paired-spectrum-eqn}]
F\,u = \mu{}\,D\,u \iff\allowbreak (J - \mu{})\,u + S_{\omega{}}^{\dagger}\,(J - \mu{})\,S_{\omega{}}\,u = 0.
\end{dmath}
\end{formalized}%
\noindent
The candidate eigenvector at \(-\mu\) is \(\Theta_{\mu}u\), with the
pairing map
{}
\begin{formalized}{eq:mu-pairing-map-definition}{Rendered from Lean 4 theorem RmtNlin.manuscriptAntiunitaryPairingMap (RmtNlin/AntiunitaryPairing.lean:149). 5 direct premises; rendered forms and their uses are in the generated assumption ledger. 25 project definitions are exposed in the assumptions and definitions ledger. Zero direct premises does not mean unrestricted model objects.}
\begin{dmath}[compact,label={eq:mu-pairing-map-definition}]
\Theta_{\mu{}}(u) = \frac{1}{\sqrt{1 - \mu{}^{2}}}\,(J - \mu{})\,S_{\omega{}}^{\dagger}\,u^{\ast} = -\eta{}\,\frac{1}{\sqrt{1 - \mu{}^{2}}}\,(J - \mu{})\,(S_{\omega{}}\,u)^{\ast}.
\end{dmath}
\end{formalized}%
\noindent
Applying \(F+\mu D\) to \(\Theta_{\mu}u\), we obtain
{}
\begin{formalized}{eq:mu-pairing-residual-chain}{Rendered from Lean 4 theorem RmtNlin.manuscriptNormalizedAntiunitaryPairingResidualChain (RmtNlin/AntiunitaryPairing.lean:361). 7 direct premises; rendered forms and their uses are in the generated assumption ledger. 31 project definitions are exposed in the assumptions and definitions ledger. Zero direct premises does not mean unrestricted model objects.}
\begin{equation}\begin{aligned}
(F + \mu{}\,D)\,\Theta_{\mu{}}(u) &= \frac{1}{\sqrt{1 - \mu{}^{2}}}\,[(J + \mu{})\,(J - \mu{})\,S_{\omega{}}^{\dagger} + S_{\omega{}}^{\dagger}\,(J + \mu{})\,S_{\omega{}}\,(J - \mu{})\,S_{\omega{}}^{\dagger}]\,u^{\ast} \\
&= \frac{1}{\sqrt{1 - \mu{}^{2}}}\,S_{\omega{}}^{\dagger}\,(1 - \mu{}^{2} + (J + \mu{})\,S_{\omega{}}\,(J - \mu{})\,S_{\omega{}}^{\dagger})\,u^{\ast} = 0,
\end{aligned}\label{eq:mu-pairing-residual-chain}\end{equation}
\end{formalized}%
\noindent
where the last equality uses the conjugate of
Eq.~\eqref{eq:paired-spectrum-eqn}. Hence \(\Theta_{\mu}u\) is a
generalized eigenvector at \(-\mu\). This pairing is one-to-one because
composition with the map at \(-\mu\) gives
{}
\begin{formalized}{eq:mu-pairing-composition-chain}{Rendered from Lean 4 theorem RmtNlin.manuscriptNormalizedAntiunitaryPairingCompositionChain (RmtNlin/AntiunitaryPairing.lean:526). 8 direct premises; rendered forms and their uses are in the generated assumption ledger. 31 project definitions are exposed in the assumptions and definitions ledger. Zero direct premises does not mean unrestricted model objects.}
\begin{equation}\begin{aligned}
\Theta_{-\mu{}}(\Theta_{\mu{}}(u)) &= \frac{1}{1 - \mu{}^{2}}\,(J + \mu{})\,S_{\omega{}}^{\dagger}\,[(J - \mu{})\,S_{\omega{}}^{\dagger}\,u^{\ast}]^{\ast} \\
&= -\eta{}\,\frac{1}{1 - \mu{}^{2}}\,(J + \mu{})\,S_{\omega{}}^{\dagger}\,(J - \mu{})\,S_{\omega{}}\,u \\
&= \eta{}\,\frac{1}{1 - \mu{}^{2}}\,(J + \mu{})\,(J - \mu{})\,u = \eta{}\,u,
\end{aligned}\label{eq:mu-pairing-composition-chain}\end{equation}
\end{formalized}%
\noindent
where the second equality uses Eq.~\eqref{eq:trs-symmetry}, and the
third uses Eq.~\eqref{eq:paired-spectrum-eqn}. Thus the nonendpoint
generalized eigenvalues come in opposite-sign pairs for both
\(\eta=\pm1\).

The endpoints \(\mu=\pm1\) only produce \(T=0\) factors; the factor from
\(\mu=-1\) cancels against the prefactor in Eq.~\eqref{eq:det-ratio-df}.
We therefore keep the nontrivial pairs \(\pm\mu_n(\omega)\), of which
there are at most \(N=\min(r_L,r_R)\), and set \(T_n(\omega)=0\) on
the remaining channels:
{}
\begin{formalized}{eq:transmission-eigenvalue}{Rendered from Lean 4 theorem RmtNlin.manuscriptTwoTerminalTransmission_mem_unitInterval (RmtNlin/PencilPositivity.lean:171). 3 direct premises; rendered forms and their uses are in the generated assumption ledger. 4 project definitions are exposed in the assumptions and definitions ledger. Zero direct premises does not mean unrestricted model objects.}
\begin{dmath}[compact,label={eq:transmission-eigenvalue}]
T_{n}(\omega{}) = 1 - \mu{}_{n}(\omega{})^{2},\qquad 0 \le T_{n}(\omega{}) \le 1.
\end{dmath}
\end{formalized}%
\noindent
We call \(T_n(\omega)\) the Matsubara transmission eigenvalues. The next
section shows that \(T_n(0)\) equals the ordinary transmission eigenvalues
\(\tau_n\) of the normal-state scattering matrix. Using
the opposite-sign pairing, each nontrivial pair \(\pm\mu_n(\omega)\)
combines into a single scalar factor,
{}
\begin{formalized}{eq:paired-scalar-transmission-factor}{Rendered from Lean 4 theorem RmtNlin.manuscriptPairedPhaseFactor_exp_eq_transmission (RmtNlin/TransmissionRepresentation.lean:63). 4 project definitions are exposed in the assumptions and definitions ledger. Zero direct premises does not mean unrestricted model objects.}
\begin{dmath}[compact,label={eq:paired-scalar-transmission-factor}]
e^{-i{}\varphi{}}\,\frac{1 + e^{i{}\varphi{}} + (1 - e^{i{}\varphi{}})\,\mu{}_{n}(\omega{})}{2}\,\frac{1 + e^{i{}\varphi{}} - (1 - e^{i{}\varphi{}})\,\mu{}_{n}(\omega{})}{2} = 1 - T_{n}(\omega{})\,\sin^{2} \frac{\varphi{}}{2}.
\end{dmath}
\end{formalized}%
\noindent
Padding the paired channels with zero transmissions gives a phase-independent
family \(T_n(\omega)\), \(n=1,\ldots,r_L\), and the determinant ratio
{}
\begin{formalized}{eq:matsubara-product}{Rendered from Lean 4 theorem RmtNlin.AllChannels.manuscriptPhysicalSortedTransmissionProduct (RmtNlin/AllChannelViews.lean:78). 9 direct premises; rendered forms and their uses are in the generated assumption ledger. 65 project definitions are exposed in the assumptions and definitions ledger. Zero direct premises does not mean unrestricted model objects.}
\begin{dmath}[compact,label={eq:matsubara-product}]
\frac{\det[i{}\,\omega{} - H(\varphi{})]}{\det[i{}\,\omega{} - H(0)]} = \prod_{n}\left(1 - T_{n}(\omega{})\,\sin^{2} \frac{\varphi{}}{2}\right).
\end{dmath}
\end{formalized}%
\noindent
Substituting Eq.~\eqref{eq:matsubara-product} into
Eq.~\eqref{eq:free-energy-matsubara} gives the transmission
representation for the phase-dependent free energy:
{}
\begin{formalized}{eq:transmission-representation-beta}{Rendered from Lean 4 theorem RmtNlin.AllChannels.manuscriptPhysicalFreeEnergyDifference_eq_logSum (RmtNlin/AllChannelViews.lean:94). 9 direct premises; rendered forms and their uses are in the generated assumption ledger. 74 project definitions are exposed in the assumptions and definitions ledger. Zero direct premises does not mean unrestricted model objects.}
\begin{dmath}[compact,label={eq:transmission-representation-beta}]
F_{\beta{}}(H(\varphi{})) - F_{\beta{}}(H(0)) = -\frac{2}{\beta{}}\,\sum_{m=0}^{\infty}\sum_{n}\log\left(1 - T_{n}(\omega_{m})\,\sin^{2} \frac{\varphi{}}{2}\right).
\end{dmath}
\end{formalized}%
\noindent
Here the channel sum runs over \(n=1,\ldots,r_L\), and at most \(N\) of
these transmissions are nonzero.
Equation~\eqref{eq:transmission-representation-beta} is the central
result of the manuscript. It shows that all \(\varphi\)-dependence again
enters through \(\sin^2(\varphi/2)\), now weighted by frequency-dependent
transmission eigenvalues. Because \(0\le T_n(\omega_m)\le1\), each
summand is a monotone increasing function of \(\sin^2(\varphi/2)\).
Therefore the phase-dependent free energy, and in particular the
zero-temperature energy, is itself a monotone increasing function of
\(\sin^2(\varphi/2)\), so \(\varphi=0\) modulo \(2\pi\) is a global
free-energy minimum.

\section{Short-junction limit}\label{short-junction-limit}

The short-junction limit recovers Beenakker's expression: the dwell time
through the normal region is
negligible on the scale \(\Delta^{-1}\), so we approximate the scattering
matrix by
{}
\begin{formalized}{eq:short-scattering-limit}{Rendered from Lean 4 definition RmtNlin.shortJunctionScattering (RmtNlin/ShortJunctionChannels.lean:368). 2 project definitions are exposed in the assumptions and definitions ledger. Zero direct premises does not mean unrestricted model objects.}
\begin{dmath}[compact,label={eq:short-scattering-limit}]
S_{\omega{}} = a\left(\frac{\omega{}}{\Delta{}}\right)\,S_{0}.
\end{dmath}
\end{formalized}%
\noindent
Here \(S_0\) is the fixed unitary normal-state scattering matrix, which
we write in block form as
{}
\begin{formalized}{eq:short-junction-scattering-blocks}{Rendered from Lean 4 theorem RmtNlin.manuscriptShortJunctionScatteringBlocks (RmtNlin/ShortJunction.lean:291). 2 project definitions are exposed in the assumptions and definitions ledger. Zero direct premises does not mean unrestricted model objects.}
\begin{dmath}[compact,label={eq:short-junction-scattering-blocks}]
S_{0} = \begin{pmatrix}r&t'\\t&r'\end{pmatrix},\qquad a\left(\frac{\omega{}}{\Delta{}}\right) = \sqrt{1 + \left(\frac{\omega{}}{\Delta{}}\right)^{2}} - \frac{\omega{}}{\Delta{}}.
\end{dmath}
\end{formalized}%
\noindent
This approximation makes the frequency dependence explicit:
{}
\begin{formalized}{eq:short-junction-d-and-f}{Rendered from Lean 4 theorem RmtNlin.manuscriptShortJunctionDAndF (RmtNlin/ShortJunction.lean:303). 3 project definitions are exposed in the assumptions and definitions ledger. Zero direct premises does not mean unrestricted model objects.}
\begin{dmath}[compact,label={eq:short-junction-d-and-f}]
D = 1 + a\left(\frac{\omega{}}{\Delta{}}\right)^{2},\qquad F = J + a\left(\frac{\omega{}}{\Delta{}}\right)^{2}\,S_{0}^{\dagger}\,J\,S_{0}.
\end{dmath}
\end{formalized}%
\noindent
Because \(D\) is scalar in this approximation, the generalized
eigenvalue problem for a fixed channel, with \(\mu_\omega=\mu_n(\omega)\),
is an ordinary eigenvalue problem for \(F\):
{}
\begin{formalized}{eq:short-junction-generalized-eigenproblem}{Rendered from Lean 4 definition RmtNlin.manuscriptShortJunctionGeneralizedEigenproblem (RmtNlin/ShortJunction.lean:317). 2 project definitions are exposed in the assumptions and definitions ledger. Zero direct premises does not mean unrestricted model objects.}
\begin{dmath}[compact,label={eq:short-junction-generalized-eigenproblem}]
F\,v = \mu{}_{\omega{}}\,D\,v.
\end{dmath}
\end{formalized}%
\noindent
To identify the
relevant channels with the usual normal-state transmission channels, it
is convenient to square \(F\). Since \((S_0^\dagger J S_0)^2=1\),
{}
\begin{formalized}{eq:short-junction-f-squared-expansion}{Rendered from Lean 4 theorem RmtNlin.manuscriptShortJunctionF_sq (RmtNlin/ShortJunction.lean:141). 1 direct premise; rendered forms and their uses are in the generated assumption ledger. 8 project definitions are exposed in the assumptions and definitions ledger. Zero direct premises does not mean unrestricted model objects.}
\begin{dmath}[compact,label={eq:short-junction-f-squared-expansion}]
F^{2} = \left(J + a\left(\frac{\omega{}}{\Delta{}}\right)^{2}\,S_{0}^{\dagger}\,J\,S_{0}\right)^{2} = 1 + a\left(\frac{\omega{}}{\Delta{}}\right)^{4} + a\left(\frac{\omega{}}{\Delta{}}\right)^{2}\,(J\,S_{0}^{\dagger}\,J\,S_{0} + S_{0}^{\dagger}\,J\,S_{0}\,J).
\end{dmath}
\end{formalized}%
\noindent
In block form, this identity is
{}
\begin{formalized}{eq:short-junction-block-transmission-identity}{Rendered from Lean 4 theorem RmtNlin.manuscriptShortJunctionCrossTerm_eq_transmission (RmtNlin/ShortJunction.lean:225). 1 direct premise; rendered forms and their uses are in the generated assumption ledger. 9 project definitions are exposed in the assumptions and definitions ledger. Zero direct premises does not mean unrestricted model objects.}
\begin{dmath}[compact,label={eq:short-junction-block-transmission-identity}]
J\,S_{0}^{\dagger}\,J\,S_{0} + S_{0}^{\dagger}\,J\,S_{0}\,J = 2\,\begin{pmatrix}r^{\dagger}\,r - t^{\dagger}\,t&0\\0&{r'}^{\dagger}\,r' - {t'}^{\dagger}\,t'\end{pmatrix} = 2 - 4\,\begin{pmatrix}t^{\dagger}\,t&0\\0&{t'}^{\dagger}\,t'\end{pmatrix}.
\end{dmath}
\end{formalized}%
\noindent
Substituting this identity into Eq.~\eqref{eq:short-junction-f-squared-expansion} gives
{}
\begin{formalized}{eq:short-junction-f-squared-transmission-block}{Rendered from Lean 4 theorem RmtNlin.manuscriptShortJunctionF_sq_eq_transmission (RmtNlin/ShortJunction.lean:271). 1 direct premise; rendered forms and their uses are in the generated assumption ledger. 11 project definitions are exposed in the assumptions and definitions ledger. Zero direct premises does not mean unrestricted model objects.}
\begin{dmath}[compact,label={eq:short-junction-f-squared-transmission-block}]
F^{2} = \left[1 + a\left(\frac{\omega{}}{\Delta{}}\right)^{2}\right]^{2} - 4\,a\left(\frac{\omega{}}{\Delta{}}\right)^{2}\,\begin{pmatrix}t^{\dagger}\,t&0\\0&{t'}^{\dagger}\,t'\end{pmatrix}.
\end{dmath}
\end{formalized}%
\noindent
Hence
{}
\begin{formalized}{eq:short-transmission-law}{Rendered from Lean 4 theorem RmtNlin.manuscriptShortTransmissionLaw (RmtNlin/ShortJunctionChannels.lean:389). 5 direct premises; rendered forms and their uses are in the generated assumption ledger. 44 project definitions are exposed in the assumptions and definitions ledger. Zero direct premises does not mean unrestricted model objects.}
\begin{dmath}[compact,label={eq:short-transmission-law}]
T_{n}(\omega{}) = \frac{4\,a\left(\frac{\omega{}}{\Delta{}}\right)^{2}}{\left[1 + a\left(\frac{\omega{}}{\Delta{}}\right)^{2}\right]^{2}}\,\tau_{n} = \frac{\tau_{n}\,\Delta{}^{2}}{\omega{}^{2} + \Delta{}^{2}}.
\end{dmath}
\end{formalized}%
\noindent
Here \(\tau_n\) are the eigenvalues of \(t^\dagger t\), and therefore
the transmission eigenvalues of \(S_0\). At \(\omega=0\), the same
\(t^\dagger t\) block results from the zero-frequency scattering matrix
of any junction, and Eq.~\eqref{eq:short-transmission-law} gives
\(T_n(0)=\tau_n\). In the short-junction approximation, the scalar
prefactor supplies the remaining \(\omega\)-dependence, and substituting
Eq.~\eqref{eq:short-transmission-law} into the zero-temperature limit of
Eq.~\eqref{eq:transmission-representation-beta} reproduces
Eq.~\eqref{eq:beenakker-short} after integration over \(\omega\). If the
two superconductors have unequal gaps, replacing the single prefactor in
Eq.~\eqref{eq:short-scattering-limit} by a separate prefactor for each lead
reproduces the known result for a point contact with unequal gaps
\citep{Zaitsev84, Yip03}.

\begin{figure}[t]
\centering
\noindent\includegraphics[width=\linewidth]{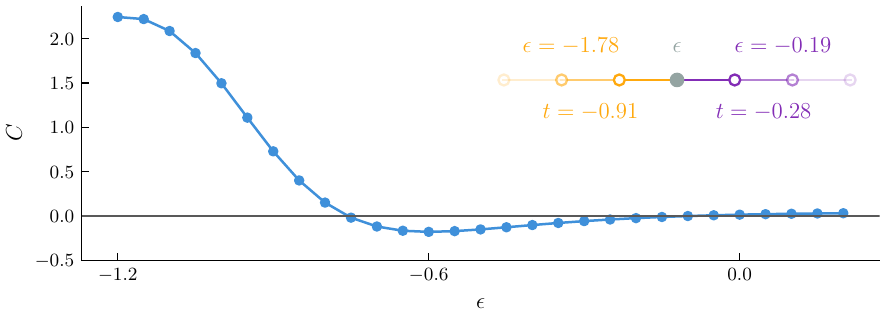}%
\caption{The coefficient \(C\) changes sign as the onsite energy
\(\epsilon\) of the central site varies. The inset shows the central site and
the two semi-infinite one-dimensional leads; their decreasing opacity indicates
that they continue beyond the displayed sites. The labels give the onsite
energy and hopping in each lead.}
\label{fig:log-correction-sign}
\end{figure}

The transmission representation also lets us derive the general form of
the finite-length correction to the energy-phase relation, despite the
singular \(\Delta\to0\) behavior caused by continuum states above the
gap \citep{Levchenko06}. For equal gaps, and as long as the normal-state
scattering depends smoothly on energy near the Fermi level (in particular,
the Fermi level does not sit at a band edge), we obtain the zero-temperature
expansion (see App.~\ref{appendix-next-to-leading-short-junction-expansion}):
{}
\begin{formalized}{eq:short-junction-next-order}{Rendered from Lean 4 theorem RmtNlin.manuscriptPhysicalMatchedExpansion (RmtNlin/MatchedEnergyViews.lean:44). 4 direct premises; rendered forms and their uses are in the generated assumption ledger. 51 project definitions are exposed in the assumptions and definitions ledger. Zero direct premises does not mean unrestricted model objects.}
\begin{dmath}[compact,label={eq:short-junction-next-order}]
U(\varphi{}) - U(0) = U_{\mathrm{short}}(\varphi{}) - U_{\mathrm{short}}(0) + \Delta{}^{2}\,(x\,C\,\log \Delta{} + B_{\mathrm{low}}(x) + x\,B_{\mathrm{high}}) + o\left(\Delta{}^{2}\right).
\end{dmath}
\end{formalized}%
\noindent
Here \(x=\sin^2(\varphi/2)\), and \(U_{\rm short}\) denotes the
short-junction energy of Eq.~\eqref{eq:beenakker-short}. The expansion is
asymptotic in \(\Delta\) relative to the energy scale on which the
normal-state scattering changes, and it holds at fixed phase provided
\(x\tau_n<1\) for every channel. The singular part
of the finite-length correction is linear in \(\sin^2(\varphi/2)\), while \(C\), \(B_{\rm low}\), and \(B_{\rm high}\)
encode the energy dependence of the normal scattering problem. Our result
extends prior work by including the nondiverging terms \(B_{\rm low}\) and
\(B_{\rm high}\) for any such normal region. In the previously studied
examples, the logarithmic correction suppresses the short-junction
response: a short point contact \citep{Beenakker23}, a ballistic SNS
junction \citep{Levchenko25}, and resonant-level and finite-length
weak-link models \citep{Fatemi25}. We find that this suppression is not
universal: a one-orbital model with different lead dispersions and a
single-site scattering region realizes either sign of \(C\), as
Fig.~\ref{fig:log-correction-sign} shows.

\section{Phase coefficients and
bounds}\label{phase-coefficients-and-bounds}

The transmission representation gives two levels of bounds. Positivity of
each Matsubara summand gives immediate constraints at every temperature.
Relations between different frequencies give a tighter upper bound on the
quartic coefficient.
We treat the finite- and zero-temperature arguments in
Sec.~\ref{zero-temperature-upper-bound}.

We first expand Eq.~\eqref{eq:transmission-representation-beta} channel by
channel:
{}
\begin{formalized}{eq:coefficient-expansion}{Rendered from Lean 4 theorem RmtNlin.AllChannels.manuscriptPhysicalCoefficientExpansion (RmtNlin/AllChannelViews.lean:148). 9 direct premises; rendered forms and their uses are in the generated assumption ledger. 81 project definitions are exposed in the assumptions and definitions ledger. Zero direct premises does not mean unrestricted model objects.}
\begin{equation}\begin{aligned}
c_{2} &= \frac{1}{2\,\beta{}}\,\sum_{m=0}^{\infty}\sum_{n}T_{n}(\omega_{m}),\qquad c_{4} = \frac{1}{16\,\beta{}}\,\sum_{m=0}^{\infty}\left[\sum_{n}T_{n}(\omega_{m})^{2} - \frac{2}{3}\,\sum_{n}T_{n}(\omega_{m})\right], \\
c_{6} &= \frac{1}{1440\,\beta{}}\,\sum_{m=0}^{\infty}\sum_{n}T_{n}(\omega_{m})\,(15\,T_{n}(\omega_{m})^{2} - 15\,T_{n}(\omega_{m}) + 2).
\end{aligned}\label{eq:coefficient-expansion}\end{equation}
\end{formalized}%
\noindent
The transmission eigenvalues satisfy \(0\le T_n(\omega_m)\le1\), so
every product \(T_n^p(\omega_m)[1-T_n(\omega_m)]^q\), with \(p,q\ge0\),
is nonnegative term by term. We combine coefficients so that their
summands have this form. The expression for \(c_2\) is a sum of
\(T_n(\omega_m)\), so \(c_2\ge0\), and adding \(c_2/12\) to \(c_4\)
cancels the linear term in \(c_4\) and gives
{}
\begin{formalized}{eq:quartic-lower-identity}{Rendered from Lean 4 theorem RmtNlin.AllChannels.manuscriptPhysicalMatsubaraQuarticLower (RmtNlin/AllChannelViews.lean:160). 9 direct premises; rendered forms and their uses are in the generated assumption ledger. 79 project definitions are exposed in the assumptions and definitions ledger. Zero direct premises does not mean unrestricted model objects.}
\begin{dmath}[compact,label={eq:quartic-lower-identity}]
c_{4} + \frac{c_{2}}{12} = \frac{1}{16\,\beta{}}\,\sum_{m=0}^{\infty}\sum_{n}T_{n}(\omega_{m})^{2} \ge 0.
\end{dmath}
\end{formalized}%
\noindent
Subtracting \(c_4\) from \(c_2/24\) instead gives a factor
\(T_n(\omega_m)[1-T_n(\omega_m)]\):
{}
\begin{formalized}{eq:quartic-upper-identity}{Rendered from Lean 4 theorem RmtNlin.AllChannels.manuscriptPhysicalMatsubaraQuarticUpper (RmtNlin/AllChannelViews.lean:168). 9 direct premises; rendered forms and their uses are in the generated assumption ledger. 79 project definitions are exposed in the assumptions and definitions ledger. Zero direct premises does not mean unrestricted model objects.}
\begin{dmath}[compact,label={eq:quartic-upper-identity}]
\frac{c_{2}}{24} - c_{4} = \frac{1}{16\,\beta{}}\,\sum_{m=0}^{\infty}\sum_{n}T_{n}(\omega_{m})\,[1 - T_{n}(\omega_{m})] \ge 0.
\end{dmath}
\end{formalized}%
\noindent
Combining these two inequalities yields
{}
\begin{formalized}{eq:quartic-window}{Rendered from Lean 4 theorem RmtNlin.AllChannels.physicalMatsubara_quartic_window (RmtNlin/AllChannelViews.lean:176). 9 direct premises; rendered forms and their uses are in the generated assumption ledger. 79 project definitions are exposed in the assumptions and definitions ledger. Zero direct premises does not mean unrestricted model objects.}
\begin{dmath}[compact,label={eq:quartic-window}]
-\frac{c_{2}}{12} \le c_{4} \le \frac{c_{2}}{24}.
\end{dmath}
\end{formalized}%
\noindent
For \(c_2>0\), normalizing by the quadratic coefficient gives
{}
\begin{formalized}{eq:normalized-quartic-ratio}{Rendered from Lean 4 theorem RmtNlin.AllChannels.manuscriptPhysicalNormalizedQuarticRatio (RmtNlin/AllChannelViews.lean:184). 10 direct premises; rendered forms and their uses are in the generated assumption ledger. 80 project definitions are exposed in the assumptions and definitions ledger. Zero direct premises does not mean unrestricted model objects.}
\begin{dmath}[compact,label={eq:normalized-quartic-ratio}]
-1 \le \alpha_{r}(\beta{}) = \frac{12\,c_{4}}{c_{2}} \le \frac{1}{2}.
\end{dmath}
\end{formalized}%
\noindent
The inequality \(c_2\ge0\) reproduces the nonnegative curvature at
\(\varphi=0\) implied by the scattering-theory result of Titov, Jacquod,
and Beenakker \citep{Titov01}. The lower bound \(\alpha_r\ge-1\) matches
the bound observed in our earlier numerical study \citep{Fatemi25} and is approached when
\(T_n(\omega_m)\ll1\) for the contributing channels and frequencies. The
upper bound \(\alpha_r\le1/2\) is weaker than the conjectured
\(\alpha_r\le0\), and the tighter bound below supersedes it. The
positive example in
Sec.~\ref{sns-junction-with-a-positive-quartic-nonlinearity} shows that
the conjectured sign constraint does not hold.

The same termwise construction extends to sixth order. We form three
nonnegative cubic products:
{}
\begin{formalized}{eq:sixth-order-nonnegative-identities}{Rendered from Lean 4 theorem RmtNlin.AllChannels.manuscriptPhysicalSixthNonnegativeIdentities (RmtNlin/AllChannelViews.lean:194). 9 direct premises; rendered forms and their uses are in the generated assumption ledger. 81 project definitions are exposed in the assumptions and definitions ledger. Zero direct premises does not mean unrestricted model objects.}
\begin{equation}\begin{aligned}
c_{6} + \frac{c_{4}}{6} + \frac{c_{2}}{90} &= \frac{1}{96\,\beta{}}\,\sum_{m=0}^{\infty}\sum_{n}T_{n}(\omega_{m})^{3} \ge 0, \\
\frac{c_{2}}{360} - c_{6} &= \frac{1}{96\,\beta{}}\,\sum_{m=0}^{\infty}\sum_{n}T_{n}(\omega_{m})^{2}\,[1 - T_{n}(\omega_{m})] \ge 0, \\
c_{6} - \frac{c_{4}}{6} + \frac{c_{2}}{240} &= \frac{1}{96\,\beta{}}\,\sum_{m=0}^{\infty}\sum_{n}T_{n}(\omega_{m})\,[1 - T_{n}(\omega_{m})]^{2} \ge 0.
\end{aligned}\label{eq:sixth-order-nonnegative-identities}\end{equation}
\end{formalized}%
\noindent
The last identity is algebraically the difference between the identity in
Eq.~\eqref{eq:quartic-upper-identity} and the \(T_n^2(\omega_m)[1-T_n(\omega_m)]\)
identity:
{}
\begin{formalized}{eq:sixth-order-difference-identity}{Rendered from Lean 4 theorem RmtNlin.sixth_difference_identity (RmtNlin/PhaseCoefficients.lean:277). 6 project definitions are exposed in the assumptions and definitions ledger. Zero direct premises does not mean unrestricted model objects.}
\begin{dmath}[compact,label={eq:sixth-order-difference-identity}]
c_{6} - \frac{c_{4}}{6} + \frac{c_{2}}{240} = \frac{1}{6}\,\left(\frac{c_{2}}{24} - c_{4}\right) - \left(\frac{c_{2}}{360} - c_{6}\right).
\end{dmath}
\end{formalized}%
\noindent
The last inequality follows directly from the nonnegative summand
\(T_n(\omega_m)[1-T_n(\omega_m)]^2\). The three sixth-order identities give two lower bounds and
one upper bound:
{}
\begin{formalized}{eq:sixth-order-coefficient-bounds}{Rendered from Lean 4 theorem RmtNlin.AllChannels.manuscriptPhysicalSixthCoefficientBounds (RmtNlin/AllChannelViews.lean:209). 9 direct premises; rendered forms and their uses are in the generated assumption ledger. 81 project definitions are exposed in the assumptions and definitions ledger. Zero direct premises does not mean unrestricted model objects.}
\begin{dmath}[compact,label={eq:sixth-order-coefficient-bounds}]
c_{6} \ge -\frac{c_{4}}{6} - \frac{c_{2}}{90},\qquad c_{6} \ge \frac{c_{4}}{6} - \frac{c_{2}}{240},\qquad c_{6} \le \frac{c_{2}}{360}.
\end{dmath}
\end{formalized}%
\noindent

\subsection{Upper bounds from the frequency dependence}\label{zero-temperature-upper-bound}

The termwise estimate treats the transmission at each frequency independently,
but the transmissions at different frequencies are related because the
scattering matrix depends analytically on frequency. In
App.~\ref{appendix-zero-temperature-quartic-bound}, we show that the reduced
scattering matrix \(S_\omega\) of Eq.~\eqref{eq:reduced-scattering} extends
analytically to the right half of the complex frequency plane and remains
subunitary there. For such a function, the Schwarz--Pick lemma bounds the rate
of change by the distance from unitarity. Applied to the generalized
eigenvalues of the pair \((F,D)\) in Eq.~\eqref{eq:d-and-f-definitions}, this bound becomes a speed limit for
every channel: the rapidity \(q_n=\operatorname{artanh}\sqrt{T_n}\) changes by
at most one per unit of \(\log\omega\).

The speed limit prevents a well-transmitting channel from switching off
abruptly, and therefore limits the second transmission moment relative to the
first. At zero temperature, we compare each channel with a trajectory that
decays at the maximal speed and obtain
{}
\begin{formalized}{eq:zero-temperature-normalized-quartic-bound}{Rendered from Lean 4 theorem RmtNlin.physical_zeroTemperature_momentRatioBound (RmtNlin/PhysicalZeroTemperatureBound.lean:199). 4 direct premises; rendered forms and their uses are in the generated assumption ledger. 74 project definitions are exposed in the assumptions and definitions ledger. Zero direct premises does not mean unrestricted model objects.}
\begin{dmath}[compact,label={eq:zero-temperature-normalized-quartic-bound}]
-1 \le -1 + \frac{3}{2}\,\frac{M_{2}}{M_{1}} \le \frac{16 - 5\,\pi}{4\,(4 - \pi)},
\end{dmath}
\end{formalized}%
\noindent
where the middle expression is \(\alpha_r\) at zero temperature and
\(M_p=\int_0^\infty\sum_nT_n(\omega)^p\,d\omega\) is the \(p\)th transmission
moment. At finite temperature, the speed limit relates the transmissions at
consecutive Matsubara frequencies, and a backward induction over the Matsubara
sum gives
{}
\begin{formalized}{eq:finite-temperature-normalized-quartic-bound}{Rendered from Lean 4 theorem RmtNlin.AllChannels.manuscriptPhysicalFiniteTemperatureNormalizedQuarticBound (RmtNlin/AllChannelViews.lean:138). 10 direct premises; rendered forms and their uses are in the generated assumption ledger. Additional axioms: RmtNlin.FiniteTemperatureCertificate.derivativeSum_compact_certificate._native.native_decide.ax_1_1, RmtNlin.FiniteTemperatureCertificate.finiteSum_compact_certificate._native.native_decide.ax_1_1 151 project definitions are exposed in the assumptions and definitions ledger. Zero direct premises does not mean unrestricted model objects.}
\begin{dmath}[compact,label={eq:finite-temperature-normalized-quartic-bound}]
-1 \le \alpha_{r}(\beta{}) \le \frac{7}{80}.
\end{dmath}
\end{formalized}%
\noindent

The finite-temperature bound is slightly weaker, and \(\alpha_r\) need not
decrease as the temperature rises. In the three-orbital example of
App.~\ref{appendix-finite-temperature-increase}, \(\alpha_r\) at
\(\beta=2\pi\) exceeds its zero-temperature value. Both values lie inside the
general bounds, so the example shows only that temperature can increase
\(\alpha_r\) and does not test either upper bound.

\section{SNS junction with a positive quartic
nonlinearity}\label{sns-junction-with-a-positive-quartic-nonlinearity}

\begin{figure}[t]
\centering
\noindent\includegraphics[width=\linewidth]{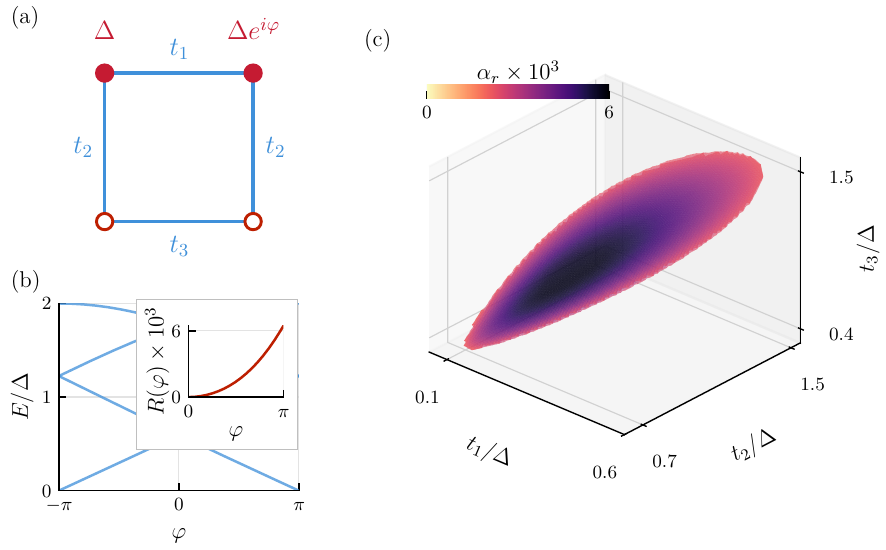}%
\caption{The four-site junction with positive quartic nonlinearity. Panel
(a) shows the square geometry. Panel (b) shows its positive-energy BdG branches;
the inset plots \(R(\varphi)=[U(\varphi)-U(0)]/(c_2\varphi^2)-1\). Panel (c) shows
the lower and upper \(t_3\) boundaries of the region with \(\alpha_r>0\)
at each \((t_1,t_2)\). Color indicates the maximum \(\alpha_r\) along
that \(t_3\) interval.}
\label{fig:positive-quartic-example}
\end{figure}

Both upper bounds allow \(c_4>0\). We now show that a conventional SNS
junction realizes this possibility and therefore refutes the conjectured sign
constraint. The coefficient formulas suggest concentrating the Matsubara
response in one transmitting channel and tuning that channel so that its
\(T^2\) weight is large relative to its \(T\) weight. Rank-one pairing blocks
\(\Delta_L\) and \(\Delta_R\) enforce a single scalar transmission function
\(T(\omega)\). The
smallest closed-form example that we found is the four-site square
junction shown in Fig.~\ref{fig:positive-quartic-example}(a):
{}
\begin{formalized}{eq:4x4-square-family-matrices}{Rendered from Lean 4 theorem RmtNlin.manuscriptSquareFamilyMatrices (RmtNlin/Counterexample.lean:76). 5 project definitions are exposed in the assumptions and definitions ledger. Zero direct premises does not mean unrestricted model objects.}
\begin{dmath}[compact,label={eq:4x4-square-family-matrices}]
h = \begin{pmatrix}0&t_{1}&0&t_{2}\\t_{1}&0&t_{2}&0\\0&t_{2}&0&t_{3}\\t_{2}&0&t_{3}&0\end{pmatrix},\qquad \Delta_{L} = \operatorname{diag}(\Delta{}, 0, 0, 0),\qquad \Delta_{R} = \operatorname{diag}(0, \Delta{}, 0, 0).
\end{dmath}
\end{formalized}%
\noindent
The transmission representation then has one nontrivial channel, with
Matsubara transmission
{}
\begin{formalized}{eq:square-family-transmission}{Rendered from Lean 4 theorem RmtNlin.manuscriptSquareFamilyTransmission (RmtNlin/SquareFamilyChannels.lean:134). 2 direct premises; rendered forms and their uses are in the generated assumption ledger. 27 project definitions are exposed in the assumptions and definitions ledger. Zero direct premises does not mean unrestricted model objects.}
\begin{dmath}[compact,label={eq:square-family-transmission}]
T_{n}(\omega{}) = \frac{4\,\Delta{}^{2}\,[t_{1}\,\omega{}^{2} + t_{3}\,(t_{1}\,t_{3} - t_{2}^{2})]^{2}}{[\omega{}^{4} + (\Delta{}^{2} + t_{1}^{2} + 2\,t_{2}^{2} + t_{3}^{2})\,\omega{}^{2} + \Delta{}^{2}\,t_{3}^{2} + (t_{1}\,t_{3} - t_{2}^{2})^{2}]^{2}}.
\end{dmath}
\end{formalized}%
\noindent
A numerical search in this family selects \(\Delta=2\), \(t_1=1/2\),
\(t_2=\sqrt{15}/2\), and \(t_3=3/2\), within about \(0.13\%\) of the
best value on the search grid. At this point, the coefficients are
{}
\begin{formalized}{eq:4x4-exact-coefficients}{Rendered from Lean 4 theorem RmtNlin.manuscriptCounterexampleExactCoefficients (RmtNlin/Counterexample.lean:825). 14 project definitions are exposed in the assumptions and definitions ledger. Zero direct premises does not mean unrestricted model objects.}
\begin{equation}\begin{aligned}
c_{2} &= \frac{(186 - 71\,s_0)\,r_{-} + (186 + 71\,s_0)\,r_{+}}{30752}, \\
c_{4} &= \frac{(844533 + 279412\,s_0)\,r_{-} + (844533 - 279412\,s_0)\,r_{+}}{11348226048} > 0.
\end{aligned}\label{eq:4x4-exact-coefficients}\end{equation}
\end{formalized}%
\noindent
Here \(s_0=\sqrt{31}\) and \(r_\pm=\sqrt{7\pm s_0}\). Numerically,
\(c_2\approx0.059\) and
\(c_4\approx3.1\times10^{-5}
\approx5.3\times10^{-4}\,c_2\).
A fourth-order perturbative calculation with \texttt{pymablock}
\citep{ArayaDay25} gives the same coefficients.

For this four-site junction,
\(\alpha_r\approx
0.0063>0\).
A larger realization of the same rank-one construction gives
\(\alpha_r\approx
0.021\).
Both values remain below the zero-temperature upper bound
\(0.085\); Sec.~\ref{conclusion}
mentions tighter bounds that we have not yet verified.

\section{Conclusion}\label{conclusion}

We expressed the phase-dependent free energy of a noninteracting
time-reversal-symmetric \(s\)-wave Josephson junction as a Matsubara sum over
frequency-dependent transmission eigenvalues. This representation keeps the
channel structure of Beenakker's short-junction formula at arbitrary junction
length, with only the transmission eigenvalues acquiring a dependence on
frequency, and it proves that \(\varphi=0\) is a global minimum of the free
energy. Near the short-junction limit, the same representation generalizes the
next-to-leading correction to any normal region whose scattering depends
smoothly on energy near the Fermi level, and a one-orbital example shows that its logarithmic
term can enhance as well as suppress the response.

Positivity of the individual Matsubara terms bounds the normalized quartic
coefficient by \(-1\le\alpha_r\le1/2\) and constrains the higher-order
coefficients. Using the analytic structure of the scattering matrix as a
function of frequency, we established a bound on how fast the transmission
eigenvalues decay with frequency, and this bound tightens the upper limit to
\(\alpha_r\le7/80\)
at finite temperature and to
\(\alpha_r\le(16 - 5\,\pi)/[4\,(4 - \pi)]
\approx0.085\)
at zero temperature. Both bounds leave room for a positive quartic coefficient,
and the four-site junction with \(\alpha_r>0\) disproves the upper bound
\(\alpha_r\le0\) conjectured in our earlier numerical study \citep{Fatemi25}.
Computer-assisted extensions of the comparison argument, which we have not yet
verified to the standard of this manuscript, lower the zero-temperature bound
to \(\alpha_r\le0.028\) for any number of channels and to \(0.022\) for a
single channel, close to the value
\(0.021\) of our larger
rank-one junction. These extensions come from a ChatGPT conversation with
GPT-6 Pro (App.~\ref{appendix-development-workflow}), archived in
Ref.~\citep{Akhmerov26Reproduction}. Whether the finite-temperature bound
admits a comparable improvement remains open.

A related open question is whether a positive quartic coefficient occurs in
real junctions. Flat-band Josephson junctions are a promising system in which
to search for it, because their Josephson response departs strongly from that
of conventional weak links: quantum-metric interface states can make the
critical current increase while the normal-state conductance decreases
\citep{Li26}. Beyond junctions, our approach may also constrain the nonlinear response of
bulk superconductors to current. This nonlinear Meissner effect
\citep{Yip92}, or nonlinear kinetic inductance \citep{Clem12}, sets the Kerr
nonlinearity of superconducting resonators and kinetic-inductance amplifiers
\citep{Zmuidzinas12, Eom12}. In a bulk superconductor, however, the phase
varies continuously instead of jumping between two contacts, so the
transmission representation would first need to be extended.

Most of the derivations, code, and text behind these results, including several
of the key ideas, came from AI models, as App.~\ref{appendix-development-workflow}
describes. Over the course of the project, the results of the models became
mostly reliable once we formulated the task correctly, but their own judgment
of whether a task was solved remained untrustworthy. Most of our effort
therefore went into making the results checkable. Every numbered equation in
this manuscript is printed from the statement of a theorem or definition
checked in the Lean proof assistant \citep{Moura21, Community20}, and the
Zenodo record \citep{Akhmerov26Reproduction} includes the code that generates
a website listing each result together with its hypotheses. To trust a
result, a reader then only needs to check that its hypotheses, such as
time-reversal symmetry and \(s\)-wave pairing, describe the junction of
interest, without repeating the derivation. We expect the same
approach to apply to other analytic results in theoretical physics whose
derivations are long but rely on standard mathematics.

\section*{Code availability}

The code, the Lean formalization, the manuscript sources, and the agent
session logs are available on Zenodo \citep{Akhmerov26Reproduction}.

\section*{Acknowledgements}

I thank V. Fatemi, P. W. Brouwer, and C. W. J. Beenakker for useful discussions, and I. Araya Day for comments on the
manuscript. I acknowledge
access to ChatGPT through OpenAI's ChatGPT for Academic Researchers program and
access to Claude through Anthropic's Claude Team plan for scientists.

\appendix

\section{Short-junction expansion to next-to-leading order}\label{appendix-next-to-leading-short-junction-expansion}

The leading term of the short-junction expansion neglects the energy
dependence of the normal scattering problem. To derive the first correction, we retain this
dependence in the zero-temperature limit of
Eq.~\eqref{eq:transmission-representation-beta} and write
\(x=\sin^2(\varphi/2)\). The correction requires the thermodynamic limit of
the superconducting leads before \(\Delta\to0\): once \(\Delta\) falls below
the level spacing of a finite lead, ordinary perturbation theory gives an
analytic response that starts at order \(\Delta^2\). We therefore eliminate
the semi-infinite leads exactly and obtain the transmissions from finite
matrices. Two expansions of these transmissions cover complementary
frequency ranges. The expansion at fixed \(\omega/\Delta\) describes low
frequencies, but the integral of its first correction diverges at large
\(\omega/\Delta\). The expansion at fixed \(\omega\) describes the remaining
range, but the integral of its second-order term diverges as \(\omega\to0\).
We separate the two ranges at a frequency \(\Lambda\) with
\(\Delta\ll\Lambda\) below the normal-state energy scale and show that the
divergences cancel, so that the final result is independent of \(\Lambda\).

\subsection{Finite transmission
problem}\label{finite-transmission-problem}

To eliminate the leads, we separate the finite central region from the
superconducting reservoirs. Instead of the electron-hole basis of the main
text, we replace each hole component \(c_X^{\dagger T}\) by its time reverse
\(U_{T,X}c_X^{\dagger T}\), where \(U_{T,X}\) is the unitary part of time
reversal on the electron space of region \(X=C,S\), with the same dimension
as \(h_X\). The matrix \(U_{T,X}\) is the identity in the real class and a
direct sum of \(i\sigma_y\) in the spinful class. In this basis,
{}
\begin{formalized}{eq:next-order-central-lead-split}{Rendered from Lean 4 theorem RmtNlin.manuscriptNextOrderCentralLeadSplit (RmtNlin/NextOrderMatrix.lean:64). 5 project definitions are exposed in the assumptions and definitions ledger. Zero direct premises does not mean unrestricted model objects.}
\begin{equation}\begin{aligned}
H_{\mathrm{BdG}} &= \begin{pmatrix}H_{C}&V^{\dagger}\\V&H_{S}(0)\end{pmatrix},\qquad H_{C} = \begin{pmatrix}h_{C}&0\\0&-h_{C}\end{pmatrix},\qquad H_{S}(0) = \begin{pmatrix}h_{S}&\Delta{}\\\Delta{}&-h_{S}\end{pmatrix}, \\
V &= \begin{pmatrix}v&0\\0&-v\end{pmatrix},
\end{aligned}\label{eq:next-order-central-lead-split}\end{equation}
\end{formalized}%
\noindent
where \(h_C\) is the normal Hamiltonian of the central region,
\(H_S(\varphi)\) contains both superconducting reservoirs, and the hopping
matrix \(v\) couples central-region orbitals to reservoir orbitals at the
junctions. The displayed \(H_S(0)\) has
zero phase; we restore the two phases after computing the lead
self-energies. The block-diagonal coupling \(V\) does not mix electrons and
holes, and the time-reversed hole basis turns the scalar pairing into
\(\Delta\) times the identity on the lead orbitals.

We now restore the phases; they enter \(H_{\rm BdG}(\varphi)\) only
through the reservoir block \(H_S(\varphi)\). A Schur complement at
positive Matsubara frequency then reduces the determinant ratio in
Eq.~\eqref{eq:free-energy-matsubara} to the central region:
{}
\begin{formalized}{eq:next-order-lead-schur-complement}{Rendered from Lean 4 theorem RmtNlin.manuscriptPhaseLeadSchur (RmtNlin/NextOrderMatrix.lean:144). 3 project definitions are exposed in the assumptions and definitions ledger. Zero direct premises does not mean unrestricted model objects.}
\begin{dmath}[compact,label={eq:next-order-lead-schur-complement}]
\frac{\det[i{}\,\omega{} - H_{\mathrm{BdG}}(\varphi{})]}{\det[i{}\,\omega{} - H_{S}(\varphi{})]} = \det(i{}\,\omega{} - H_{C} - V^{\dagger}\,[i{}\,\omega{} - H_{S}(\varphi{})]^{-1}\,V).
\end{dmath}
\end{formalized}%
\noindent
The two reservoirs are decoupled in \(H_S(\varphi)\), so a gauge
transformation of each reservoir removes its phase and
\(\det[i\omega-H_S(\varphi)]\) does not depend on \(\varphi\). For finite
leads, the determinant ratio in Eq.~\eqref{eq:free-energy-matsubara}
therefore equals the right-hand side of
Eq.~\eqref{eq:next-order-lead-schur-complement} divided by its value at
\(\varphi=0\). For semi-infinite leads, this ratio of determinants of finite
matrices acting on the central electron and hole states defines the
phase-dependent relative determinant.

The finite matrix contains the reservoir self-energy
\(V^\dagger[i\omega-H_S(\varphi)]^{-1}V\), which we compute lead by lead.
Because the pairing in each lead is \(\Delta\) times the identity, the
superconducting self-energy of lead \(l\in\{L,R\}\) follows from its normal
self-energy,
{}
\begin{formalized}{eq:next-order-normal-selfenergy}{Rendered from Lean 4 theorem RmtNlin.manuscriptNormalLeadSelfEnergyAt (RmtNlin/LeadGreenFunction.lean:412). 6 project definitions are exposed in the assumptions and definitions ledger. Zero direct premises does not mean unrestricted model objects.}
\begin{dmath}[compact,label={eq:next-order-normal-selfenergy}]
\Sigma_{\mathrm{N},l}(z) = v_{l}^{\dagger}\,(z - h_{l})^{-1}\,v_{l},
\end{dmath}
\end{formalized}%
\noindent
where \(h_l\) is the normal Hamiltonian of the lead and \(v_l\) is its part
of the contact hopping. To obtain the superconducting self-energy, we gauge
the phase of lead \(l\) to zero and invert its BdG Hamiltonian
\(H_{S,l}\). The electron-electron and electron-hole blocks of
\((i\omega-H_{S,l})^{-1}\) are
{}
\begin{formalized}{eq:next-order-lead-green-blocks}{Rendered from Lean 4 theorem RmtNlin.manuscriptPairedLeadGreenBlocks (RmtNlin/LeadGreenFunction.lean:379). 2 direct premises; rendered forms and their uses are in the generated assumption ledger. 15 project definitions are exposed in the assumptions and definitions ledger. Zero direct premises does not mean unrestricted model objects.}
\begin{equation}\begin{aligned}
G_{\mathrm{ee},l} &= \frac{1 + \frac{\omega{}}{\sqrt{\omega{}^{2} + \Delta{}^{2}}}}{2}\,G_{+,l} + \frac{1 - \frac{\omega{}}{\sqrt{\omega{}^{2} + \Delta{}^{2}}}}{2}\,G_{-,l}, \\
G_{\mathrm{eh},l} &= \frac{\Delta{}}{2\,i{}\,\sqrt{\omega{}^{2} + \Delta{}^{2}}}\,(G_{+,l} - G_{-,l}),
\end{aligned}\label{eq:next-order-lead-green-blocks}\end{equation}
\end{formalized}%
\noindent
where \(G_{\pm,l}=(\pm i\sqrt{\omega^2+\Delta^2}-h_l)^{-1}\) are the Green
functions of the normal lead at the imaginary energies
\(\pm i\sqrt{\omega^2+\Delta^2}\). Sandwiching these blocks between the
contact hoppings gives the self-energy of the superconducting lead,
{}
\begin{formalized}{eq:next-order-exact-paired-selfenergy}{Rendered from Lean 4 theorem RmtNlin.manuscriptPairedLeadSelfEnergyBcs (RmtNlin/LeadGreenFunction.lean:725). 3 direct premises; rendered forms and their uses are in the generated assumption ledger. 17 project definitions are exposed in the assumptions and definitions ledger. Zero direct premises does not mean unrestricted model objects.}
\begin{equation}\begin{aligned}
\Sigma_{\mathrm{ee},l}(\omega{},\Delta{}) &= \frac{1}{1 + a\left(\frac{\omega{}}{\Delta{}}\right)^{2}}\,\Sigma_{\mathrm{N},l}\left(i{}\,\sqrt{\omega{}^{2} + \Delta{}^{2}}\right) + \frac{a\left(\frac{\omega{}}{\Delta{}}\right)^{2}}{1 + a\left(\frac{\omega{}}{\Delta{}}\right)^{2}}\,\Sigma_{\mathrm{N},l}\left(-i{}\,\sqrt{\omega{}^{2} + \Delta{}^{2}}\right), \\
\Sigma_{\mathrm{eh},l}(\omega{},\Delta{}) &= i{}\,\frac{a\left(\frac{\omega{}}{\Delta{}}\right)}{1 + a\left(\frac{\omega{}}{\Delta{}}\right)^{2}}\,\left[\Sigma_{\mathrm{N},l}\left(i{}\,\sqrt{\omega{}^{2} + \Delta{}^{2}}\right) - \Sigma_{\mathrm{N},l}\left(-i{}\,\sqrt{\omega{}^{2} + \Delta{}^{2}}\right)\right], \\
a\left(\frac{\omega{}}{\Delta{}}\right) &= \sqrt{1 + \left(\frac{\omega{}}{\Delta{}}\right)^{2}} - \frac{\omega{}}{\Delta{}}.
\end{aligned}\label{eq:next-order-exact-paired-selfenergy}\end{equation}
\end{formalized}%
\noindent
Particle-hole symmetry fixes the two remaining blocks: in the
time-reversed hole basis, \(\Sigma_{\mathrm{he},l}=\Sigma_{\mathrm{eh},l}\)
and \(\Sigma_{\mathrm{hh},l}=-\Sigma_{\mathrm{ee},l}^\dagger\).
For \(\Delta>0\), Eq.~\eqref{eq:next-order-exact-paired-selfenergy} gives
the superconducting self-energy without expanding in \(\Delta\), for finite
and semi-infinite leads alike. The two cases differ in the analytic behavior of the normal self-energy near the Fermi
energy, which controls the expansion below.

To obtain the transmissions rather than only the determinant, we use that
the electron-hole self-energy is positive semidefinite:
{}
\begin{formalized}{eq:next-order-anomalous-positivity}{Rendered from Lean 4 theorem RmtNlin.manuscriptAnomalousLeadSelfEnergyPositivity (RmtNlin/LeadGreenFunction.lean:789). 2 direct premises; rendered forms and their uses are in the generated assumption ledger. 7 project definitions are exposed in the assumptions and definitions ledger. Zero direct premises does not mean unrestricted model objects.}
\begin{dmath}[compact,label={eq:next-order-anomalous-positivity}]
\Sigma_{\mathrm{eh},l}(\omega{},\Delta{}) = \Delta{}\,v_{l}^{\dagger}\,(\omega{}^{2} + \Delta{}^{2} + h_{l}^{2})^{-1}\,v_{l} \succeq 0.
\end{dmath}
\end{formalized}%
\noindent
Its factorization moves the superconducting phases into the spaces on
which the lead self-energies act:
{}
\begin{formalized}{eq:next-order-anomalous-factorization}{Rendered from Lean 4 theorem RmtNlin.manuscriptNextOrderAnomalousFactorization (RmtNlin/LeadGreenFunction.lean:770). 2 direct premises; rendered forms and their uses are in the generated assumption ledger. 11 project definitions are exposed in the assumptions and definitions ledger. Zero direct premises does not mean unrestricted model objects.}
\begin{dmath}[compact,label={eq:next-order-anomalous-factorization}]
K_{l}(\omega{},\Delta{})\,K_{l}(\omega{},\Delta{})^{\dagger} = \Pi{}_{l}\,\Sigma_{\mathrm{eh},l}(\omega{},\Delta{})\,\Pi{}_{l}^{\dagger},\qquad K = \begin{pmatrix}K_{L}(\omega{},\Delta{})&K_{R}(\omega{},\Delta{})\end{pmatrix},
\end{dmath}
\end{formalized}%
\noindent
where \(\Pi_l\) embeds the space on which the self-energy of lead \(l\)
acts into the electron space of the central region. The electron-electron self-energies of both leads determine the
electron Green function of the central region,
{}
\begin{formalized}{eq:next-order-central-diagonal-blocks}{Rendered from Lean 4 theorem RmtNlin.manuscriptCentralElectronGreen (RmtNlin/LeadGreenFunction.lean:578). 1 project definitions are exposed in the assumptions and definitions ledger. Zero direct premises does not mean unrestricted model objects.}
\begin{dmath}[compact,label={eq:next-order-central-diagonal-blocks}]
G_{C}(\omega{}) = [i{}\,\omega{} - h_{C} - (\Pi{}_{L}\,\Sigma{}_{\mathrm{ee},L}\,\Pi{}_{L}^{\dagger} + \Pi{}_{R}\,\Sigma{}_{\mathrm{ee},R}\,\Pi{}_{R}^{\dagger})]^{-1},
\end{dmath}
\end{formalized}%
\noindent
and particle-hole symmetry relates the hole block to \(G_C^\dagger\). With
the superconducting phases restored, the finite matrix in
Eq.~\eqref{eq:next-order-lead-schur-complement} becomes
{}
\begin{formalized}{eq:next-order-finite-central-matrix}{Rendered from Lean 4 theorem RmtNlin.manuscriptFiniteCentralBdGBlock (RmtNlin/NextOrderMatrix.lean:176). 2 project definitions are exposed in the assumptions and definitions ledger. Zero direct premises does not mean unrestricted model objects.}
\begin{dmath}[compact,label={eq:next-order-finite-central-matrix}]
M_C(\Phi{}_{\varphi{}}) = \begin{pmatrix}G_{C}^{-1}&-K\,\Phi{}_{\varphi{}}^{\dagger}\,K^{\dagger}\\-K\,\Phi{}_{\varphi{}}\,K^{\dagger}&-(G_{C}^{\dagger})^{-1}\end{pmatrix}.
\end{dmath}
\end{formalized}%
\noindent
A Schur complement with respect to the electron block, followed by
Sylvester's determinant identity, moves the phase dependence into a
determinant over these lead spaces:
{}
\begin{formalized}{eq:next-order-finite-green-determinant}{Rendered from Lean 4 theorem RmtNlin.manuscriptCentralBlockDeterminantToContact (RmtNlin/NextOrderMatrix.lean:248). 3 project definitions are exposed in the assumptions and definitions ledger. Zero direct premises does not mean unrestricted model objects.}
\begin{dmath}[compact,label={eq:next-order-finite-green-determinant}]
\det M_C(\Phi{}_{\varphi{}}) = (\det G_{C}^{-1})\,\det[-(G_{C}^{\dagger})^{-1}]\,\det(1 + \Phi{}_{\varphi{}}^{\dagger}\,G^{\dagger}\,\Phi{}_{\varphi{}}\,G),\qquad G = K^{\dagger}\,G_{C}\,K.
\end{dmath}
\end{formalized}%
\noindent
The first two factors do not depend on \(\varphi\), so dividing
\(\det M_C(\Phi_\varphi)\) by its value at \(\varphi=0\) leaves the
phase-dependent determinant ratio entering
Eq.~\eqref{eq:free-energy-matsubara}.

For two terminals, \(\Phi_\varphi\) is a linear combination of \(1\) and
\(P_L-P_R\), and the determinant takes the generalized-eigenvalue form of
Eq.~\eqref{eq:det-ratio-df}:
{}
\begin{formalized}{eq:next-order-contact-pencil}{Rendered from Lean 4 theorem RmtNlin.manuscriptNextOrderContactPencil (RmtNlin/TwoTerminalDeterminant.lean:280). 13 project definitions are exposed in the assumptions and definitions ledger. Zero direct premises does not mean unrestricted model objects.}
\begin{equation}\begin{aligned}
\frac{\det(1 + \Phi(e^{-i{}\varphi{}})\,G^{\dagger}\,\Phi(e^{i{}\varphi{}})\,G)}{\det(1 + G^{\dagger}\,G)} &= e^{-i{}\,r_R\,\varphi{}}\,\frac{\det\left(\frac{1 + e^{i{}\varphi{}}}{2}\,D + \frac{1 - e^{i{}\varphi{}}}{2}\,F\right)}{\det D},\qquad D = 1 + G^{\dagger}\,G, \\
F &= J + G^{\dagger}\,J\,G,\qquad J = P_{L} - P_{R}.
\end{aligned}\label{eq:next-order-contact-pencil}\end{equation}
\end{formalized}%
\noindent
As in Eqs.~\eqref{eq:transmission-eigenvalue}
and~\eqref{eq:paired-scalar-transmission-factor}, the nonendpoint
generalized eigenvalues of \((F,D)\) give the factors
\(1-xT_n(\omega,\Delta)\). We expand \(F\) and \(D\) by substituting
Eq.~\eqref{eq:next-order-exact-paired-selfenergy}.

\subsection{Expansion at fixed \(\omega/\Delta\)}\label{expansion-at-fixed-omegadelta}

The energy at leading order and the nonanalytic part of its correction
require the transmissions through first order in \(\Delta\) at fixed
\(\omega/\Delta\). We use the rescaled frequency \(y=\omega/\Delta\) and
the transmissions \(\widetilde T_n(y,\Delta)=T_n(\Delta y,\Delta)\) in this
coordinate.

We assume that each lead is a continuum BCS superconductor whose normal
self-energy has analytic boundary values at the Fermi level, meaning that its
retarded and advanced limits depend analytically on energy there, and a
positive-semidefinite broadening on the nonnegative imaginary-frequency axis.
The superconducting self-energy in
Eq.~\eqref{eq:next-order-exact-paired-selfenergy} evaluates the normal
self-energy at \(\pm i\Delta\sqrt{1+y^2}\), which approaches the Fermi level
from both sides as \(\Delta\to0\). We therefore expand the normal
self-energy around its two boundary values there.

\begin{samepage}
In the expansion below,
\(\widehat\Sigma_{\mathrm{N},l}(s)=\Sigma_{\mathrm{N},l}(is)\) is the normal
self-energy as a function of the imaginary part of the energy,
\(\Sigma_{R,l}\) and \(\Sigma_{A,l}=\Sigma_{R,l}^\dagger\) are its two
boundary values at the Fermi level, and the slopes \(\dot\Sigma_{R,l}\) and
\(\dot\Sigma_{A,l}=\dot\Sigma_{R,l}^\dagger\) are derivatives with respect to
the energy \(z=\pm is\):
\end{samepage}
{}
\begin{formalized}{eq:next-order-normal-boundary-expansion}{Rendered from Lean 4 theorem RmtNlin.manuscriptNextOrderNormalBoundaryExpansion (RmtNlin/NextOrderCorrespondence.lean:1149). 24 project definitions are exposed in the assumptions and definitions ledger. Zero direct premises does not mean unrestricted model objects.}
\begin{equation}\begin{aligned}
\widehat{\Sigma}_{\mathrm{N},l}\left(\sqrt{1 + y^{2}}\,\Delta{}\right) &= \Sigma_{R,l} + i{}\,\Delta{}\,\sqrt{1 + y^{2}}\,\dot{\Sigma}_{R,l} + O\left(\Delta{}^{2}\right), \\
\widehat{\Sigma}_{\mathrm{N},l}\left(\sqrt{1 + y^{2}}\,\Delta{}\right)^{\dagger} &= \Sigma_{A,l} - i{}\,\Delta{}\,\sqrt{1 + y^{2}}\,\dot{\Sigma}_{A,l} + O\left(\Delta{}^{2}\right).
\end{aligned}\label{eq:next-order-normal-boundary-expansion}\end{equation}
\end{formalized}%
\noindent
For a time-reversal-symmetric lead in the real basis, the advanced boundary
values and their derivatives are the complex conjugates of the retarded
ones. Substituting Eq.~\eqref{eq:next-order-normal-boundary-expansion} into
Eq.~\eqref{eq:next-order-exact-paired-selfenergy} at fixed \(y\) gives the
self-energies \(\Sigma_{\mathrm{ee},l}^{(y)}(\Delta)\) and
\(\Sigma_{\mathrm{eh},l}^{(y)}(\Delta)\):
{}
\begin{formalized}{eq:next-order-paired-boundary-expansion}{Rendered from Lean 4 theorem RmtNlin.manuscriptNextOrderPairedBoundaryExpansion (RmtNlin/NextOrderCorrespondence.lean:1182). 1 direct premise; rendered forms and their uses are in the generated assumption ledger. 31 project definitions are exposed in the assumptions and definitions ledger. Zero direct premises does not mean unrestricted model objects.}
\begin{equation}\begin{aligned}
\Sigma_{\mathrm{ee},l}^{(y)}(\Delta{}) &= \frac{1}{1 + a(y)^{2}}\,(\Sigma_{R,l} + a(y)^{2}\,\Sigma_{A,l}) \\
&\quad{}+ i{}\,\frac{\Delta{}}{2}\,(a(y)^{-1}\,\dot{\Sigma}_{R,l} - a(y)\,\dot{\Sigma}_{A,l}) + O\left(\Delta{}^{2}\right), \\
\Sigma_{\mathrm{eh},l}^{(y)}(\Delta{}) &= \frac{a(y)}{1 + a(y)^{2}}\,\Gamma{}_{l} - \frac{\Delta{}}{2}\,(\dot{\Sigma}_{R,l} + \dot{\Sigma}_{A,l}) + O\left(\Delta{}^{2}\right), \\
\Gamma{}_{l} &= i{}\,(\Sigma_{R,l} - \Sigma_{A,l}).
\end{aligned}\label{eq:next-order-paired-boundary-expansion}\end{equation}
\end{formalized}%
\noindent
To carry this expansion over to the transmissions, we also expand the factor
\(K_l\) of Eq.~\eqref{eq:next-order-anomalous-factorization} to first order in
\(\Delta\):
{}
\begin{formalized}{eq:next-order-factor-expansion}{Rendered from Lean 4 theorem RmtNlin.manuscriptNextOrderFactorExpansion (RmtNlin/NextOrderCorrespondence.lean:1254). 2 direct premises; rendered forms and their uses are in the generated assumption ledger. 27 project definitions are exposed in the assumptions and definitions ledger. Zero direct premises does not mean unrestricted model objects.}
\begin{equation}\begin{aligned}
K_{l}(\Delta{}) &= K_{0,l} + \Delta{}\,K_{1,l} + O\left(\Delta{}^{2}\right),\qquad K_{0,l}\,K_{0,l}^{\dagger} = \Pi{}_{l}\,(\Sigma_{\mathrm{eh},l}^{(y)})_{0}\,\Pi{}_{l}^{\dagger}, \\
K_{1,l}\,K_{0,l}^{\dagger} + K_{0,l}\,K_{1,l}^{\dagger} &= \Pi{}_{l}\,(\Sigma_{\mathrm{eh},l}^{(y)})_{1}\,\Pi{}_{l}^{\dagger}.
\end{aligned}\label{eq:next-order-factor-expansion}\end{equation}
\end{formalized}%
\noindent
The subscripts \(0\) and \(1\) denote Taylor coefficients in \(\Delta\) at
fixed \(y\), both for \(K\) and for the self-energy. Substituting the
expansions of \(\Sigma_{\mathrm{ee},l}\) and \(K_l\) into \(G_C\) and \(G\) expands
both matrices \(D\) and \(F\) of Eq.~\eqref{eq:next-order-contact-pencil}:
{}
\begin{formalized}{eq:next-order-contact-pencil-expansion}{Rendered from Lean 4 theorem RmtNlin.manuscriptNextOrderContactPencilExpansion (RmtNlin/NextOrderCorrespondence.lean:608). 2 direct premises; rendered forms and their uses are in the generated assumption ledger. 7 project definitions are exposed in the assumptions and definitions ledger. Zero direct premises does not mean unrestricted model objects.}
\begin{dmath}[compact,label={eq:next-order-contact-pencil-expansion}]
D(\Delta{}) = D_{0} + \Delta{}\,D_{1} + O\left(\Delta{}^{2}\right),\qquad F(\Delta{}) = F_{0} + \Delta{}\,F_{1} + O\left(\Delta{}^{2}\right).
\end{dmath}
\end{formalized}%
\noindent
For \(\Delta\to0^+\) at fixed \(y\), the determinant ratio in
Eq.~\eqref{eq:next-order-contact-pencil} reduces to its normal-state form
with \(x\) replaced by \(x/(1+y^2)\), so the leading transmissions are
{}
\begin{formalized}{eq:next-order-leading-transmissions}{Rendered from Lean 4 theorem RmtNlin.NextOrderJunction.IsReciprocal.scaledTransmission_zero (RmtNlin/NextOrderModel.lean:456). 3 direct premises; rendered forms and their uses are in the generated assumption ledger. 77 project definitions are exposed in the assumptions and definitions ledger. Zero direct premises does not mean unrestricted model objects.}
\begin{dmath}[compact,label={eq:next-order-leading-transmissions}]
\widetilde{T}_{n}(y,0) = \frac{\tau_{n}}{1 + y^{2}},
\end{dmath}
\end{formalized}%
\noindent
where \(\tau_n\) are the normal-state transmission eigenvalues at the Fermi
level. For an isolated generalized eigenvalue, first-order perturbation
theory gives
{}
\begin{formalized}{eq:next-order-transmission-perturbation}{Rendered from Lean 4 theorem RmtNlin.manuscriptGeneralizedEigenvalueFirstOrder (RmtNlin/NextOrder.lean:174). 7 direct premises; rendered forms and their uses are in the generated assumption ledger. 0 project definitions are exposed in the assumptions and definitions ledger. Zero direct premises does not mean unrestricted model objects.}
\begin{equation}\begin{aligned}
\mu{}_{1n} &= \nu{}_{n}\cdot [(F_{1} - \mu{}(0)\,D_{1})\,v(0)],\qquad F(0)\,v(0) = \mu{}(0)\,D(0)\,v(0), \\
\nu{}_{n}\cdot [D(0)\,v(0)] &= 1,
\end{aligned}\label{eq:next-order-transmission-perturbation}\end{equation}
\end{formalized}%
\noindent
where \(v(0)\) is the right eigenvector of the leading-order generalized
eigenvalue problem and \(\nu_n\)
is its left eigenvector, written as a row. The energy, however, does not
require isolated eigenvalues, because it depends only on the logarithm of
the determinant ratio in Eq.~\eqref{eq:next-order-contact-pencil}. The
first-order term of this logarithm is a trace, which we evaluate in an
eigenbasis of \(t^\dagger t\) and \({t'}^\dagger t'\). Each basis vector
contributes a coefficient \(T_{1n}\) with the weight
\(x/[1-x\tau_n/(1+y^2)]\) that appears in
Eq.~\eqref{eq:next-order-low-frequency-integrands}. The apparent powers
\(a^{-1}\) in Eq.~\eqref{eq:next-order-paired-boundary-expansion} cancel in
this trace, and collecting the remaining powers of \(a\) gives the frequency
dependence needed for the integral,
{}
\begin{formalized}{eq:next-order-first-transmission-correction}{Rendered from Lean 4 theorem RmtNlin.NextOrderJunction.manuscriptFirstTransmissionCorrection (RmtNlin/NextOrderLowFrequencyModel.lean:857). 1 direct premise; rendered forms and their uses are in the generated assumption ledger. 41 project definitions are exposed in the assumptions and definitions ledger. Zero direct premises does not mean unrestricted model objects.}
\begin{dmath}[compact,label={eq:next-order-first-transmission-correction}]
T_{1n}(y) = \frac{2\,a(y)}{[1 + a(y)^{2}]^{3}}\,[\gamma_{n}\,[1 - a(y)^{2}]^{2} + 4\,\beta_{n}\,a(y)^{2}] = \frac{\beta_{n} + \gamma_{n}\,y^{2}}{\left(1 + y^{2}\right)^{3/2}},
\end{dmath}
\end{formalized}%
\noindent
where \(\beta_n=T_{1n}(0)\) and
\(\gamma_n=\lim_{\omega/\Delta\to\infty}(\omega/\Delta)T_{1n}(\omega/\Delta)\).
The BCS self-energy for equal gaps fixes this frequency dependence, and the
normal junction enters only through \(\tau_n\), \(\beta_n\), and
\(\gamma_n\). The matrix elements of \(\dot\Sigma_{R,l}\) and \(\dot\Sigma_{A,l}\) in
the leading eigenbasis determine \(\beta_n\) and \(\gamma_n\). Within a
degenerate eigenspace, the individual \(T_{1n}\) depend on the chosen
eigenbasis, but their sum, and hence the energy, does not.

Integrating the fixed-\(\omega/\Delta\) expansion up to \(\Lambda\) gives the
leading energy, the logarithmic correction, and the part of the finite
correction determined by \(\tau_n\), \(\beta_n\), and \(\gamma_n\). The
logarithm in Eq.~\eqref{eq:transmission-representation-beta} expands as
{}
\begin{formalized}{eq:next-order-low-frequency-integrands}{Rendered from Lean 4 theorem RmtNlin.NextOrderJunction.IsReciprocal.manuscriptLowFrequencyIntegrands (RmtNlin/NextOrderMatchedModel.lean:66). 9 direct premises; rendered forms and their uses are in the generated assumption ledger. 121 project definitions are exposed in the assumptions and definitions ledger. Zero direct premises does not mean unrestricted model objects.}
\begin{equation}\begin{aligned}
-\sum_{n}\log[1 - x\,\widetilde{T}_{n}(y,\Delta{})] &= f_{0}(y, x) + \Delta{}\,f_{1}(y, x) + O\left(\Delta{}^{2}\right), \\
f_{0}(y, x) &= -\sum_{n}\log\left(1 - \frac{x\,\tau_{n}}{1 + y^{2}}\right),\qquad f_{1}(y, x) = \sum_{n}\frac{x\,T_{1n}(y)}{1 - \frac{x\,\tau_{n}}{1 + y^{2}}}.
\end{aligned}\label{eq:next-order-low-frequency-integrands}\end{equation}
\end{formalized}%
\noindent
The leading term \(f_0\) needs no subtraction: integrating it over all
frequencies reproduces the short-junction energy of
Eq.~\eqref{eq:beenakker-short},
{}
\begin{formalized}{eq:next-order-leading-energy-integral}{Rendered from Lean 4 theorem RmtNlin.AllChannels.manuscriptNextOrderLeadingEnergyIntegral (RmtNlin/AllChannelViews.lean:34). 2 direct premises; rendered forms and their uses are in the generated assumption ledger. 6 project definitions are exposed in the assumptions and definitions ledger. Zero direct premises does not mean unrestricted model objects.}
\begin{dmath}[compact,label={eq:next-order-leading-energy-integral}]
-\frac{1}{\pi}\,\int_{0}^{\infty}\sum_{n}\log\left(1 - \frac{\tau{}_{n}\,\Delta{}^{2}}{\omega{}^{2} + \Delta{}^{2}}\,x\right)\,\mathrm d\omega{} = \Delta{}\,\sum_{n}\left(1 - \sqrt{1 - \tau{}_{n}\,x}\right).
\end{dmath}
\end{formalized}%
\noindent
The terms to subtract from \(f_1\) are those that the two expansions share in
their common frequency range. Expanding \(f_0\) and \(f_1\) at large
\(\omega/\Delta\) gives
{}
\begin{formalized}{eq:next-order-low-frequency-tails}{Rendered from Lean 4 theorem RmtNlin.manuscriptNextOrderLowFrequencyTails (RmtNlin/NextOrderCorrespondence.lean:815). 1 direct premise; rendered forms and their uses are in the generated assumption ledger. 10 project definitions are exposed in the assumptions and definitions ledger. Zero direct premises does not mean unrestricted model objects.}
\begin{dmath}[compact,label={eq:next-order-low-frequency-tails}]
f_{0}(y, x) = x\,y^{-2}\,\sum_{n}\tau{}_{n} + O\left(y^{-4}\right),\qquad f_{1}(y, x) = -\pi\,x\,C\,y^{-1} + O\left(y^{-3}\right),
\end{dmath}
\end{formalized}%
\noindent
with \(C=-\pi^{-1}\sum_n\gamma_n\). The \((\omega/\Delta)^{-1}\) tail produces
the logarithm and is linear in \(x\) because every transmission is small
where the two expansions overlap. Subtracting this tail defines the finite
contribution
{}
\begin{formalized}{eq:next-order-low-finite-part-definition}{Rendered from Lean 4 theorem RmtNlin.manuscriptNextOrderLowFinitePartDefinition (RmtNlin/BCSBoundaryIntegration.lean:120). 1 direct premise; rendered forms and their uses are in the generated assumption ledger. 11 project definitions are exposed in the assumptions and definitions ledger. Zero direct premises does not mean unrestricted model objects.}
\begin{dmath}[compact,label={eq:next-order-low-finite-part-definition}]
\lim_{Y\to\infty}\left(C\,x\,\log Y + \frac{1}{\pi}\,\int_{0}^{Y}f_{1}(y, x)\,\mathrm dy\right)=B_{\mathrm{low}}(x).
\end{dmath}
\end{formalized}%
\noindent
Substituting Eq.~\eqref{eq:next-order-first-transmission-correction}, we
evaluate the integral in closed form:
{}
\begin{formalized}{eq:next-order-low-finite-part}{Rendered from Lean 4 theorem RmtNlin.manuscriptNextOrderLowFinitePartInterior (RmtNlin/BCSBoundaryIntegration.lean:154). 1 direct premise; rendered forms and their uses are in the generated assumption ledger. 2 project definitions are exposed in the assumptions and definitions ledger. Zero direct premises does not mean unrestricted model objects.}
\begin{dmath}[compact,label={eq:next-order-low-finite-part}]
B_{\mathrm{low}}(x) = \frac{x}{\pi}\,\sum_{n}\left(\gamma{}_{n}\,\log 2 + (\beta{}_{n} - \gamma{}_{n} + \gamma{}_{n}\,x\,\tau{}_{n})\,\frac{\arcsin \sqrt{x\,\tau{}_{n}}}{\sqrt{x\,\tau{}_{n}\,(1 - x\,\tau{}_{n})}}\right).
\end{dmath}
\end{formalized}%
\noindent
At \(x\tau_n=0\), the fraction equals one by continuity, and we exclude the
endpoint \(x\tau_n=1\), where the weight \(x/[1-x\tau_n/(1+y^2)]\) diverges at
\(y=0\) and the expansion in \(\Delta\) breaks down. To combine this
result with the fixed-\(\omega\) expansion, we restore the upper limit
\(\Lambda/\Delta\). The contribution of frequencies below the cutoff is then
{}
\begin{formalized}{eq:next-order-low-frequency-result}{Rendered from Lean 4 theorem RmtNlin.manuscriptPhysicalLowFrequencyResult (RmtNlin/MatchedEnergyViews.lean:114). 5 direct premises; rendered forms and their uses are in the generated assumption ledger. 42 project definitions are exposed in the assumptions and definitions ledger. Zero direct premises does not mean unrestricted model objects.}
\begin{equation}\begin{aligned}
U_{<\Lambda{}}(\varphi{}) - U_{<\Lambda{}}(0) &= U_{\mathrm{short}}(\varphi{}) - U_{\mathrm{short}}(0) + \Delta{}^{2}\,\left[\frac{-x}{\pi\,\Lambda{}}\,\sum_{n}\tau{}_{n} - x\,C\,\log \frac{\Lambda{}}{\Delta{}} + B_{\mathrm{low}}(x)\right] \\
&\quad{}+ o\left(\Delta{}^{2}\right) + O\left(M\,\Lambda{}\,\Delta{}^{2}\right).
\end{aligned}\label{eq:next-order-low-frequency-result}\end{equation}
\end{formalized}%
\noindent
Here \(M\) is a single bound on the second derivative with respect to
\(\Delta\) of the left-hand side of
Eq.~\eqref{eq:next-order-low-frequency-integrands} divided by \(\pi\), valid
for all \(y\geq0\) and all sufficiently small \(\Delta>0\).
Integrating the Taylor remainder up to \(\Lambda/\Delta\) gives
\(O(M\Lambda\Delta^2)\), while the leading tail and the first-order integral, which converges once
regularized, give \(o(\Delta^2)\) at fixed
\(\Lambda>0\).

\subsection{Expansion at fixed \(\omega\)}\label{expansion-at-fixed-omega}

The expansion at fixed \(\omega/\Delta\) fails when \(\omega\) reaches the
normal-state energy scale. At fixed \(\omega\), the anomalous self-energy of
lead \(l\) instead starts at first order in \(\Delta\):
{}
\begin{formalized}{eq:next-order-fixed-frequency-selfenergy}{Rendered from Lean 4 theorem RmtNlin.manuscriptNextOrderFixedFrequencySelfEnergy (RmtNlin/NextOrderCorrespondence.lean:1217). 2 direct premises; rendered forms and their uses are in the generated assumption ledger. 12 project definitions are exposed in the assumptions and definitions ledger. Zero direct premises does not mean unrestricted model objects.}
\begin{dmath}[compact,label={eq:next-order-fixed-frequency-selfenergy}]
\Sigma{}_{l}(\Delta{}) = \Delta{}\,\Sigma{}_{1,l} + O\left(\Delta{}^{3}\right),\qquad \Sigma{}_{1,l} = v_{l}^{\dagger}\,(\omega{}^{2} + h_{l}^{2})^{-1}\,v_{l},\qquad \Sigma{}_{1,l} \succeq 0,
\end{dmath}
\end{formalized}%
\noindent
where the expression for the coefficient follows from
Eq.~\eqref{eq:next-order-normal-selfenergy}. Substituting this expansion
into \(G\) gives \(G=\Delta G_1+o(\Delta)\), provided the matrix left after
removing the explicit factor of \(\Delta\) is continuous at zero gap.
The restriction of \(\Sigma_{1,l}\) to the contact orbitals of lead \(l\) is
\(\sigma_l\), and the left-to-right block of \(G_1\) is
{}
\begin{formalized}{eq:next-order-fixed-frequency-left-right-block}{Rendered from Lean 4 theorem RmtNlin.manuscriptNextOrderFixedFrequencyLeftRightBlock (RmtNlin/NextOrderCorrespondence.lean:1061). 3 direct premises; rendered forms and their uses are in the generated assumption ledger. 10 project definitions are exposed in the assumptions and definitions ledger. Zero direct premises does not mean unrestricted model objects.}
\begin{dmath}[compact,label={eq:next-order-fixed-frequency-left-right-block}]
(1 - P_{L})\,G_{1}\,P_{L} = \sigma_{R}(\omega{})^{1/2}\,\Pi{}_{R}^{\dagger}\,G_{C}\,\Pi{}_{L}\,\sigma_{L}(\omega{})^{1/2},
\end{dmath}
\end{formalized}%
\noindent
where \(G_C\) is evaluated at \(\Delta=0\). This block,
\(X(\omega)=P_RG_1(\omega)P_L\), determines the leading \(\Delta^2\)
contribution to the transmissions at fixed \(\omega>0\). Because \(T_n(\omega,\Delta)=O(\Delta^2)\), the logarithm depends
at this order only on the sum of the transmissions:
{}
\begin{formalized}{eq:next-order-fixed-frequency-integrand}{Rendered from Lean 4 theorem RmtNlin.NextOrderJunction.IsReciprocal.manuscriptFixedFrequencyIntegrand (RmtNlin/NextOrderFixedFrequencyModel.lean:910). 3 direct premises; rendered forms and their uses are in the generated assumption ledger. 82 project definitions are exposed in the assumptions and definitions ledger. Zero direct premises does not mean unrestricted model objects.}
\begin{equation}\begin{aligned}
-\sum_{n}\log[1 - x\,T_{n}(\omega{}, \Delta{})] &= x\,\sum_{n}T_{n}(\omega{}, \Delta{}) + O\left(\Delta{}^{4}\right), \\
x\,\sum_{n}T_{n}(\omega{}, \Delta{}) &= 4\,\Delta{}^{2}\,x\,\operatorname{Tr}[X(\omega{})^{\dagger}\,X(\omega{})] + o\left(\Delta{}^{2}\right).
\end{aligned}\label{eq:next-order-fixed-frequency-integrand}\end{equation}
\end{formalized}%
\noindent
The trace runs over the space on which the self-energy of the left lead
acts. The integral of the \(\Delta^2\) term diverges at its lower limit.
Substituting the boundary values
of the normal self-energy at small \(\omega\) into
Eqs.~\eqref{eq:next-order-fixed-frequency-selfenergy}
and~\eqref{eq:next-order-fixed-frequency-left-right-block} gives
{}
\begin{formalized}{eq:next-order-x-laurent}{Rendered from Lean 4 theorem RmtNlin.NextOrderJunction.manuscriptXLaurent (RmtNlin/NextOrderFixedFrequencyModel.lean:1489). 3 direct premises; rendered forms and their uses are in the generated assumption ledger. 65 project definitions are exposed in the assumptions and definitions ledger. Zero direct premises does not mean unrestricted model objects.}
\begin{dmath}[compact,label={eq:next-order-x-laurent}]
\Sigma{}_{1,l}(\omega{}) = \frac{1}{2\,\omega{}}\,\Gamma_{l} - \frac{1}{2}\,(\dot{\Sigma}_{R,l} + \dot{\Sigma}_{A,l}) + O\left(\omega{}\right),\qquad X(\omega{}) = \omega{}^{-1}\,X_{-1} + X_{0} + o\left(1\right).
\end{dmath}
\end{formalized}%
\noindent
Analyticity of the normal self-energy gives the \(O(\omega)\) remainder of
\(\Sigma_{1,l}(\omega)\), and differentiability at \(\omega=0\) of the square
roots of \(\omega\sigma_l(\omega)\) gives the \(o(1)\) remainder of
\(X(\omega)\). The quadratic term in \(X_{-1}\) gives the total normal-state transmission
at zero energy, as in
Eq.~\eqref{eq:short-junction-block-transmission-identity}, and its cross
terms with \(X_0\) give the sum of the coefficients \(\gamma_n\) in
Eq.~\eqref{eq:next-order-first-transmission-correction}:
{}
\begin{formalized}{eq:next-order-tau-gamma-from-x}{Rendered from Lean 4 theorem RmtNlin.NextOrderJunction.IsReciprocal.manuscriptTauGammaFromX (RmtNlin/NextOrderFixedFrequencyModel.lean:1583). 3 direct premises; rendered forms and their uses are in the generated assumption ledger. 65 project definitions are exposed in the assumptions and definitions ledger. Zero direct premises does not mean unrestricted model objects.}
\begin{dmath}[compact,label={eq:next-order-tau-gamma-from-x}]
4\,\operatorname{Tr}(X_{-1}^{\dagger}\,X_{-1}) = \sum_{n}\tau_{n},\qquad 4\,\operatorname{Tr}(X_{-1}^{\dagger}\,X_{0} + X_{0}^{\dagger}\,X_{-1}) = \sum_{n}\gamma_{n} = -\pi\,C.
\end{dmath}
\end{formalized}%
\noindent
The small-\(\omega\) integrand therefore has the singular terms
{}
\begin{formalized}{eq:next-order-common-frequency-range}{Rendered from Lean 4 theorem RmtNlin.NextOrderJunction.manuscriptCommonFrequencyRange (RmtNlin/NextOrderFixedFrequencyModel.lean:1623). 3 direct premises; rendered forms and their uses are in the generated assumption ledger. 42 project definitions are exposed in the assumptions and definitions ledger. Zero direct premises does not mean unrestricted model objects.}
\begin{dmath}[compact,label={eq:next-order-common-frequency-range}]
4\,x\,\operatorname{Tr}[X(\omega{})^{\dagger}\,X(\omega{})] = x\,\left[\omega{}^{-2}\,\sum_{n}\tau_{n} - \frac{\pi\,C}{\omega{}} + O\left(1\right)\right].
\end{dmath}
\end{formalized}%
\noindent
Writing \(\tilde{X}(\omega)=\omega X(\omega)\) removes the pole at
\(\omega=0\). Subtracting the two singular terms from the integral leaves the
finite contribution
{}
\begin{formalized}{eq:next-order-high-finite-part}{Rendered from Lean 4 theorem RmtNlin.manuscriptNextOrderHighFinitePart (RmtNlin/HighFrequencyFinitePart.lean:121). 3 direct premises; rendered forms and their uses are in the generated assumption ledger. 6 project definitions are exposed in the assumptions and definitions ledger. Zero direct premises does not mean unrestricted model objects.}
\begin{dmath}[compact,label={eq:next-order-high-finite-part}]
\lim_{Y\to 0^{+}}\left(\frac{4\,x}{\pi}\,\int_{Y}^{\infty}\frac{\operatorname{Tr} [\tilde{X}(\omega{})^{\dagger}\,\tilde{X}(\omega{})]}{\omega{}^{2}}\,\mathrm d\omega{} - \frac{x}{\pi\,Y} \cdot 4\,\operatorname{Tr} [\tilde{X}(0)^{\dagger}\,\tilde{X}(0)] - x\,C\,\log Y\right)=x\,B_{\mathrm{high}}.
\end{dmath}
\end{formalized}%
\noindent
If \(g(\omega)=\operatorname{Tr}[\tilde{X}^\dagger(\omega)\tilde{X}(\omega)]\)
is analytic, the subtractions \(\sum_n\tau_n=4g(0)\) and
\(C=-4g'(0)/\pi\) leave an integrand that is continuous at \(\omega=0\). The limit
in Eq.~\eqref{eq:next-order-high-finite-part} therefore exists when
\(g(\omega)/\omega^2\) is integrable at large \(\omega\). Because
the fixed-\(\omega\) integrand is linear in \(x\), \(B_{\rm high}\) is
independent of \(x\). Restoring the lower cutoff \(\Lambda\) gives the
contribution of frequencies above the cutoff,
{}
\begin{formalized}{eq:next-order-high-frequency-result}{Rendered from Lean 4 theorem RmtNlin.manuscriptPhysicalHighFrequencyResult (RmtNlin/MatchedEnergyViews.lean:86). 3 direct premises; rendered forms and their uses are in the generated assumption ledger. 21 project definitions are exposed in the assumptions and definitions ledger. Zero direct premises does not mean unrestricted model objects.}
\begin{equation}\begin{aligned}
U_{>\Lambda{}}(\varphi{}) - U_{>\Lambda{}}(0) &= \Delta{}^{2}\,\left(\frac{4\,x\,g(0)}{\pi\,\Lambda{}} + x\,C\,\log \Lambda{} + x\,B_{\mathrm{high}}\right) + o\left(\Delta{}^{2}\right) \\
&\quad{}+ \Delta{}^{2}\,x\,(B_{\mathrm{ch}}(\Lambda{}) - B_{\mathrm{high}}), \\
B_{\mathrm{ch}}(\Lambda{}) &= \frac{4}{\pi}\,\int_{\Lambda{}}^{\infty}\frac{g(w)}{w^{2}}\,\mathrm dw - \frac{4\,g(0)}{\pi\,\Lambda{}} - C\,\log \Lambda{},
\end{aligned}\label{eq:next-order-high-frequency-result}\end{equation}
\end{formalized}%
\noindent
where \(B_{\rm ch}(\Lambda)\) is the regularized coefficient at finite
cutoff, so that \(B_{\rm ch}(\Lambda)\to B_{\rm high}\) as
\(\Lambda\to0^+\). The \(o(\Delta^2)\) remainder requires taking the limit
\(\Delta\to0\) inside the frequency integral. Dominated convergence justifies
this step when \(T_n(\omega,\Delta)/\Delta^2\) is continuous at \(\Delta=0\)
and, above each positive cutoff, is bounded by a \(\Delta\)-independent
function with a finite integral, and when the argument of the logarithm stays
above a positive constant for all these frequencies and all small \(\Delta\).

\subsection{Combined result}\label{combined-result}

The contributions of the two frequency ranges contain opposite terms that
depend on the arbitrary cutoff \(\Lambda\). Adding
Eqs.~\eqref{eq:next-order-low-frequency-result}
and~\eqref{eq:next-order-high-frequency-result} cancels both the
\(1/\Lambda\) and the \(\log\Lambda\) terms. After division by
\(\Delta^2\), the remaining cutoff errors are bounded by
\(M\Lambda+|x[B_{\rm ch}(\Lambda)-B_{\rm high}]|\). We first choose
\(\Lambda\) to make this bound small, then choose \(\Delta\) to control the
remaining \(o(\Delta^2)\) terms. For fixed
\(x=\sin^2(\varphi/2)\) with \(x\tau_n<1\) in every channel, this gives
{}
\begin{formalized}{eq:next-order-final-expansion}{Rendered from Lean 4 theorem RmtNlin.manuscriptPhysicalMatchedExpansion (RmtNlin/MatchedEnergyViews.lean:44). 4 direct premises; rendered forms and their uses are in the generated assumption ledger. 51 project definitions are exposed in the assumptions and definitions ledger. Zero direct premises does not mean unrestricted model objects.}
\begin{dmath}[compact,label={eq:next-order-final-expansion}]
U(\varphi{}) - U(0) = U_{\mathrm{short}}(\varphi{}) - U_{\mathrm{short}}(0) + \Delta{}^{2}\,(x\,C\,\log \Delta{} + B_{\mathrm{low}}(x) + x\,B_{\mathrm{high}}) + o\left(\Delta{}^{2}\right),
\end{dmath}
\end{formalized}%
\noindent
where the two finite contributions combine into
{}
\begin{formalized}{eq:next-order-finite-term}{Rendered from Lean 4 definition RmtNlin.nextOrderFiniteTerm (RmtNlin/NextOrder.lean:206). 1 project definitions are exposed in the assumptions and definitions ledger. Zero direct premises does not mean unrestricted model objects.}
\begin{dmath}[compact,label={eq:next-order-finite-term}]
B(x) = B_{\mathrm{low}}(x) + x\,B_{\mathrm{high}}.
\end{dmath}
\end{formalized}%
\noindent
These coefficients are those of Eq.~\eqref{eq:short-junction-next-order}:
the coefficient \(C\) sets the logarithmic term, the function
\(B_{\rm low}(x)\) depends only on \(\tau_n\), \(\beta_n\), and \(\gamma_n\),
and \(B_{\rm high}\) contains the energy dependence of the normal system
away from the Fermi level.

The logarithm \(\log\Delta\) is taken in the same microscopic energy units
as \(B\). Replacing \(\log\Delta\) by \(\log(\Delta/E_*)\) shifts \(B(x)\)
by \(xC\log E_*\), so only their combination in
Eq.~\eqref{eq:next-order-final-expansion} is independent of this choice.

\begin{figure}[t]
\centering
\noindent\includegraphics[width=\linewidth]{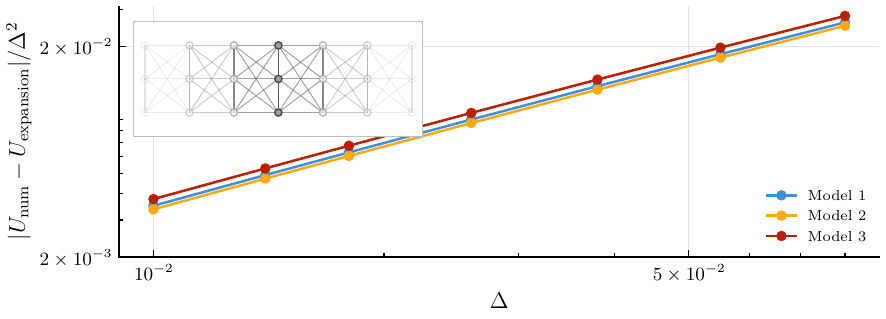}%
\caption{Numerical validation of the short-junction expansion to
next-to-leading order for three random three-orbital realizations. The plotted
residual is the difference between the finite-gap energy and the asymptotic
energy in Eq.~\eqref{eq:next-order-final-expansion}, divided by \(\Delta^2\). Its decay
checks the stated \(o(\Delta^2)\) remainder. The inset sketches the
three-orbital central site and semi-infinite leads.}
\label{fig:next-order-validation}
\end{figure}

We validate the complete expansion in three random three-orbital models
with one central site and two semi-infinite leads. The reference
calculation integrates the finite-gap transmissions without expanding in
\(\Delta\), while the comparison uses the coefficients and finite parts
derived above. We also check that the integrand obtained from the
lead-space determinant in Eq.~\eqref{eq:next-order-finite-green-determinant}
agrees with the one obtained from the full determinant of \(M_C\), and that
the lead self-energies agree with solutions of their boundary equations and
with self-energies computed with Kwant \citep{Groth14}. The code that generates the model parameters and these data is archived
in Ref.~\citep{Akhmerov26Reproduction}.
Figure~\ref{fig:next-order-validation} shows the difference between the
finite-gap and asymptotic energies divided by \(\Delta^2\) at fixed phase;
its decrease towards zero checks the \(o(\Delta^2)\) remainder.

\section{Quartic bounds from the frequency dependence}
\label{appendix-zero-temperature-quartic-bound}

We prove the finite- and zero-temperature bounds from one local constraint on
the frequency dependence of each transmission channel. We first derive this
constraint and then apply it to the frequency integral at zero temperature
and to the Matsubara sum at finite temperature.

\subsection{Channel velocity}

The constraint follows from the analytic structure of the reduced scattering
matrix. We continue it from the positive frequency axis to \(z\) in the right
half-plane:
{}
\begin{formalized}{eq:zero-bound-schur-family}{Rendered from Lean 4 theorem RmtNlin.manuscriptZeroBoundSchurFamily (RmtNlin/ResolventVelocity.lean:659). 4 project definitions are exposed in the assumptions and definitions ledger. Zero direct premises does not mean unrestricted model objects.}
\begin{dmath}[compact,label={eq:zero-bound-schur-family}]
S_{z} = (1 - 2\,i{}\,L^{\dagger}\,A_{z}^{-1}\,L)\,Q,\qquad A_{z} = i{}\,z - h + i{}\,L\,L^{\dagger}.
\end{dmath}
\end{formalized}%
\noindent
For every nonzero \(\psi\) and \(\operatorname{Re}z>0\),
{}
\begin{formalized}{eq:zero-bound-schur-invertibility}{Rendered from Lean 4 theorem RmtNlin.manuscriptZeroBoundSchurInvertibility (RmtNlin/ResolventVelocity.lean:225). 3 direct premises; rendered forms and their uses are in the generated assumption ledger. 5 project definitions are exposed in the assumptions and definitions ledger. Zero direct premises does not mean unrestricted model objects.}
\begin{dmath}[compact,label={eq:zero-bound-schur-invertibility}]
\operatorname{Im} \left\langle \psi{}, A_{z}\,\psi{}\right\rangle = (\operatorname{Re} z)\,\lVert \psi{}\rVert^{2} + \lVert L^{\dagger}\,\psi{}\rVert^{2} > 0,
\end{dmath}
\end{formalized}%
\noindent
so \(A_z\) is invertible and \(S_z\) is holomorphic. On the positive real
axis, the right and left contractivity defects of \(S_z\), which measure its
deviation from unitarity, are
{}
\begin{formalized}{eq:zero-bound-schur-defect}{Rendered from Lean 4 theorem RmtNlin.manuscriptZeroBoundSchurDefects (RmtNlin/ResolventVelocity.lean:449). 1 direct premise; rendered forms and their uses are in the generated assumption ledger. 12 project definitions are exposed in the assumptions and definitions ledger. Zero direct premises does not mean unrestricted model objects.}
\begin{dmath}[compact,label={eq:zero-bound-schur-defect}]
1 - S^{\dagger}\,S = 4\,\omega{}\,Q^{\dagger}\,L^{\dagger}\,(A_{\omega{}}^{-1})^{\dagger}\,A_{\omega{}}^{-1}\,L\,Q,\qquad 1 - S\,S^{\dagger} = 4\,\omega{}\,L^{\dagger}\,A_{\omega{}}^{-1}\,(A_{\omega{}}^{-1})^{\dagger}\,L.
\end{dmath}
\end{formalized}%
\noindent
Both right-hand sides are positive semidefinite. At complex \(z\), the same
calculation holds with \(\omega\) replaced by \(\operatorname{Re}z\) and \(h\)
by the Hermitian matrix \(h+(\operatorname{Im}z)1\), which leaves \(A_z\)
unchanged, and the prefactors become \(4\operatorname{Re}z>0\). Thus \(S_z\) is
holomorphic and contractive throughout the right half-plane, which makes it
an operator-valued Schur function, the matrix analogue of an analytic function
with modulus at most one \citep{Ball15}. We now derive the
corresponding Schwarz--Pick bound \citep{Anderson06} from the Cauchy--Schwarz
inequality.

For a positive generalized eigenvalue \(\mu\) of the pair \((F,D)\) in
Eq.~\eqref{eq:d-and-f-definitions}, we choose an eigenvector \(u\) normalized
by \(\langle u,D_\omega u\rangle=1\) and set
{}
\begin{formalized}{eq:zero-bound-eigenchannel-vectors}{Rendered from Lean 4 theorem RmtNlin.manuscriptZeroBoundEigenchannelVectors (RmtNlin/EigenchannelVelocity.lean:275). 6 project definitions are exposed in the assumptions and definitions ledger. Zero direct premises does not mean unrestricted model objects.}
\begin{dmath}[compact,label={eq:zero-bound-eigenchannel-vectors}]
v = S_{\omega{}}\,u(\omega{}),\qquad a = [J - \mu{}(\omega{})]\,v.
\end{dmath}
\end{formalized}%
\noindent
In terms of these vectors, the eigenvalue equation reads
\(S_\omega^\dagger a=-(J-\mu)u\). Differentiating the resolvent
\(R=A_\omega^{-1}\) along the positive real axis gives
{}
\begin{formalized}{eq:zero-bound-resolvent-derivative}{Rendered from Lean 4 theorem RmtNlin.manuscriptZeroBoundResolventDerivative (RmtNlin/ResolventVelocity.lean:683). 2 direct premises; rendered forms and their uses are in the generated assumption ledger. 7 project definitions are exposed in the assumptions and definitions ledger. Zero direct premises does not mean unrestricted model objects.}
\begin{dmath}[compact,label={eq:zero-bound-resolvent-derivative}]
\frac{\mathrm d A_{z}^{-1}}{\mathrm d z} = -i{}\,A_{z}^{-2},\qquad \frac{\mathrm d S_{z}}{\mathrm d z} = -2\,L^{\dagger}\,A_{z}^{-2}\,L\,Q.
\end{dmath}
\end{formalized}%
\noindent
With the two resolvent vectors
{}
\begin{formalized}{aux:zero-bound-resolvent-vectors}{Rendered from Lean 4 theorem RmtNlin.manuscriptZeroBoundResolventVectors (RmtNlin/ResolventVelocity.lean:879). 2 project definitions are exposed in the assumptions and definitions ledger. Zero direct premises does not mean unrestricted model objects.}
\begin{dmath}[compact,label={aux:zero-bound-resolvent-vectors}]
x = R\,L\,Q\,u,\qquad y = R^{\dagger}\,L\,a,
\end{dmath}
\end{formalized}%
\noindent
the contractivity defects and the derivative matrix element factor as
{}
\begin{formalized}{eq:zero-bound-defect-factorization}{Rendered from Lean 4 theorem RmtNlin.manuscriptZeroBoundDefectFactorization (RmtNlin/ResolventVelocity.lean:890). 1 direct premise; rendered forms and their uses are in the generated assumption ledger. 18 project definitions are exposed in the assumptions and definitions ledger. Zero direct premises does not mean unrestricted model objects.}
\begin{equation}\begin{aligned}
\left\langle u, (1 - S^{\dagger}\,S)\,u\right\rangle &= 4\,\omega{}\,\lVert x\rVert^{2},\qquad \left\langle a, (1 - S\,S^{\dagger})\,a\right\rangle = 4\,\omega{}\,\lVert y\rVert^{2}, \\
\left\langle a, S'\,u\right\rangle &= -2\,\left\langle y, x\right\rangle,\qquad \lvert \left\langle a, S'\,u\right\rangle\rvert = 2\,\lvert \left\langle y, x\right\rangle\rvert.
\end{aligned}\label{eq:zero-bound-defect-factorization}\end{equation}
\end{formalized}%
\noindent
The Cauchy--Schwarz inequality for \(x\) and \(y\) gives
{}
\begin{formalized}{aux:zero-bound-resolvent-cauchy-schwarz}{Rendered from Lean 4 theorem RmtNlin.manuscriptScatteringDerivativeMatrixElement_le (RmtNlin/ResolventVelocity.lean:923). 3 project definitions are exposed in the assumptions and definitions ledger. Zero direct premises does not mean unrestricted model objects.}
\begin{dmath}[compact,label={aux:zero-bound-resolvent-cauchy-schwarz}]
\lvert \left\langle a, S'\,u\right\rangle\rvert \le 2\,\lVert y\rVert\,\lVert x\rVert.
\end{dmath}
\end{formalized}%
\noindent
Because \(\omega>0\), taking the square roots of the two defect
factorizations yields
{}
\begin{formalized}{eq:zero-bound-derivative-defect-bound}{Rendered from Lean 4 theorem RmtNlin.manuscriptScatteringDerivative_defect_bound (RmtNlin/ResolventVelocity.lean:1016). 1 direct premise; rendered forms and their uses are in the generated assumption ledger. 19 project definitions are exposed in the assumptions and definitions ledger. Zero direct premises does not mean unrestricted model objects.}
\begin{dmath}[compact,label={eq:zero-bound-derivative-defect-bound}]
2\,\omega{}\,\lvert \left\langle a, S'\,u\right\rangle\rvert \le \sqrt{\left\langle u, (1 - S^{\dagger}\,S)\,u\right\rangle}\,\sqrt{\left\langle a, (1 - S\,S^{\dagger})\,a\right\rangle}.
\end{dmath}
\end{formalized}%
\noindent
The frequency derivatives of \(F\) and \(D\) enter through the quadratic-form
identity
{}
\begin{formalized}{aux:zero-bound-hellmann-feynman-derivative}{Rendered from Lean 4 theorem RmtNlin.manuscriptZeroBoundPointwiseDerivative (RmtNlin/EigenchannelVelocity.lean:506). 1 direct premise; rendered forms and their uses are in the generated assumption ledger. 3 project definitions are exposed in the assumptions and definitions ledger. Zero direct premises does not mean unrestricted model objects.}
\begin{dmath}[compact,label={aux:zero-bound-hellmann-feynman-derivative}]
\left\langle u, (F' - \mu{}\,D')\,u\right\rangle = 2\,\operatorname{Re} \left\langle (J - \mu{})\,S\,u, S'\,u\right\rangle.
\end{dmath}
\end{formalized}%
\noindent
Together with \(|\operatorname{Re}w|\le|w|\),
Eq.~\eqref{eq:zero-bound-derivative-defect-bound} bounds this quadratic form
by the product of the two defect factors. Both factors depend on the
eigenvector through
{}
\begin{formalized}{eq:zero-bound-norm-defect}{Rendered from Lean 4 theorem RmtNlin.manuscriptZeroBoundNormDefect (RmtNlin/EigenchannelVelocity.lean:291). 6 project definitions are exposed in the assumptions and definitions ledger. Zero direct premises does not mean unrestricted model objects.}
\begin{dmath}[compact,label={eq:zero-bound-norm-defect}]
\delta = \lVert u(\omega{})\rVert^{2} - \lVert v\rVert^{2}.
\end{dmath}
\end{formalized}%
\noindent
The normalization and the eigenvalue equation give
{}
\begin{formalized}{eq:zero-bound-normalization-identities}{Rendered from Lean 4 theorem RmtNlin.manuscriptZeroBoundNormalizationIdentities (RmtNlin/EigenchannelVelocity.lean:242). 1 direct premise; rendered forms and their uses are in the generated assumption ledger. 12 project definitions are exposed in the assumptions and definitions ledger. Zero direct premises does not mean unrestricted model objects.}
\begin{dmath}[compact,label={eq:zero-bound-normalization-identities}]
\lVert u(\omega{})\rVert^{2} + \lVert S_{\omega{}}\,u(\omega{})\rVert^{2} = 1,\qquad \left\langle u(\omega{}), J\,u(\omega{})\right\rangle + \left\langle S_{\omega{}}\,u(\omega{}), J\,S_{\omega{}}\,u(\omega{})\right\rangle = \mu{}(\omega{}).
\end{dmath}
\end{formalized}%
\noindent
Because the spectrum of \(J\) is \(\{-1,1\}\), we have
\(\langle u,Ju\rangle\le\lVert u\rVert^2\) and
\(\langle v,Jv\rangle\ge-\lVert v\rVert^2\). Combining each inequality with
Eq.~\eqref{eq:zero-bound-normalization-identities} gives
{}
\begin{formalized}{eq:zero-bound-signature-bounds}{Rendered from Lean 4 theorem RmtNlin.manuscriptZeroBoundSignatureBounds (RmtNlin/EigenchannelVelocity.lean:304). 1 direct premise; rendered forms and their uses are in the generated assumption ledger. 14 project definitions are exposed in the assumptions and definitions ledger. Zero direct premises does not mean unrestricted model objects.}
\begin{equation}\begin{aligned}
\left\langle u(\omega{}), J\,u(\omega{})\right\rangle - \left\langle v, J\,v\right\rangle &\le 1 + \delta - \mu{}(\omega{}), \\
\left\langle u(\omega{}), J\,u(\omega{})\right\rangle - \left\langle v, J\,v\right\rangle &\le 1 + \mu{}(\omega{}) - \delta,
\end{aligned}\label{eq:zero-bound-signature-bounds}\end{equation}
\end{formalized}%
\noindent
and the smaller right-hand side yields
{}
\begin{formalized}{eq:zero-bound-signature-inequality}{Rendered from Lean 4 theorem RmtNlin.manuscriptSignatureDifference_le (RmtNlin/EigenchannelVelocity.lean:322). 1 direct premise; rendered forms and their uses are in the generated assumption ledger. 14 project definitions are exposed in the assumptions and definitions ledger. Zero direct premises does not mean unrestricted model objects.}
\begin{dmath}[compact,label={eq:zero-bound-signature-inequality}]
\left\langle u(\omega{}), J\,u(\omega{})\right\rangle - \left\langle v, J\,v\right\rangle \le 1 - \lvert \delta - \mu{}(\omega{})\rvert.
\end{dmath}
\end{formalized}%
\noindent

The quadratic form under the first square root in
Eq.~\eqref{eq:zero-bound-derivative-defect-bound} equals \(\delta\), and the
choice of \(a\) expresses the second quadratic form through \(\delta\) and
\(\mu\):
{}
\begin{formalized}{eq:zero-bound-second-defect}{Rendered from Lean 4 theorem RmtNlin.manuscriptZeroBoundSecondDefect (RmtNlin/EigenchannelVelocity.lean:388). 1 direct premise; rendered forms and their uses are in the generated assumption ledger. 19 project definitions are exposed in the assumptions and definitions ledger. Zero direct premises does not mean unrestricted model objects.}
\begin{equation}\begin{aligned}
\left\langle a, (1 - S_{\omega{}}\,S_{\omega{}}^{\dagger})\,a\right\rangle &= -[1 + \mu{}(\omega{})^{2}]\,\delta + 2\,\mu{}(\omega{})\,\left(\left\langle u(\omega{}), J\,u(\omega{})\right\rangle - \left\langle v, J\,v\right\rangle\right) \\
&\le -[1 + \mu{}(\omega{})^{2}]\,\delta + 2\,\mu{}(\omega{})\,(1 - \lvert \delta - \mu{}(\omega{})\rvert).
\end{aligned}\label{eq:zero-bound-second-defect}\end{equation}
\end{formalized}%
\noindent
Contractivity and normalization restrict \(\delta\) to \(0\le\delta\le1\). On
this interval, the product of \(\delta\) with the upper bound in
Eq.~\eqref{eq:zero-bound-second-defect} increases up to \(\delta=\mu\) and then
decreases, so its maximum gives
{}
\begin{formalized}{eq:zero-bound-defect-product}{Rendered from Lean 4 theorem RmtNlin.manuscriptZeroBoundDefectProduct (RmtNlin/EigenchannelVelocity.lean:423). 7 direct premises; rendered forms and their uses are in the generated assumption ledger. 12 project definitions are exposed in the assumptions and definitions ledger. Zero direct premises does not mean unrestricted model objects.}
\begin{dmath}[compact,label={eq:zero-bound-defect-product}]
\delta\,\left\langle a, (1 - S\,S^{\dagger})\,a\right\rangle \le \mu{}^{2}\,(1 - \mu{}^{2}).
\end{dmath}
\end{formalized}%
\noindent
Substituting this bound into
Eq.~\eqref{eq:zero-bound-derivative-defect-bound} bounds the quadratic form
\(\langle u,(F'-\mu D')u\rangle\). The sorted eigenvalues can have kinks where
they cross, so we use the Courant--Fischer theorem. With it, the same estimate bounds from both sides how fast each
sorted eigenvalue changes as \(\omega\) increases, also at a kink; formally, it
bounds the right Dini derivatives, the upper and lower limits of the
difference quotients from the right. Wherever the sorted eigenvalues are
differentiable, the estimate gives
{}
\begin{formalized}{eq:zero-bound-mu-velocity}{Rendered from Lean 4 theorem RmtNlin.manuscriptZeroBoundMuVelocity (RmtNlin/SortedChannels.lean:442). 5 direct premises; rendered forms and their uses are in the generated assumption ledger. 35 project definitions are exposed in the assumptions and definitions ledger. Zero direct premises does not mean unrestricted model objects.}
\begin{dmath}[compact,label={eq:zero-bound-mu-velocity}]
\lvert \omega{}\,\dot{\mu}_{n}(\omega{})\rvert \le \lvert \mu_{n}(\omega{})\rvert\,\sqrt{T_{n}(\omega{})}.
\end{dmath}
\end{formalized}%
\noindent

The factor \(\lvert\mu_n\rvert\sqrt{T_n}\) makes this velocity bound depend on the
channel. The transmission rapidity removes this dependence:
{}
\begin{formalized}{eq:zero-bound-rapidity}{Rendered from Lean 4 theorem RmtNlin.manuscriptZeroBoundRapidity (RmtNlin/SortedChannels.lean:484). 5 direct premises; rendered forms and their uses are in the generated assumption ledger. 28 project definitions are exposed in the assumptions and definitions ledger. Zero direct premises does not mean unrestricted model objects.}
\begin{dmath}[compact,label={eq:zero-bound-rapidity}]
q_{n}(\omega{}) = \operatorname{arcosh} \frac{1}{\lvert \mu_{n}(\omega{})\rvert} = \operatorname{artanh} \sqrt{T_{n}(\omega{})}.
\end{dmath}
\end{formalized}%
\noindent
The bounds on the right Dini derivatives imply that each sorted rapidity changes
by at most \(\lvert\log(\omega_2/\omega_1)\rvert\) between any two frequencies
\(\omega_1,\omega_2>0\). Each sorted rapidity is therefore locally absolutely
continuous on \(\omega>0\): on every closed interval of positive frequencies,
it is differentiable except on a set of zero total length, and its change
equals the integral of its derivative. Wherever the rapidity is
differentiable, the chain rule gives
{}
\begin{formalized}{eq:zero-bound-logarithmic-velocity}{Rendered from Lean 4 theorem RmtNlin.manuscriptPhysicalChannelLogarithmicVelocity (RmtNlin/SortedChannels.lean:711). 6 direct premises; rendered forms and their uses are in the generated assumption ledger. 43 project definitions are exposed in the assumptions and definitions ledger. Zero direct premises does not mean unrestricted model objects.}
\begin{dmath}[compact,label={eq:zero-bound-logarithmic-velocity}]
\lvert \omega{}\,\dot{q}_{n}(\omega{})\rvert \le 1.
\end{dmath}
\end{formalized}%
\noindent
Because the argument works with the sorted channels, it holds through
eigenvalue crossings. Since the change of each rapidity equals the integral of
its derivative, the integral comparisons below use the chain rule only at
frequencies where the rapidity is differentiable. At
\(\omega>0\), the quadratic form of \(F\) is positive on the channels of the
left superconductor and negative on those of the right one, while \(D\) is
positive definite. By the Courant--Fischer theorem, no generalized eigenvalue
vanishes, so \(\lvert\mu_n(\omega)\rvert>0\) and
\(T_n(\omega)<1\): perfect transmission can occur only at \(\omega=0\). We
therefore integrate from a lower cutoff \(\varepsilon\) and let
\(\varepsilon\downarrow0\); the cutoff changes each moment by at most
\(N\varepsilon\), because \(0\le T_n\le1\) and at most \(N\) channels are
nonzero.

\subsection{Zero-temperature moment bound}

The velocity bound turns the moment inequality into a one-dimensional
comparison problem. We prove for every integer \(p\ge2\) that the \(p\)th
moment of each channel is at most \(C_p\) times its \((p-1)\)st moment. For one
channel, this amounts to showing that the frequency integral of the signed
density \(f_p(T)=T^p-C_pT^{p-1}\) is nonpositive. The density is positive at
high transmission and negative at low transmission, so a pointwise estimate
cannot prove the result. We compare instead the accumulated positive and
negative contributions along the channel trajectory.

The natural comparison trajectory decreases at the maximum speed allowed by
Eq.~\eqref{eq:zero-bound-logarithmic-velocity}. Its rapidity is
\(q_\star(\omega)=\log(\Omega/\omega)\) until it reaches zero at
\(\omega=\Omega\), and its transmission vanishes thereafter. On the nonzero
part of the trajectory, \(r=\omega/\Omega=e^{-q_\star}\) ranges from zero to
one, and we denote the comparison transmission by \(\tau(r)\). Requiring this
trajectory to have zero total signed moment fixes \(C_p\), independently of
\(\Omega\):
{}
\begin{formalized}{eq:zero-bound-calibration-constant}{Rendered from Lean 4 theorem RmtNlin.manuscriptZeroBoundCalibrationConstant (RmtNlin/ZeroTemperatureMoment.lean:44). 3 project definitions are exposed in the assumptions and definitions ledger. Zero direct premises does not mean unrestricted model objects.}
\begin{dmath}[compact,label={eq:zero-bound-calibration-constant}]
C_{p} = \frac{I_{p}}{I_{p - 1}},\qquad I_{p} = \int_{0}^{1}\tau{}(r)^{p}\,\mathrm dr,\qquad \tau{}(r) = \left(\frac{1 - r^{2}}{1 + r^{2}}\right)^{2}.
\end{dmath}
\end{formalized}%
\noindent

We next attach to every point \((\omega,q)\) the signed moment accumulated by
the comparison trajectory through that point. This trajectory has scale
\(\Omega=\omega e^q\), and its contribution between zero and \(\omega\) is
\(\omega G_p(q)\), where
{}
\begin{formalized}{eq:zero-bound-calibration}{Rendered from Lean 4 theorem RmtNlin.manuscriptZeroBoundCalibrationDefinition (RmtNlin/ZeroTemperatureCalibration.lean:91). 5 project definitions are exposed in the assumptions and definitions ledger. Zero direct premises does not mean unrestricted model objects.}
\begin{dmath}[compact,label={eq:zero-bound-calibration}]
G_{p}(q) = e^{q}\,\int_{0}^{e^{-q}}[\tau(r)^{p} - C_{p}\,\tau(r)^{p - 1}]\,\mathrm dr.
\end{dmath}
\end{formalized}%
\noindent

The corresponding integral over \(0\le r\le1\) vanishes by the definition of
\(C_p\). Because \(\tau(r)\) decreases from one to zero, the signed moment
density changes sign once, from positive to negative, so every partial
integral starting at zero is nonnegative.

The function \(G_p(q)\) is the average of the signed density over the
comparison trajectory up to the point \((\omega,q)\). Before the sign change,
this average is at least the value at the endpoint, because the density
decreases. After the sign change, the average remains nonnegative while the
endpoint value is nonpositive. Differentiating and using
\(T=\tau(e^{-q})\) gives
{}
\begin{formalized}{eq:zero-bound-calibration-identities}{Rendered from Lean 4 theorem RmtNlin.manuscriptZeroBoundCalibrationIdentities (RmtNlin/ZeroTemperatureCalibration.lean:469). 2 direct premises; rendered forms and their uses are in the generated assumption ledger. 6 project definitions are exposed in the assumptions and definitions ledger. Zero direct premises does not mean unrestricted model objects.}
\begin{equation}\begin{aligned}
G_{p}'(q) &= G_{p}(q) - [\tau(e^{-q})^{p} - C_{p}\,\tau(e^{-q})^{p - 1}] \ge 0, \\
G_{p}(q) - G_{p}'(q) &= \tau(e^{-q})^{p} - C_{p}\,\tau(e^{-q})^{p - 1}.
\end{aligned}\label{eq:zero-bound-calibration-identities}\end{equation}
\end{formalized}%
\noindent

We now compare the signed moment accumulated by a sorted channel with that of
its comparison trajectory. Wherever \(q_n\) is differentiable, the chain rule,
\(G_p'\ge0\), and the velocity bound imply that the total derivative of
\(\omega G_p\) is smallest at \(dq/d\log\omega=-1\):
{}
\begin{formalized}{eq:zero-bound-total-derivative}{Rendered from Lean 4 theorem RmtNlin.manuscriptZeroBoundTotalDerivative (RmtNlin/ZeroTemperatureSortedInputs.lean:372). 7 direct premises; rendered forms and their uses are in the generated assumption ledger. 51 project definitions are exposed in the assumptions and definitions ledger. Zero direct premises does not mean unrestricted model objects.}
\begin{equation}\begin{aligned}
\frac{\mathrm d}{\mathrm d \omega{}}[\omega{}\,G_{p}(q_{n}(\omega{}))] &= G_{p}(q_{n}(\omega{})) + G_{p}'(q_{n}(\omega{}))\,\omega{}\,\dot{q}_{n}(\omega{}) \\
&\ge G_{p}(q_{n}(\omega{})) - G_{p}'(q_{n}(\omega{})) \\
&= T_{n}(\omega{})^{p} - C_{p}\,T_{n}(\omega{})^{p - 1}.
\end{aligned}\label{eq:zero-bound-total-derivative}\end{equation}
\end{formalized}%
\noindent
The last line is the signed moment density of the sorted channel, which is
therefore bounded by the total derivative of \(\omega G_p(q_n(\omega))\).
For a finite-dimensional realization, \(\omega G_p(q_n(\omega))\) vanishes at
both integration endpoints. Because \(q_n\) is locally absolutely continuous,
integrating Eq.~\eqref{eq:zero-bound-total-derivative} therefore gives
{}
\begin{formalized}{eq:zero-bound-one-channel-moment}{Rendered from Lean 4 theorem RmtNlin.manuscriptPhysicalPowerMoment_le (RmtNlin/PhysicalZeroTemperatureBound.lean:118). 4 direct premises; rendered forms and their uses are in the generated assumption ledger. 65 project definitions are exposed in the assumptions and definitions ledger. Zero direct premises does not mean unrestricted model objects.}
\begin{dmath}[compact,label={eq:zero-bound-one-channel-moment}]
\int_{0}^{\infty}T_{n}(\omega{})^{p}\,\mathrm d\omega{} \le C_{p}\,\int_{0}^{\infty}T_{n}(\omega{})^{p - 1}\,\mathrm d\omega{}.
\end{dmath}
\end{formalized}%
\noindent

Summing over channels bounds the total moments
\(M_{p} = \int_{0}^{\infty}\sum_{n}T_{n}(\omega{})^{p}\,\mathrm d\omega{}\):
{}
\begin{formalized}{eq:zero-bound-moment-hierarchy}{Rendered from Lean 4 theorem RmtNlin.manuscriptPhysicalZeroBoundMomentHierarchy (RmtNlin/PhysicalZeroTemperatureBound.lean:161). 4 direct premises; rendered forms and their uses are in the generated assumption ledger. 66 project definitions are exposed in the assumptions and definitions ledger. Zero direct premises does not mean unrestricted model objects.}
\begin{dmath}[compact,label={eq:zero-bound-moment-hierarchy}]
M_{p} \le C_{p}\,M_{p - 1}.
\end{dmath}
\end{formalized}%
\noindent
For the quartic coefficient, only \(p=2\) is needed. The two integrals that
fix \(C_2\) are
\(\int_{0}^{1}\tau(r)\,\mathrm dr = 2 - \pi/2,\ \int_{0}^{1}\tau(r)^{2}\,\mathrm dr = 8/3 - 3\,\pi/4\),
and their ratio yields
{}
\begin{formalized}{eq:zero-bound-moment-ratio}{Rendered from Lean 4 theorem RmtNlin.manuscriptPhysicalZeroBoundMomentRatio (RmtNlin/PhysicalZeroTemperatureBound.lean:181). 4 direct premises; rendered forms and their uses are in the generated assumption ledger. 69 project definitions are exposed in the assumptions and definitions ledger. Zero direct premises does not mean unrestricted model objects.}
\begin{dmath}[compact,label={eq:zero-bound-moment-ratio}]
\frac{M_{2}}{M_{1}} \le C_{2} = \frac{32 - 9\,\pi}{6\,(4 - \pi)}.
\end{dmath}
\end{formalized}%
\noindent
Together with the coefficient formulas in
Eq.~\eqref{eq:coefficient-expansion}, this moment ratio gives the upper bound
in Eq.~\eqref{eq:zero-temperature-normalized-quartic-bound}.

\subsection{Finite-temperature increase}
\label{appendix-finite-temperature-increase}

The zero-temperature moment bound does not carry over by simply replacing the
frequency integral with a Matsubara sum, because the Matsubara grid can
sample frequencies at which a channel transmits more than it does on average
in the continuous integral. A three-orbital junction shows this effect. Its
matrices are
{}
\begin{formalized}{eq:finite-temperature-example-matrix-elements}{Rendered from Lean 4 theorem RmtNlin.manuscriptFiniteTemperatureExampleMatrixElements (RmtNlin/FiniteTemperatureExample.lean:50). 5 project definitions are exposed in the assumptions and definitions ledger. Zero direct premises does not mean unrestricted model objects.}
\begin{dmath}[compact,label={eq:finite-temperature-example-matrix-elements}]
h = \begin{pmatrix}0&\frac{1}{2}&\frac{1}{2}\\\frac{1}{2}&0&0\\\frac{1}{2}&0&0\end{pmatrix},\qquad \Delta_{L} = \operatorname{diag}(1, 0, 0),\qquad \Delta_{R} = \operatorname{diag}(0, 1, 0).
\end{dmath}
\end{formalized}%
\noindent
The first and second orbitals couple to the left and right superconductors,
respectively, and the third is side-coupled to the first. The only nonzero
transmission is
{}
\begin{formalized}{eq:finite-temperature-example-transmission}{Rendered from Lean 4 theorem RmtNlin.manuscriptFiniteTemperatureExampleTransmission (RmtNlin/FiniteTemperatureExample.lean:137). 1 direct premise; rendered forms and their uses are in the generated assumption ledger. 5 project definitions are exposed in the assumptions and definitions ledger. Zero direct premises does not mean unrestricted model objects.}
\begin{dmath}[compact,label={eq:finite-temperature-example-transmission}]
T_{\star}(\omega{}) = \frac{16\,\omega{}^{2}}{16\,\omega{}^{6} + 48\,\omega{}^{4} + 36\,\omega{}^{2} + 1},
\end{dmath}
\end{formalized}%
\noindent
which vanishes at \(\omega=0\) and takes the value
\(T_\star(1/2)=16/53\). At
\(\beta=2\pi\), the fermionic Matsubara frequencies are the half-integers, and
the continuous moments and the corresponding discrete sums give
{}
\begin{formalized}{eq:finite-temperature-example-alpha-values}{Rendered from Lean 4 theorem RmtNlin.manuscriptFiniteTemperatureExampleAlphaValues (RmtNlin/FiniteTemperatureExampleEnclosure.lean:674). Additional axioms: RmtNlin.FiniteTemperatureExampleCertificate.rational_grid_certificate._native.native_decide.ax_1_1 33 project definitions are exposed in the assumptions and definitions ledger. Zero direct premises does not mean unrestricted model objects.}
\begin{dmath}[compact,label={eq:finite-temperature-example-alpha-values}]
\alpha_{r,\star}(\beta\to\infty) \approx -0.70 < -0.64 \approx \alpha_{r,\star}(\beta=2\pi).
\end{dmath}
\end{formalized}%
\noindent
Finite temperature thus increases \(\alpha_r\) by about
\(0.063\) in
this example. The executable calculation reproduces the discrete-sum value
independently from the finite-temperature BdG free energy. Because the
integral comparison does not transfer directly to Matsubara sums, we give a
separate finite-temperature argument.

\subsection{Finite-temperature moment bound}\label{finite-temperature-moment-bound}

At finite temperature, the same comparison problem is sampled on the
Matsubara grid \(\omega_m=r_m\pi/\beta\) with \(r_m=2m+1\). We write
\(q_m=q_n(\omega_m)\) for the rapidity of a fixed channel at these nodes. The
bound \(\lvert\log(\omega_2/\omega_1)\rvert\) on the change of the rapidity
between two frequencies, applied to consecutive Matsubara frequencies, gives
{}
\begin{formalized}{eq:finite-bound-discrete-velocity}{Rendered from Lean 4 theorem RmtNlin.manuscriptFiniteBoundDiscreteVelocity (RmtNlin/PhysicalQuarticBound.lean:65). 4 direct premises; rendered forms and their uses are in the generated assumption ledger. 50 project definitions are exposed in the assumptions and definitions ledger. Zero direct premises does not mean unrestricted model objects.}
\begin{dmath}[compact,label={eq:finite-bound-discrete-velocity}]
\lvert q_{n}(\omega_{m + 1}) - q_{n}(\omega_{m})\rvert \le \log \frac{r_{m + 1}}{r_{m}}.
\end{dmath}
\end{formalized}%
\noindent
The quartic bound follows if the discrete second moment of every channel is at
most \(29/40\) times its first moment. Because \(T=\tanh^2q\), the difference
between the two sides has the summand
{}
\begin{formalized}{eq:finite-bound-signed-density}{Rendered from Lean 4 definition RmtNlin.thermalSignedDensity (RmtNlin/FiniteTemperatureBellman.lean:25). 1 project definitions are exposed in the assumptions and definitions ledger. Zero direct premises does not mean unrestricted model objects.}
\begin{dmath}[compact,label={eq:finite-bound-signed-density}]
f(q) = \tanh^{4} q - \frac{29}{40}\,\tanh^{2} q,
\end{dmath}
\end{formalized}%
\noindent
and it remains to prove \(\sum_m f(q_m)\le0\).

The discrete counterpart of the zero-temperature comparison trajectory
descends by the largest step allowed between consecutive nodes: starting from
\(q\) at node \(m\), its rapidity at node \(j\) is
\(\max\{q-\log(r_j/r_m),0\}\). With \(h(z)=\tau(z)^2-(29/40)\tau(z)\), where
\(\tau\) from Eq.~\eqref{eq:zero-bound-calibration-constant} is extended by
zero for \(z\ge1\), the tail sum along this sequence is
{}
\begin{formalized}{eq:finite-bound-descending-tail}{Rendered from Lean 4 theorem RmtNlin.manuscriptFiniteBoundDescendingTail (RmtNlin/FiniteTemperatureTail.lean:773). 4 project definitions are exposed in the assumptions and definitions ledger. Zero direct premises does not mean unrestricted model objects.}
\begin{dmath}[compact,label={eq:finite-bound-descending-tail}]
H_{m}(q) = \sum_{k=0}^{\infty}h\left(\frac{r_{m + k}}{r_{m}\,e^{q}}\right).
\end{dmath}
\end{formalized}%
\noindent
The tail sum \(H_m\) is the discrete analogue of the accumulated comparison
moment \(\omega G_p\), except that it collects the remaining rather than the
preceding contribution. Separating the current node gives the recursion
{}
\begin{formalized}{eq:finite-bound-tail-recursion}{Rendered from Lean 4 theorem RmtNlin.manuscriptFiniteBoundTailRecursion (RmtNlin/FiniteTemperatureTail.lean:781). 1 direct premise; rendered forms and their uses are in the generated assumption ledger. 6 project definitions are exposed in the assumptions and definitions ledger. Zero direct premises does not mean unrestricted model objects.}
\begin{dmath}[compact,label={eq:finite-bound-tail-recursion}]
H_{m}(q) = f(q) + H_{m + 1}\left(\max\left(q - \log \frac{r_{m + 1}}{r_{m}}, 0\right)\right).
\end{dmath}
\end{formalized}%
\noindent

The derivative of the tail sum involves \(g(z)=-zh'(z)\): differentiating the
finite sum of nonzero terms in \(H_m\) gives
\(H_m'(q)=\sum_{k\ge0}g(r_{m+k}/[r_me^q])\). Two properties make \(H_m\) an
upper bound for every admissible tail:
{}
\begin{formalized}{eq:finite-bound-tail-properties}{Rendered from Lean 4 theorem RmtNlin.manuscriptFiniteBoundTailProperties (RmtNlin/FiniteTemperatureTransfer.lean:1484). 2 direct premises; rendered forms and their uses are in the generated assumption ledger. Additional axioms: RmtNlin.FiniteTemperatureCertificate.derivativeSum_compact_certificate._native.native_decide.ax_1_1, RmtNlin.FiniteTemperatureCertificate.finiteSum_compact_certificate._native.native_decide.ax_1_1 32 project definitions are exposed in the assumptions and definitions ledger. Zero direct premises does not mean unrestricted model objects.}
\begin{dmath}[compact,label={eq:finite-bound-tail-properties}]
H_{0}(q) \le 0,\qquad \sum_{k=0}^{\infty}g\left(\frac{r_{m + k}}{r_{m}\,e^{q}}\right) \le 0,\qquad m \ge 1.
\end{dmath}
\end{formalized}%
\noindent
The first states that even the complete comparison sequence has a
nonpositive sum. The second is the nontrivial comparison step: increasing the
rapidity at the next node cannot increase the remaining tail sum. This is a
statement about the whole tail, not a pointwise ordering of \(f\).

Equation~\eqref{eq:finite-bound-discrete-velocity} gives
\(q_{m+1}\ge\max\{q_m-\log(r_{m+1}/r_m),0\}\). Because \(H_{m+1}\) is
nonincreasing, replacing the actual next value by the comparison value can
only increase the bound on the remaining tail sum. Backward induction with a
finite cutoff, followed by the convergent limit of the cutoff, therefore
gives
{}
\begin{formalized}{eq:finite-bound-bellman-result}{Rendered from Lean 4 theorem RmtNlin.manuscriptFiniteBoundBellmanResult (RmtNlin/FiniteTemperatureTransfer.lean:1496). 4 direct premises; rendered forms and their uses are in the generated assumption ledger. Additional axioms: RmtNlin.FiniteTemperatureCertificate.derivativeSum_compact_certificate._native.native_decide.ax_1_1, RmtNlin.FiniteTemperatureCertificate.finiteSum_compact_certificate._native.native_decide.ax_1_1 34 project definitions are exposed in the assumptions and definitions ledger. Zero direct premises does not mean unrestricted model objects.}
\begin{dmath}[compact,label={eq:finite-bound-bellman-result}]
\sum_{m=0}^{\infty}f(q_{m}) \le H_{0}(q_{0}) \le 0.
\end{dmath}
\end{formalized}%
\noindent

We verify the two properties in Eq.~\eqref{eq:finite-bound-tail-properties}
with the substitution \(a=r_me^q\), which writes the comparison sequence as
\(\tau(r_j/a)\). The first property then becomes the midpoint sum
{}
\begin{formalized}{eq:finite-bound-F-definition}{Rendered from Lean 4 theorem RmtNlin.manuscriptFiniteBoundFDefinition (RmtNlin/FiniteTemperatureTail.lean:790). 5 project definitions are exposed in the assumptions and definitions ledger. Zero direct premises does not mean unrestricted model objects.}
\begin{dmath}[compact,label={eq:finite-bound-F-definition}]
H_{0}(q) = F(e^{q}),\qquad F(a) = \sum_{k=0}^{\infty}h\left(\frac{2\,k + 1}{a}\right).
\end{dmath}
\end{formalized}%
\noindent
For the derivative bound, we compare \(H_m'(q)\), for \(m\ge1\), with the sum
of \(g\) over all Matsubara nodes except \(r_0=1\):
{}
\begin{formalized}{eq:finite-bound-G-definition}{Rendered from Lean 4 theorem RmtNlin.manuscriptFiniteBoundGDefinition (RmtNlin/FiniteTemperatureTail.lean:798). 3 project definitions are exposed in the assumptions and definitions ledger. Zero direct premises does not mean unrestricted model objects.}
\begin{dmath}[compact,label={eq:finite-bound-G-definition}]
G(a) = \sum_{k=0}^{\infty}g\left(\frac{2\,k + 3}{a}\right).
\end{dmath}
\end{formalized}%
\noindent
Here \(g\) is extended by zero for \(z\geq1\), and on \(0\leq z<1\) it equals
\(z^2(1-z^2)(51z^4-218z^2+51)/[5(1+z^2)^5]\). Consequently, \(F(a)\le0\) for
\(a\ge1\) proves that the complete comparison sequence has a nonpositive sum.
To prove that the remaining tail sum is nonincreasing, we use that \(g\)
changes sign once on \((0,1)\). If the first term of \(H_m'\) lies before this
sign change, the omitted lower nodes add positive terms to \(G\); if it lies
after the sign change, every retained term is already nonpositive. Hence
\(G(a)\le0\) for \(a\ge3\) implies \(H_m'\le0\) for every \(m\ge1\).

We prove these two midpoint-sum inequalities with analytic bounds at large
\(a\) and exact rational checks on the remaining finite intervals.
Polynomial root counting on \([0,1]\) gives
{}
\begin{formalized}{eq:finite-bound-derivative-bounds}{Rendered from Lean 4 theorem RmtNlin.FiniteTemperatureCertificate.manuscriptFiniteBoundDerivativeBounds (RmtNlin/FiniteTemperatureCertificate.lean:775). 2 direct premises; rendered forms and their uses are in the generated assumption ledger. 7 project definitions are exposed in the assumptions and definitions ledger. Zero direct premises does not mean unrestricted model objects.}
\begin{equation}\begin{aligned}
\lvert h''(z)\rvert &\le \frac{51}{5},\qquad \lvert g(z)\rvert \le \frac{17}{50},\qquad \lvert z\,g'(z)\rvert \le \frac{29}{20}, \\
\lvert g''(z)\rvert &\le \frac{102}{5}.
\end{aligned}\label{eq:finite-bound-derivative-bounds}\end{equation}
\end{formalized}%
\noindent
The first and last bounds control the midpoint-rule errors for \(F\) and
\(G\), and the middle two control their variation on the finite intervals.
At large \(a\), both midpoint sums are governed by the same negative integral,
{}
\begin{formalized}{aux:finite-bound-integral}{Rendered from Lean 4 theorem RmtNlin.FiniteTemperatureCertificate.manuscriptFiniteBoundIntegral (RmtNlin/FiniteTemperatureCertificate.lean:789). 6 project definitions are exposed in the assumptions and definitions ledger. Zero direct premises does not mean unrestricted model objects.}
\begin{dmath}[compact,label={aux:finite-bound-integral}]
I = \int_{0}^{1}h(z)\,\mathrm dz = \int_{0}^{1}g(z)\,\mathrm dz = \frac{73}{60} - \frac{31\,\pi}{80},\qquad I < 0,
\end{dmath}
\end{formalized}%
\noindent
so the composite midpoint estimate and
Eq.~\eqref{eq:finite-bound-derivative-bounds} prove negativity for large
\(a\):
{}
\begin{formalized}{eq:finite-bound-tail-estimates}{Rendered from Lean 4 theorem RmtNlin.manuscriptFiniteBoundTailEstimates (RmtNlin/FiniteTemperatureTransfer.lean:1517). 2 direct premises; rendered forms and their uses are in the generated assumption ledger. 18 project definitions are exposed in the assumptions and definitions ledger. Zero direct premises does not mean unrestricted model objects.}
\begin{equation}\begin{aligned}
F(a_{F}) &\le \frac{a_{F}}{2}\,\left(\frac{73}{60} - \frac{31\,\pi}{80}\right) + \frac{17}{20\,a_{F}}\,\left(1 + \frac{2}{a_{F}}\right) < 0,\qquad a_{F} \ge 51, \\
G(a_{G}) &\le \frac{a_{G}}{2}\,\left(\frac{73}{60} - \frac{31\,\pi}{80}\right) + \frac{11}{2\,a_{G}} < 0,\qquad a_{G} \ge 127.
\end{aligned}\label{eq:finite-bound-tail-estimates}\end{equation}
\end{formalized}%
\noindent
The estimate for \(G\) uses the derivative of the zero extension of \(g\),
whose total variation, the summed size of all its increases and decreases, is
at most \(22\).

On \(1\le a\le3\), \(F\) contains one nonpositive term, and the same holds for
\(G\) on \(3\le a\le5\). The executable certificate partitions \([3,51]\) into
rational cells of width \(1/512\) and \([5,127]\) into cells of width
\(1/64\). It evaluates every midpoint exactly and uses the middle two bounds in
Eq.~\eqref{eq:finite-bound-derivative-bounds} to cover the complete cells.
Every resulting rational upper bound is strictly negative, which proves
Eq.~\eqref{eq:finite-bound-tail-properties}.

Applying Eq.~\eqref{eq:finite-bound-bellman-result} to every channel and
restoring the Matsubara scale yields
{}
\begin{formalized}{eq:finite-bound-moment-ratio}{Rendered from Lean 4 theorem RmtNlin.AllChannels.manuscriptPhysicalFiniteBoundMomentRatio (RmtNlin/AllChannelViews.lean:113). 4 direct premises; rendered forms and their uses are in the generated assumption ledger. Additional axioms: RmtNlin.FiniteTemperatureCertificate.derivativeSum_compact_certificate._native.native_decide.ax_1_1, RmtNlin.FiniteTemperatureCertificate.finiteSum_compact_certificate._native.native_decide.ax_1_1 98 project definitions are exposed in the assumptions and definitions ledger. Zero direct premises does not mean unrestricted model objects.}
\begin{dmath}[compact,label={eq:finite-bound-moment-ratio}]
\sum_{n}\sum_{m=0}^{\infty}T_{n}(\omega_{m})^{2} \le \frac{29}{40}\,\sum_{n}\sum_{m=0}^{\infty}T_{n}(\omega_{m}).
\end{dmath}
\end{formalized}%
\noindent
Together with Eq.~\eqref{eq:coefficient-expansion}, this gives
{}
\begin{formalized}{eq:finite-temperature-bound-conclusion}{Rendered from Lean 4 theorem RmtNlin.AllChannels.manuscriptPhysicalFiniteTemperatureBoundConclusion (RmtNlin/AllChannelViews.lean:125). 5 direct premises; rendered forms and their uses are in the generated assumption ledger. Additional axioms: RmtNlin.FiniteTemperatureCertificate.derivativeSum_compact_certificate._native.native_decide.ax_1_1, RmtNlin.FiniteTemperatureCertificate.finiteSum_compact_certificate._native.native_decide.ax_1_1 98 project definitions are exposed in the assumptions and definitions ledger. Zero direct premises does not mean unrestricted model objects.}
\begin{dmath}[compact,label={eq:finite-temperature-bound-conclusion}]
-1 + \frac{3}{2}\,\frac{\sum_{n}\sum_{m=0}^{\infty}T_{n}(\omega_{m})^{2}}{\sum_{n}\sum_{m=0}^{\infty}T_{n}(\omega_{m})} \le -1 + \frac{3}{2}\,\frac{29}{40} = \frac{7}{80},
\end{dmath}
\end{formalized}%
\noindent
for \(0<\beta<\infty\), which is
Eq.~\eqref{eq:finite-temperature-normalized-quartic-bound}.

\section{Development workflow and use of AI agents}\label{appendix-development-workflow}

This appendix records how the results came about, which contributions came
from AI models and which from me, and which checks support the claims. I formulated the
problem, chose which results to pursue and keep, and reviewed the derivations
and the full text. I take full responsibility for the validity of the results.

\subsection{History of the project}

\begin{figure}[t]
\centering
\noindent\includegraphics[width=\linewidth]{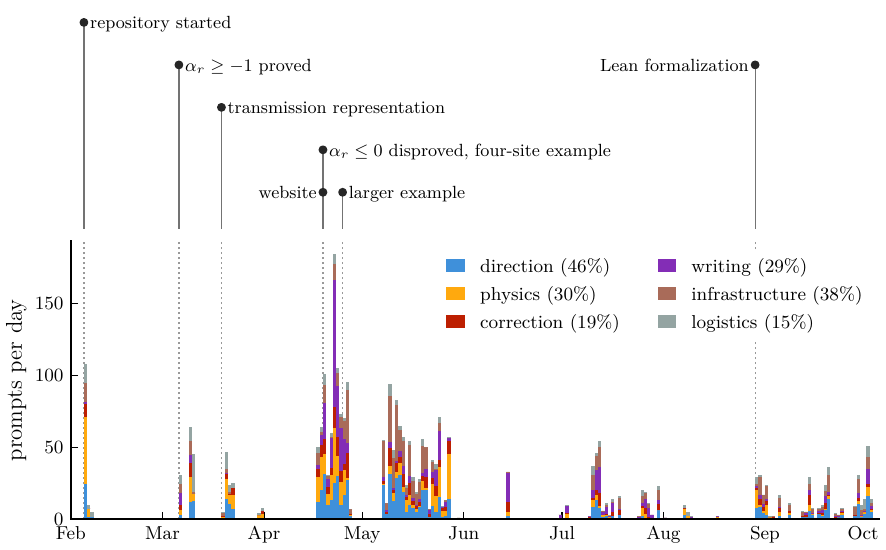}%
\caption{Timeline of the project and of my prompts to the coding agents. The
top row marks the steps described in this appendix on the day when each
result or tool entered the repository or the recorded sessions; dashed lines
repeat these days below. The bar height is the number of prompts per day.
Each bar is divided among the six labels in proportion to how many of that
day's prompts carry each label; since a prompt can carry several labels, this
division is only indicative. The legend gives the share of all prompts that
carry each label. The prompt labels are approximate: GPT-6-luna assigned them
automatically to the \(2\,863\) prompts recorded from 5 February to 3 October,
without validation against hand labels.}
\label{fig:development-timeline}
\end{figure}

The problem comes from our earlier numerical study \citep{Fatemi25}, which
conjectured \(-1\le\alpha_r\le0\). Before February 2026, I used the proof of
this conjecture as a benchmark of the mathematical ability of language models,
and GPT-5.2 made no progress on it. In February 2026, I started a repository
whose instruction file asked a coding agent to construct analytic proofs, to
pursue several approaches in parallel, to validate each step numerically, and
to keep a running log. The agent explored trace-logarithm inequalities, Schur
complements, and pinching arguments, and it recorded most of them as dead ends.
Figure~\ref{fig:development-timeline} places the following steps on a time
axis.

GPT-5.4 gave the first complete proof of the lower bound \(\alpha_r\ge-1\) in
early March, as a sum-of-squares identity that occupied more than a thousand
lines of notes. I gave these notes to GPT-5.4 Pro and asked it to prove the
missing upper bound. Its first answer rested on a lemma that it reported to
hold numerically to machine precision; when I asked whether the lemma was
true, the model refuted it with an explicit example. The model then observed
numerically that the nonzero roots of the Matsubara characteristic polynomial,
as a function of \(e^{i\varphi}\), lie on the negative real axis, and it proved
this observation. The result is the first version of the transmission
representation, from which the lower bound follows in a few lines. This result
convinced me that the problem deserved a complete study rather than serving
only as a benchmark. A coding agent then reworked the derivation under my
guidance into the form of Sec.~\ref{derivation-of-the-transmission-representation}.

The upper bound \(\alpha_r\le0\) resisted about a dozen proof strategies in
March, including moment-problem, Padé, thermal-interpolation, and
eigenvalue-monotonicity arguments. For part of this search, I asked an agent to
take over my role: to launch exploratory subagents, evaluate their results, and
choose new directions. In April, I asked GPT-5.4 Pro to prove the upper bound,
stated as the sign of a frequency integral. The model reduced the integral to
the transmission moments that determine \(\alpha_r\) and, after several
unproductive attempts, reported that the bound is false. It supported this
claim with a violation found by a numerical search over a structured family of
junctions, a result that came as a complete surprise to me. Within a day, a coding
agent confirmed the positive \(c_4\) by perturbation theory, by the
transmission-moment identity, and by a direct fit of the energy. When I asked
whether a smaller junction suffices, GPT-5.4 Pro found the four-site example of
Sec.~\ref{sns-junction-with-a-positive-quartic-nonlinearity}; the coding agent
brought it to the square form shown there and computed its coefficients
exactly. I then removed
the exploratory code and notes, about \(24\,000\) lines, and redefined the
project goal around the retained results. Later in April, I asked GPT-5.5
Pro to improve the upper bound. When this failed, I asked it to push
\(\alpha_r\) higher instead, and it constructed the larger example of
Sec.~\ref{sns-junction-with-a-positive-quartic-nonlinearity}.

Pro models also found the tighter upper bound of
Sec.~\ref{zero-temperature-upper-bound}, and GPT-6 Pro later produced the
unverified improvements of this bound mentioned in Sec.~\ref{conclusion}. Coding agents derived the
short-junction expansion to next-to-leading order in
App.~\ref{appendix-next-to-leading-short-junction-expansion}, built the
numerical checks, and wrote the formal proofs described in
App.~\ref{appendix-verification}.

\subsection{Agents and working practice}

The work used two kinds of models. In the ChatGPT interface, the Pro versions of
GPT-5.4 through GPT-6 received self-contained mathematical questions and
produced the three results above that I consider hardest. Coding agents worked
inside the repository: Codex with GPT-5.2 through GPT-6.1 throughout, and
Claude Code with Claude Fable~5.1 and Claude Opus~5.5 from September. These
agents carried out derivations, wrote numerical and verification code,
formalized the proofs, and wrote most of the text.

From April on, a project website showed the current state of the work.
Continuous integration rebuilt it after every change: first an executable
version of the manuscript with every number recomputed from the code, and later
the PDF and HTML versions of the paper together with pages generated from the
Lean formalization that list each claim with its hypotheses. I reviewed the
work on this website, reading the results as built from the repository. The
review tools of App.~\ref{appendix-writing-review} also run on this website.

Agent sessions were long and often ran in parallel in separate copies of the
repository. In the final stage, one agent coordinated others: it assigned
scoped tasks, requested reviews from a model of a different vendor, and merged
the results. Between February and October 2026, the repository accumulated
about \(1\,550\) commits, and I wrote about \(2\,900\) prompts in the recorded
agent sessions (Fig.~\ref{fig:development-timeline}). The Zenodo record \citep{Akhmerov26Reproduction} contains
the 192 recorded agent sessions, from February on, and the ChatGPT
conversations that produced the key results, with private data removed on a
best-effort basis. My own time is hard to estimate, because I usually
worked on several tasks in parallel, except during a period of detailed
feedback on the text. I estimate it at several weeks of full-time work.

Most of my prompts set directions, judged the physics, or corrected agent
errors. An approximate automatic classification assigns about 45\%, 30\%,
and 20\% of the prompts to these three kinds
(Fig.~\ref{fig:development-timeline}). The errors followed a few recurring patterns. Agents restated the
target as an assumption, presented unverified claims with confidence, made sign
errors in symmetry constraints, and wrote checks that passed without testing
the intended statement. For example, an early cross-check with Kwant used Kwant
only to diagonalize a finite matrix and did not solve the scattering problem. I
found most of these errors by following the derivations myself. The errors
became rarer with each model generation, and by the end of the project the
results were mostly reliable once I had formulated the task correctly. What did
not change is that a model's own judgment of whether it had solved a task was
not reliable, especially for the hardest tasks. Checking that a result matches
the task as I described it therefore remained a continuous part of my work.

\subsection{Verification}\label{appendix-verification}

By the time I started building verification tools, I had followed every
derivation and was confident in the results. I built the tools to raise the
standard of verifiability, so that a reader can trust the results without
repeating my checks. Each generation of tools raised this
standard. Executable notes first recomputed every reported number. A symbolic
layer then checked individual derivation steps with SymPy, a compromise between
rigor and effort. While building this layer, agents repeatedly satisfied a
check without meeting its intent; in one case, the printing code contained the
derivation that the check was supposed to establish. In August, I asked an agent
to attempt a formalization in Lean~4 \citep{Moura21} with Mathlib
\citep{Community20}. Lean checks complete statements together with their
hypotheses rather than isolated algebraic steps, and it replaced the symbolic
layer within days.

In the final pipeline, the manuscript source contains code cells that the build
runs in a Jupyter kernel \citep{Kluyver16} before typesetting; jupytertex, a
LaTeX package in the repository, replaces them with their output. These cells
print every display equation from the statement of a checked Lean
declaration. The printer maps Lean identifiers to symbols
through a single notation table. Apart from a few sign and ordering conventions, such as
printing \(x-(-y)\) as \(x+y\), it performs no algebraic rewriting, so each
display shows the checked statement itself. The printer is Python code that
Lean does not check. The build fails when a manuscript
label has no Lean declaration or a registered declaration does not appear in
the manuscript. Each main result lists its hypotheses, classified as model
assumptions, physical conditions, or technical conditions. Because the printer
can show distinct objects with the same symbol, an audit records for every such
pair either a proof that the objects are equal or a confirmation that they are
distinct. The proofs contain no unproved steps and add no axioms. Only the
rational certificate of App.~\ref{finite-temperature-moment-bound} relies on
compiled evaluation, which trusts the Lean compiler in addition to its kernel.

The formal proofs cover the displayed statements under their listed hypotheses.
These hypotheses include the energy-independent short-junction model of
Sec.~\ref{short-junction-limit} and regularity conditions on the normal-state
scattering in App.~\ref{appendix-next-to-leading-short-junction-expansion}.
The formalization works with finite matrices. In
App.~\ref{appendix-next-to-leading-short-junction-expansion}, each
semi-infinite lead enters only through its normal self-energy, which the
formal model takes as an input with the stated regularity; neither the
reduction of the infinite lead to this self-energy nor the identification of
the relative determinant with the thermodynamic limit of finite leads is
formalized. The figures and the comparisons with numerical simulations rely on Python code
using NumPy \citep{Harris20}, SciPy \citep{Virtanen20}, SymPy
\citep{Meurer17}, mpmath \citep{mpmath}, Matplotlib \citep{Hunter07},
Pymablock \citep{ArayaDay25}, and Kwant \citep{Groth14}. Only review checks the
prose between the displays.

Verification dominates the size of the code. The repository contains about
\(33\,000\) lines of Lean and \(8\,000\) lines of publication and audit code,
but only about \(1\,000\) lines of physics numerics. Each generation of tools
replaced the previous one, and about \(40\,000\) lines of discarded tooling
did not enter the final version.

\subsection{Writing and review}\label{appendix-writing-review}

Agents wrote most of the text from my instructions on its content and
structure and revised it in response to my comments, and I reviewed all of it.
To make this review efficient, I built a commenting tool for the manuscript on the project website that
attaches each comment to the corresponding source lines and passes both to an
agent. I also wrote my writing guidelines as reusable agent instructions, had
models of a different vendor review the text, and added a linter for display
punctuation and cross-references. A typical revision cycle consisted of my
comments, an agent revision, an independent agent review, and my final reading.
With later model generations, the drafts needed far fewer of my comments.
The amount of revision was nevertheless large: over the course of the project,
the agents wrote about four times as many words of manuscript text as the final
version contains.

\iflatexml

\else
    \bibliographystyle{SciPost_bibstyle}
    \bibliography{manuscript}

\begin{thebibliography}{10}
\providecommand{\url}[1]{\texttt{#1}}
\providecommand{\urlprefix}{URL }
\expandafter\ifx\csname urlstyle\endcsname\relax
  \providecommand{\doi}[1]{doi:\discretionary{}{}{}#1}\else
  \providecommand{\doi}{doi:\discretionary{}{}{}\begingroup
  \urlstyle{rm}\Url}\fi
\providecommand{\eprint}[2][]{\href{https://arxiv.org/abs/#2}{arXiv:#2}}

\bibitem{Josephson62}
B.~D. Josephson,
\newblock \emph{Possible new effects in superconductive tunnelling},
\newblock Phys. Lett. \textbf{1}(7), 251 (1962),
\newblock \doi{10.1016/0031-9163(62)91369-0}.

\bibitem{Beenakker91Universal}
C.~W.~J. Beenakker,
\newblock \emph{Universal limit of critical-current fluctuations in mesoscopic
  {Josephson} junctions},
\newblock Phys. Rev. Lett. \textbf{67}(27), 3836 (1991),
\newblock \doi{10.1103/physrevlett.67.3836}.

\bibitem{Larsen15}
T.~W. Larsen, K.~D. Petersson, F.~Kuemmeth, T.~S. Jespersen, P.~Krogstrup,
  J.~Nyg{\aa}rd and C.~M. Marcus,
\newblock \emph{Semiconductor-nanowire-based superconducting qubit},
\newblock Phys. Rev. Lett. \textbf{115}(12), 127001 (2015),
\newblock \doi{10.1103/physrevlett.115.127001}.

\bibitem{Lange15}
G.~de~Lange, B.~van Heck, A.~Bruno, D.~J. van Woerkom, A.~Geresdi, S.~R.
  Plissard, E.~P. A.~M. Bakkers, A.~R. Akhmerov and L.~DiCarlo,
\newblock \emph{Realization of microwave quantum circuits using hybrid
  superconducting-semiconducting nanowire {Josephson} elements},
\newblock Phys. Rev. Lett. \textbf{115}(12), 127002 (2015),
\newblock \doi{10.1103/physrevlett.115.127002}.

\bibitem{Kringhj18}
A.~Kringh{\o}j, L.~Casparis, M.~Hell, T.~W. Larsen, F.~Kuemmeth, M.~Leijnse,
  K.~Flensberg, P.~Krogstrup, J.~Nyg{\aa}rd, K.~D. Petersson and C.~M. Marcus,
\newblock \emph{Anharmonicity of a superconducting qubit with a few-mode
  {Josephson} junction},
\newblock Phys. Rev. B \textbf{97}(6), 060508 (2018),
\newblock \doi{10.1103/physrevb.97.060508}.

\bibitem{Furusaki91}
A.~Furusaki and M.~Tsukada,
\newblock \emph{Current-carrying states in {Josephson} junctions},
\newblock Phys. Rev. B \textbf{43}(13), 10164 (1991),
\newblock \doi{10.1103/physrevb.43.10164}.

\bibitem{Titov01}
M.~Titov, P.~Jacquod and C.~W.~J. Beenakker,
\newblock \emph{Negative superfluid density: Mesoscopic fluctuations and
  reverse of the supercurrent through a disordered {Josephson} junction},
\newblock Phys. Rev. B \textbf{65}(1), 012504 (2001),
\newblock \doi{10.1103/physrevb.65.012504},
\newblock \eprint{cond-mat/0106018}.

\bibitem{Fatemi25}
V.~Fatemi, P.~Kurilovich, A.~R. Akhmerov and B.~van Heck,
\newblock \emph{Nonlinearity of transparent {SNS} weak links decreases sharply
  with length},
\newblock SciPost Phys. \textbf{18}(3), 091 (2025),
\newblock \doi{10.21468/scipostphys.18.3.091},
\newblock \eprint{2410.01913}.

\bibitem{Moura21}
L.~de~Moura and S.~Ullrich,
\newblock \emph{The {Lean} 4 theorem prover and programming language},
\newblock In \emph{Automated Deduction -- {CADE} 28}, vol. 12699 of
  \emph{Lecture Notes in Computer Science}, pp. 625--635. Springer
  International Publishing,
\newblock \doi{10.1007/978-3-030-79876-5\_37} (2021).

\bibitem{Community20}
{The mathlib Community},
\newblock \emph{The {Lean} mathematical library},
\newblock In \emph{Proceedings of the 9th ACM SIGPLAN International Conference
  on Certified Programs and Proofs}, pp. 367--381,
\newblock \doi{10.1145/3372885.3373824} (2020).

\bibitem{Waintal24}
X.~Waintal, M.~Wimmer, A.~Akhmerov, C.~Groth, B.~K. Nikolic, M.~Istas,
  T.~{\"O}. Rosdahl and D.~Varjas,
\newblock \emph{Computational quantum transport: a scattering approach
  perspective}  (2024),
\newblock \eprint{2407.16257}.

\bibitem{Brouwer97}
P.~W. Brouwer and C.~W.~J. Beenakker,
\newblock \emph{Anomalous temperature dependence of the supercurrent through a
  chaotic {Josephson} junction},
\newblock Chaos, Solitons \& Fractals \textbf{8}(7-8), 1249 (1997),
\newblock \doi{10.1016/s0960-0779(97)00018-0}.

\bibitem{Repin19}
E.~V. Repin, Y.~Chen and Y.~V. Nazarov,
\newblock \emph{Topological properties of multiterminal superconducting
  nanostructures: Effect of a continuous spectrum},
\newblock Phys. Rev. B \textbf{99}(16), 165414 (2019),
\newblock \doi{10.1103/physrevb.99.165414}.

\bibitem{Zaitsev84}
A.~V. Zaitsev,
\newblock \emph{Quasiclassical equations of the theory of superconductivity for
  contiguous metals and the properties of constricted microcontacts},
\newblock Sov. Phys. JETP \textbf{59}(5), 1015 (1984).

\bibitem{Yip03}
S.~K. Yip,
\newblock \emph{Supercurrent and noise in point contacts between two different
  superconductors},
\newblock Phys. Rev. B \textbf{68}(2), 024511 (2003),
\newblock \doi{10.1103/physrevb.68.024511}.

\bibitem{Levchenko06}
A.~Levchenko, A.~Kamenev and L.~Glazman,
\newblock \emph{Singular length dependence of critical current in
  superconductor/normal-metal/superconductor bridges},
\newblock Phys. Rev. B \textbf{74}(21), 212509 (2006),
\newblock \doi{10.1103/physrevb.74.212509},
\newblock \eprint{cond-mat/0601177}.

\bibitem{Beenakker23}
C.~W.~J. Beenakker,
\newblock \emph{{Casimir}-{Josephson} force on a point contact between two
  superconductors},
\newblock Phys. Rev. B \textbf{108}(19), 195422 (2023),
\newblock \doi{10.1103/physrevb.108.195422},
\newblock \eprint{2310.06672}.

\bibitem{Levchenko25}
A.~Levchenko,
\newblock \emph{{Casimir} effect in {Josephson} junctions},
\newblock J. Low Temp. Phys. \textbf{218}(3-4), 268 (2025),
\newblock \doi{10.1007/s10909-024-03254-3},
\newblock \eprint{2410.08398}.

\bibitem{ArayaDay25}
I.~Araya~Day, S.~Miles, H.~Kerstens, D.~Varjas and A.~R. Akhmerov,
\newblock \emph{{Pymablock}: An algorithm and a package for quasi-degenerate
  perturbation theory},
\newblock SciPost Phys. Codebases \textbf{50} (2025),
\newblock \doi{10.21468/scipostphyscodeb.50}.

\bibitem{Akhmerov26Reproduction}
A.~R. Akhmerov,
\newblock \emph{Source code and agent logs for ``transmission representation of
  the {Josephson} free energy beyond the short-junction limit''},
\newblock \doi{10.5281/zenodo.21705015} (2026).

\bibitem{Li26}
Z.~C.~F. Li, Y.~Deng, D.~K. Efetov and K.~T. Law,
\newblock \emph{Quantum metric induced critical current anomaly in flat band
  {Josephson} junctions}  (2026),
\newblock \eprint{2608.08120}.

\bibitem{Yip92}
S.~K. Yip and J.~A. Sauls,
\newblock \emph{Nonlinear {Meissner} effect in {CuO} superconductors},
\newblock Phys. Rev. Lett. \textbf{69}(15), 2264 (1992),
\newblock \doi{10.1103/physrevlett.69.2264}.

\bibitem{Clem12}
J.~R. Clem and V.~G. Kogan,
\newblock \emph{Kinetic impedance and depairing in thin and narrow
  superconducting films},
\newblock Phys. Rev. B \textbf{86}(17) (2012),
\newblock \doi{10.1103/physrevb.86.174521}.

\bibitem{Zmuidzinas12}
J.~Zmuidzinas,
\newblock \emph{Superconducting microresonators: physics and applications},
\newblock Annu. Rev. Condens. Matter Phys. \textbf{3}(1), 169 (2012),
\newblock \doi{10.1146/annurev-conmatphys-020911-125022}.

\bibitem{Eom12}
B.~H. Eom, P.~K. Day, H.~G. LeDuc and J.~Zmuidzinas,
\newblock \emph{A wideband, low-noise superconducting amplifier with high
  dynamic range},
\newblock Nature Phys \textbf{8}(8), 623 (2012),
\newblock \doi{10.1038/nphys2356}.

\bibitem{Groth14}
C.~W. Groth, M.~Wimmer, A.~R. Akhmerov and X.~Waintal,
\newblock \emph{{Kwant}: a software package for quantum transport},
\newblock New J. Phys. \textbf{16}(6), 063065 (2014),
\newblock \doi{10.1088/1367-2630/16/6/063065}.

\bibitem{Ball15}
J.~A. Ball, M.~Kurula, O.~J. Staffans and H.~Zwart,
\newblock \emph{{De Branges}–{Rovnyak} realizations of operator-valued
  {Schur} functions on the complex right half-plane},
\newblock Complex Anal. Oper. Theory \textbf{9}(4), 723 (2015),
\newblock \doi{10.1007/s11785-014-0358-2}.

\bibitem{Anderson06}
J.~M. Anderson and J.~Rovnyak,
\newblock \emph{On generalized {Schwarz}‐{Pick} estimates},
\newblock Mathematika \textbf{53}(1), 161 (2006),
\newblock \doi{10.1112/s0025579300000085}.

\bibitem{Kluyver16}
T.~Kluyver, B.~Ragan-Kelley, F.~P{\'e}rez, B.~Granger, M.~Bussonnier,
  J.~Frederic, K.~Kelley, J.~Hamrick, J.~Grout, S.~Corlay, P.~Ivanov, D.~Avila
  \emph{et~al.},
\newblock \emph{Jupyter notebooks -- a publishing format for reproducible
  computational workflows},
\newblock In F.~Loizides and B.~Schmidt, eds., \emph{Positioning and Power in
  Academic Publishing: Players, Agents and Agendas}, pp. 87--90. IOS Press,
\newblock \doi{10.3233/978-1-61499-649-1-87} (2016).

\bibitem{Harris20}
C.~R. Harris, K.~J. Millman, S.~J. van~der Walt, R.~Gommers, P.~Virtanen,
  D.~Cournapeau, E.~Wieser, J.~Taylor, S.~Berg, N.~J. Smith, R.~Kern, M.~Picus
  \emph{et~al.},
\newblock \emph{Array programming with {NumPy}},
\newblock Nature \textbf{585}(7825), 357 (2020),
\newblock \doi{10.1038/s41586-020-2649-2}.

\bibitem{Virtanen20}
P.~Virtanen, R.~Gommers, T.~E. Oliphant, M.~Haberland, T.~Reddy, D.~Cournapeau,
  E.~Burovski, P.~Peterson, W.~Weckesser, J.~Bright, S.~J. van~der Walt,
  M.~Brett \emph{et~al.},
\newblock \emph{{SciPy} 1.0: fundamental algorithms for scientific computing in
  {Python}},
\newblock Nat Methods \textbf{17}(3), 261 (2020),
\newblock \doi{10.1038/s41592-019-0686-2}.

\bibitem{Meurer17}
A.~Meurer, C.~P. Smith, M.~Paprocki, O.~{\v{C}}ert{\'\i}k, S.~B. Kirpichev,
  M.~Rocklin, A.~Kumar, S.~Ivanov, J.~K. Moore, S.~Singh, T.~Rathnayake, S.~Vig
  \emph{et~al.},
\newblock \emph{{SymPy}: symbolic computing in {Python}},
\newblock PeerJ Computer Science \textbf{3}, e103 (2017),
\newblock \doi{10.7717/peerj-cs.103}.

\bibitem{mpmath}
{The mpmath development team},
\newblock \emph{mpmath: a {Python} library for arbitrary-precision
  floating-point arithmetic},
\newblock \url{https://mpmath.org/} (2026).

\bibitem{Hunter07}
J.~D. Hunter,
\newblock \emph{Matplotlib: A 2d graphics environment},
\newblock Comput. Sci. Eng. \textbf{9}(3), 90 (2007),
\newblock \doi{10.1109/mcse.2007.55}.

\end{thebibliography}
\fi

\end{document}